\documentclass[prd, twocolumn, amsmath, floatfix, preprintnumbers, nofootinbib,superscriptaddress,letterpaper]{revtex4-2}

\usepackage[toc,page]{appendix}
\usepackage{graphicx}
\usepackage{amsfonts}
\usepackage{amsmath}
\usepackage{amssymb}
\usepackage{multirow}
\usepackage{enumerate}
\usepackage[normalem]{ulem}
\usepackage[usenames]{xcolor}
\usepackage[colorlinks = true, linkcolor = blue, urlcolor  = blue, citecolor = blue, anchorcolor = blue]{hyperref}

\usepackage{float}
\usepackage{xspace}
\usepackage{pifont}

\usepackage{orcidlink}

\newcommand{\xmark}{\ding{55}}

\begin{document}
\title{The Gravitational-Wave Signature of Core-Collapse Supernovae}

\author{David Vartanyan\,\orcidlink{0000-0003-1938-9282}}
\email{dvartanyan@carnegiescience.edu}
\affiliation{Carnegie Observatories, 813 Santa Barbara St, Pasadena, CA 91101, USA; NASA Hubble Fellow}
\author{Adam Burrows\,\orcidlink{0000-0002-3099-5024}}
\affiliation{Department of Astrophysical Sciences, 4 Ivy Lane, Princeton University, Princeton, NJ 08544, USA}
\author{Tianshu Wang\,\orcidlink{0000-0002-0042-9873}}
\affiliation{Department of Astrophysical Sciences, 4 Ivy Lane, Princeton University, Princeton, NJ 08544, USA}
\author{Matthew S.B. Coleman\,\orcidlink{0000-0001-5939-5957}}
\affiliation{Department of Astrophysical Sciences, 4 Ivy Lane, Princeton University, Princeton, NJ 08544, USA}
\affiliation{Department of Physics and Engineering Physics, Stevens Institute of Technology,
Castle Point on the Hudson, Hoboken, NJ 07030, USA}
\author{Christopher J. White\,\orcidlink{0000-0002-3099-5024}}
\affiliation{Department of Astrophysical Sciences, 4 Ivy Lane, Princeton University, Princeton, NJ 08544, USA}
\affiliation{Center for Computational Astrophysics, Flatiron Institute, 162 5th Ave., New York, NY 10010, USA}

\begin{abstract}
 We calculate the gravitational-wave (GW) signatures
of detailed 3D core-collapse supernova simulations spanning a range of massive stars. Most of the simulations are carried out to times late enough to capture more than 95\% of the total GW emission. We find that the f/g-mode and f-mode of proto-neutron star oscillations carry away most of the GW power. The f-mode frequency inexorably rises as the proto-neutron star (PNS) core shrinks. We demonstrate that the GW emission is excited mostly by accretion plumes onto the PNS that energize modal oscillations and also high-frequency (``haze") emission correlated with the phase of violent accretion. The duration of the major phase of emission varies with exploding progenitor and there is a strong correlation between the total GW energy radiated and the compactness of the progenitor. Moreover, the total GW emissions vary by as much as three orders of magnitude from star to star. For black-hole formation, the GW signal tapers off slowly and does not manifest the haze seen for the exploding models. For such failed models, we also witness the emergence of a spiral shock motion that modulates the GW emission at a frequency near $\sim$100 Hertz that slowly increases as the stalled shock sinks. We find significant angular anisotropy of both the high- and low-frequency (memory) GW emissions, though the latter have very little power.
\end{abstract} 
\maketitle

\label{sec:int}

The theory of core-collapse supernova (CCSN) explosions
has been developed over the last six decades and is now a mature field at the interface of gravitational, particle, nuclear, statistical, and numerical physics. The majority of explosions are thought to be driven by neutrino heating behind a shock wave formed upon the collision of the rebounding inner core with the infalling mantle of the Chandrasekhar mass birthed in the center of stars more massive than $\sim$8 M$_{\odot}$\citep{CoWh66,Wilson1985,janka2012,burrows2013,2021Natur.589...29B}. After implosion ensues, this inner white dwarf core, with a mass near $\sim$ 1.5M$_{\odot}$ and a radius of only a few thousand kilometers, requires only hundreds of milliseconds of implosion to achieve a central density above that of the atomic nucleus.  At this point, the inner core stiffens, rebounds, and collides with the outer core, thereby generating a shock wave that should be the supernova explosion in its infancy.  However, detailed 3D simulations
\citep{takiwaki:12,hanke_13,lentz:15,burrows:16,melson:15a,melson:15b,TaKoSu16,roberts:16,ott2018_rel,muller2017, 2018MNRAS.477L..80K, vartanyan2018b,summa2018, oconnor_couch2018b, glas2019, burrows_2020, 2022MNRAS.510.4689V} and physical understanding dictate that this shock generally stalls into accretion, but is often reenergized into explosion by heating via the neutrinos emerging from the hot, dense, accreting proto-neutron star (PNS), aided by the effects of vigorous neutrino-driven turbulent convection behind that shock \citep{herant:92,burrows:95,2015ApJ...799....5C,muller2017,hiroki_2019, 2022MNRAS.510.4689V}.
The delay to explosion can last another few hundred milliseconds,
after which the explosion is driven to an asymptotic state in a period of from $\sim$a few to $\sim$10 seconds. An extended period of neutrino heating seems required \citep{2021Natur.589...29B,muller2017,2021ApJ...915...28B}.
The shock wave then takes a minute to a day to emerge from the massive star, and this emergence inaugurates the brilliant electromagnetic display that is the supernova. The outcomes and timescales
depend upon the progenitor core density and thermal structure at the time of collapse \citep{wang, tsang2022}, which itself is an important function of progenitor mass, metallicity, and rotational profile.
If a black hole eventually forms, the core must still go through the PNS stage, and it is still possible to launch an explosion, even when a black hole is the residue.  There is never a direct collapse to a black hole. A small fraction of supernova (hypernovae?) may be driven by magnetic jets from the PNS if the cores are rotating at millisecond periods.  Otherwise, magnetic effects are generally subdominant, but of persistent interest in the context of pulsar and magnetar birth \citep{mosta2014,2020ApJ...896..102K,nagakura_pns, Obergaulinger2021, Aloy2021,2022ApJ...926..111W,2022arXiv221200200P}.

Though this scenario is buttressed by extensive simulation and theory, and most 3D (and 2D) models now explode without artifice \citep{summa2016, lentz:15,roberts:16,TaKoSu16,vartanyan2018a,vartanyan2018b,muller2017,glas2019, hiroki_2019,burrows_2020,2022MNRAS.510.4689V}, direct verification of the details and the timeline articulated above are difficult to come by. However, the neutrino and gravitational-wave (GW) signatures of this dynamical event would allow one to follow the theoretically expected sequence of events in real time.  The detection of 19 neutrinos from SN1987A was a landmark \citep{Bionta:1987qt,Hirata:1987hu}, but little was learned, other than that a copious burst of neutrinos, whose properties are roughly in line with theory, attends CCSN and the birth of a compact object.  The real-time witnessing of the events described as they unfold is the promise of GW detection from a supernova explosion. The detection of these GWs and the simultaneous detection of the neutrinos overlapping in time is the holy grail of the discipline.

The CCSN GW signal
\citep{murphy:09,yakunin:10,2013ApJ...766...43M,kuroda:14,yakunin:15,kuroda2016,andresen2017,muller2017b,kuroda2017,tk18,2018ApJ...861...10M,radice2019,2021MNRAS.502.3066S,mezzacappa2022} from bounce through supernova explosion and into the late-time (many-second) proto-neutron star (PNS) \citep{2006ApJ...645..534D,roberts_12c,nagakura_pns} cooling phase (or black hole formation) can be decomposed into various stages with characteristic features, frequency spectra, strains and polarizations \citep{2018hayama}. GWs are generated immediately around bounce in CCSN by time-dependent rotational flattening (if initially rotating, see e.g. \citep{pajkos2019}) and prompt post-shock overturning convection (lasting tens of milliseconds) \citep{burrows:95,marek:09}, then during a low phase (lasting perhaps $\sim$50$-$200 milliseconds)
during which post-shock neutrino-driven turbulence builds, followed by vigorous accretion-plume-energized PNS modal oscillations (predominantly a mixed f/g-mode early, then a pure f-mode later, see \S\ref{description}) \citep{murphy:09,2018ApJ...861...10M}. If an explosion ensues, these components are accompanied by low-frequency ($\sim$1$-$25 Hertz (Hz)) GW ``memory" (see \S\ref{memory}) due to asymmetric emission of neutrinos and aspherical explosive mass motions \citep{burrows1996,2020ApJ...901..108V,2022PhRvD.105j3008R,2022PhRvD.106d3020M}. The duration of the more quiescent phase between prompt overturning convection and vigorous turbulence depends upon the seed perturbations in the progenitor core \citep{muller2017}, of which the duration of this phase is diagnostic. Relevant for the GW signature is the equation of state (EOS) \citep{SuYaSu05,marek:09,pan:17,richers:17}, the rate of neutronization and neutrino cooling of the core \citep{burlat86,roberts_12c}, and the stellar core's initial angular momentum and mass density distributions.  Measurable GW signatures of rotation require particular progenitors that rotate fast, while all other phases/phenomena are expected to be operative in any core-collapse context, except matter memory, which requires an asymmetric explosion.  The rotational signature is primarily through its dependence upon the ratio of rotational to gravitational energy ($T/W$) \citep{kuroda:14,tk18,summa2018}, and, for fast rotating cores, to the degree of initial differential rotation. Black hole formation will approximately recapitulate the sequence followed by neutron star formation, except that when the black hole forms long after collapse the GW signal ceases abruptly \citep{1986ApJ...300..488B} and that due to the shrinking stalled shock radius a spiral mode is excited that quasi-periodically modulates its GW signal (see \S\ref{black}). The termination of the neutrino emission is correspondingly abrupt.  Though these phases are generic, their duration, strain magnitudes, and degree of stochastic variation and episodic bursting (due in part to episodic accretion and fallback) is a function of massive star progenitor density structure, degree of rotation, and the chaoticity of the turbulence.

Though neutrino-driven convection generally overwhelms manifestations of the ``SASI" (Standing Accretion Shock Instability \citep{blondin:03,foglizzo:07}), the SASI is sometimes discernible as a near 100$-$200 Hz subdominant component.  However, when the explosion aborts and the average shock radius sinks deeper below $\sim$100 km, the so-called ``spiral SASI" \citep{blondin_shaw,kuroda2016,vartanyan2019} emerges (\S\ref{black}).  This spiral rotational mode has a frequency of $\sim$100-200 Hz, is interestingly polarized, and if present clearly modulates both the neutrino and GW signatures until the general-relativistic instability that leads to black hole formation.

Therefore, each phase of a supernova has a range of characteristic signatures in GWs that can provide diagnostic constraints on the evolution and physical parameters of a CCSN and on the dynamics of the nascent PNS.  Core bounce and rotation, the excitation of core oscillatory modes, neutrino-driven convection, explosion onset, explosion asymmetries, the magnitude and geometry of mass accretion, and black hole formation all have unique signatures that, if measured, would speak volumes about the supernova phenomenon in real time.

For this paper, we have run a broad set of 3D simulations (11 progenitors in total, with progenitors ranging from 9 to 23 M$_{\odot}$) out to the late post-bounce times (up to $\sim$6 seconds post-bounce). These late-time 3D simulations illustrate the sustained multi-second GW signal and are the longest 3D core-collapse simulations with sophisticated neutrino transport performed to date. Simulations out to one second or less do not capture the entire time evolution of the GW signal. We find that the GW signal persists out to late times for all models, is strongly correlated with the turbulence interior to the shock (see \S\ref{excitation}), and is not correlated with proto-neutron star convection (\S\ref{pns_convection}). We see a memory signature at $\le$25 Hz associated with large-scale ejecta for our exploding models, and a spiral SASI signature at $\sim$100 Hz for the non-exploding models. All our models show an early prompt-convection phase at $\sim$50 milliseconds (ms) associated with a negative entropy gradient interior to the stalled shock front at $\sim$150 km.  For most models, this is followed by a quiescent phase of duration $\le$50 ms, after which the strain grows in association with turbulent motion interior to the stalled shock. The 23-M${\odot}$ model shows an interesting exception, as the strain illustrates another feature lasting from $\sim$175 to 350 ms, coincident with the shock receding before reviving. As the accretion rate and turbulence diminish, so too does the strain. However, at late times, we see a consistent offset in the strain associated with matter and neutrino memory in the exploding models (\S\ref{memory}). 

We now proceed to a more detailed discussion of our new results that span a broad progenitor mass range, capture for the first time and for most models the entire GW signal of CCSN, and do not suffer from Nyquist sampling problems. In this paper, we focus on initially non-rotating progenitors, whose general behavior should also encompass models for slowly rotating initial cores. We note that even initially non-rotating 3D CCSN models experience core spin up due to stochastic fallback \citep{coleman}, and this effect is a normal by product of sophisticated 3D simulations. In \S\ref{methods}, we provide information on the simulation suite and its characteristics and briefly describe the various models' hydrodynamic developments. Then, in \S\ref{results} we present our comprehensive set of findings concerning the complete GW signature of initially non-rotating CCSN, partitioned into subsections that each focus on a different aspect of this signature and its import. In subsection \S\ref{description}, we lay out the basic signal behaviors as a function of progenitor. This section contains our major results and then is followed in \S\ref{black} by a digression into the GW signature of black hole formation.  In \S\ref{energy}, we present an interesting finding concerning the dependence of the total radiated GW energy on compactness \citep{2011ApJ...730...70O} and in \S\ref{avoided} we note the avoided crossing that is universally manifest in all CCSN GW spectra and seems a consequence of the presence of inner proto-neutron star (PNS) convection and its growth with time. In \S\ref{anisotropy} we discuss the solid-angle dependence of the matter-sourced GW emission, both at high and low (matter ``memory") frequencies, and in \S\ref{memory} we present our results concerning neutrino memory at low frequencies. Then, we transition in \S\ref{excitation} to a discussion of the predominant  excitation mechanism. Finally, in \S\ref{conclusions} we recapitulate our basic findings and wrap up with some observations.

\section{Setup and Hydrodynamics Summary}
\label{methods}
We present in this paper a theoretical study of the GW emission of eleven CCSN progenitors, from 9 to 23 M$_{\odot}$, evolved in three dimensions using the radiation-hydrodynamic code F{\sc{ornax}} \citep{skinner2019}. To calculate the quadrupole tensor and the GW strains we employ the formalisms of references \citep{finn} and \citep{oohara97} (see also \S\ref{appendix}) and we dump these data at high cadences near the LIGO sampling rate (Table \ref{sn_tab}). The progenitor models were selected from \cite{sukhbold2018} for the 14-, 15.01-, and 23-M$_{\odot}$ models and from \cite{swbj16} for models between 9 and 12.25 M$_{\odot}$. The radial extent of the models spans 20,000 kilometers (km) to 100,000 km, generally increasing with progenitor mass. All the models (except the 9-M$_{\odot}$ model on Blue Waters, which had 648 radial zones) were run with 1024$\times$128$\times$256 cells in radius, $\theta$, and $\phi$. We employ 12 neutrino energy groups for each of the $\nu_e$, $\bar{\nu}_e$, and ``$\nu_{\mu}$"s followed (see \cite{burrows_2020,coleman}) and the SFHo equation of state \citep{2013ApJ...774...17S}. The progenitor models are non-rotating, though some degree of rotation is naturally induced due to fallback \citep{coleman}. These simulations include two of the longest 3D CCSN simulations run to date, a 11-M$_{\odot}$ model evolved past 4.5 seconds post-bounce, and a 23-M$_{\odot}$ model evolved to $\sim$6.2 seconds post-bounce.
All of our models explode except the 12.25- and 14-M$_{\odot}$ progenitors, and all models besides the 14-M$_{\odot}$ are evolved beyond one second post-bounce. We include four simulations of the 9-M$_{\odot}$ model on different high-performance clusters (Frontera, Theta, and Blue Waters) at various stages of code evolution. The low-mass 9-M$_{\odot}$ models asymptote early on in diagnostic quantities such as explosion energy and residual mass, and one iteration has been evolved past two seconds post-bounce. Our models, run times, and explosion outcome are summarized in Table\,\ref{sn_tab}. Several of the models have been published before, including three of the four 9-M$_{\odot}$ iterations (the fourth, the longest simulation, is new) in references \citep{burrows_2019}, \citep{burrows_2020}, and \citep{coleman}.

\begin{figure*}
    \centering
    \includegraphics[width=0.87\textwidth]{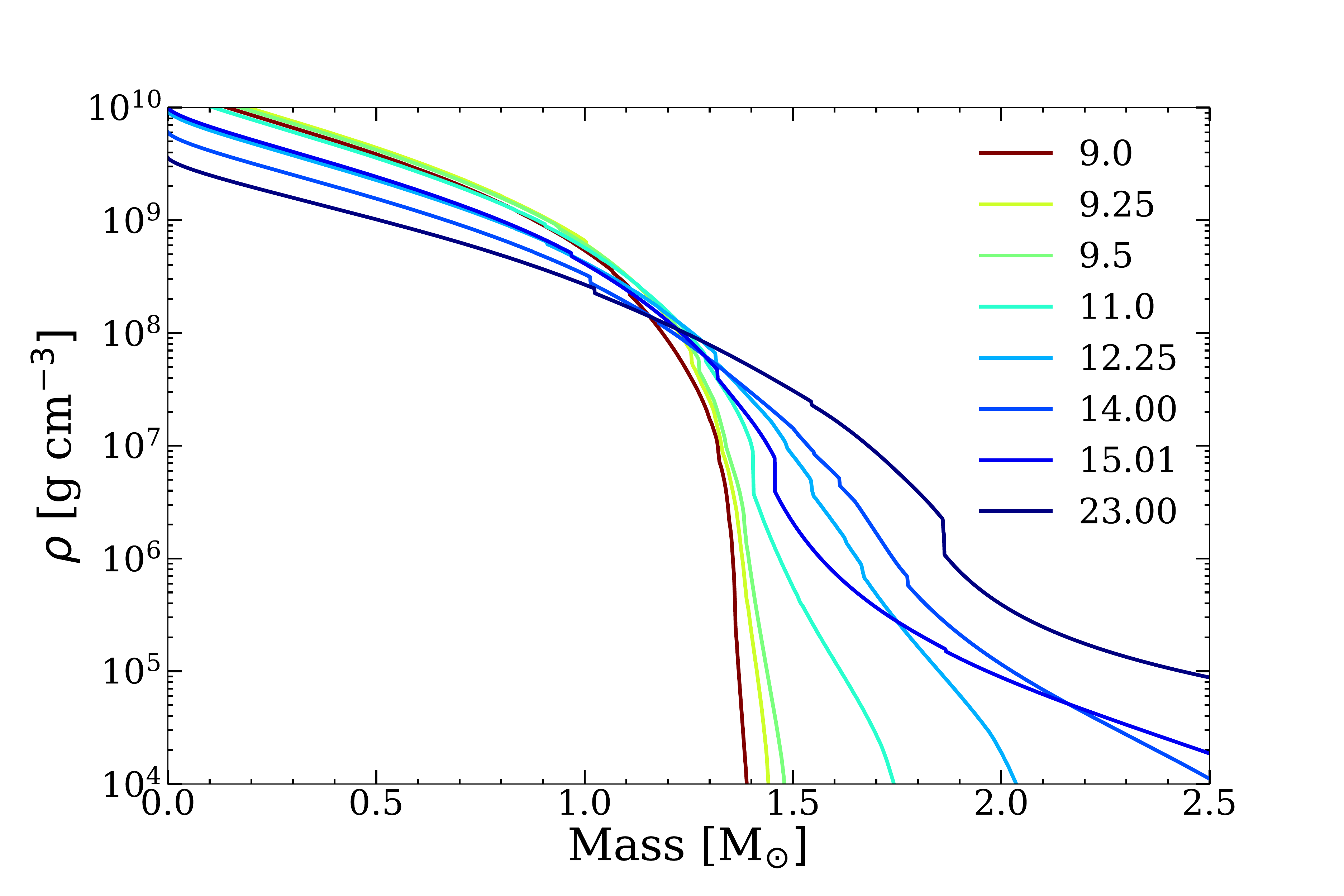}
    \caption{Mass density profiles for the various progenitors studied here, from 9.0 (reddish) to 23 (blue) M$_{\odot}$. The accretion of the Si/O interface frequently corresponds with the onset of shock revival. Note that the 12.25- and 14-M$_{\odot}$ models do not explode and are destined to produce a stellar-mass black hole.}
    \label{fig:rho}
\end{figure*}

In Figure\,\ref{fig:rho}, we plot the density profiles against enclosed mass for all eleven models studied here. Note the association of the silicon-oxygen (Si/O) interface density drop (see, e.g. \citep{fryer1999, ott2018_rel,vartanyan2018a,burrows_2020,2021ApJ...916L...5V,2021Natur.589...29B,tsang2022,wang}) with the onset of successful shock revival. The 14-M$_{\odot}$ and 12.25-M$_{\odot}$ progenitors lack such a strong interface and do not explode. Low-mass progenitors (e.g. 9-9.5 M$_{\odot}$) have a steep density profile and explode easily (though still with the aid of turbulent convection). For instance, compared to models 11-M$_{\odot}$, models 15.01-, and 23-M$_{\odot}$ have Si/O interfaces successively further out, and explode successively later.

We show the angle-averaged shock radii at early and late-times in Figure\,{\ref{fig:rs1}}. All models except the 12- and 14-M$_{\odot}$ models explode, with an approximate correlation between progenitor compactness \citep{2011ApJ...730...70O} and explosion time. The two non-exploding models experience $\sim$10 ms oscillations in the shock radii due to a spiral SASI that manifests itself after $\sim$350 ms.  We also see a longer secular timescale oscillation of $\sim$70 ms in the 12.25-M$_{\odot}$ and 14-M$_{\odot}$ black-hole formers. We summarize the eleven models in Table\,\ref{sn_tab}. The more massive 15.01- and 23-M$_{\odot}$ progenitors explode later, with the latter showing shock revival only after $\sim$0.5 seconds post-bounce. A later shock revival time, again $\sim$0.5 s, was also seen for the 25-M$_{\odot}$ progenitor in \cite{burrows_2020}. After the first $\sim$500 ms, the shock velocities settle into approximately asymptotic values that range from 7000 to 16000 km s$^{-1}$, inversely correlated crudely with the progenitor mass (see also \cite{wanajo2018}).

\section{Results}
\label{results}

\subsection{General Gravitational-Wave Signal Systematics of Core-Collapse Supernovae}\label{description}  

As stated, we highlight for this study of the GW signatures of CCSN eleven of our recent initially non-rotating 3D F{\sc{ornax}} simulations.  Care has been taken to calculate the quadrupole tensor with a high enough cadence to avoid Nyquist sampling problems, and we have been able to simulate to late enough times to capture what is effectively the entire GW signal after bounce for a large subset of the models.  Table \ref{sn_tab} provides the duration of each simulation and the minimum Nyquist frequency achieved during each run. We will discuss both matter and neutrino contributions to GW energy, and the relevant equations are summarized in Appendix\,\ref{appendix}.

In Figure \ref{fig:strain}, we plot the plus (black) and cross (blue) polarizations in the x-direction of our computational grid of the strain multiplied by the distance to the source. Other orientations yield qualitatively similar numbers for the higher frequency components that
dominate the GW power. However, there is a large variation at low frequencies ($\le$25 Hz) of the matter and neutrino memories with solid angle (see \S\ref{memory}).  Note that the x- and y-axes cover different ranges for each model.  A red star on the panels indicates the rough time of explosion, defined loosely. 
As Figure \ref{fig:strain} demonstrates, all the exploding models transition through similar phases. During the first $\sim$50 ms there is a burst of emission due to prompt overturning convection driven by the negative entropy gradient produced behind the shock wave as it stalls. The detailed time behavior of this overturn will depend on the initial accreted perturbations, which will set the number of e-folds to the non-linear phase.  However, the basic behavior and timescales are broadly similar. Figure \ref{fig:strain_early} focuses on this early first 0.25 seconds. For the lower-mass progenitors, explosion (the red star) ensues towards the end, or not long after, the prompt signal, and this is followed by the early growth of the second phase.  For the more massive progenitors (such as the 23-M$_{\odot}$ model), the onset of explosion can be much later. As Figure \ref{fig:strain} shows, the growth phase of the GW emission continues beyond what is shown in Figure \ref{fig:strain_early} to a strong peak. That peak phase is powered by the accretion of the infalling plumes during explosion. A core aspect of 3D core-collapse explosions is the breaking of spherical symmetry that allows simultaneous accretion in one direction and explosion in another \citep{vartanyan2019, burrows_2020,2021Natur.589...29B,wang}. For exploding models, the infalling plumes that strike the surface of the PNS can achieve supersonic speeds before impact. For the black hole formers (the 14-M$_{\odot}$ and 12.25 M$_{\odot}$ models here), the accretion is maintained, but impinges upon the PNS core subsonically.  This will have interesting consequences we discuss in \S\ref{black}.

Figures \ref{fig:strain} show that the lower-mass models have smaller strains and that the phase of high strain lasts for a shorter time.  For the 9-M$_{\odot}$ through 9.5-M$_{\odot}$ models, much of the GW emission subsides by $\sim$0.25-0.5 seconds, while the high phase lasts $\sim$1.2 seconds for the 23-M$_{\odot}$ model and continues beyond $\sim$1.0 and $\sim$1.5 seconds for the 11-M$_{\odot}$ and 15.01-M$_{\odot}$ models, respectively.  These differences reflect the differences in the initial density profiles (Figure \ref{fig:rho}) and the compactness (see also Figure\,\ref{fig:EGW_eta}).

After this vigorous phase, the pounding of the accretion plumes subsides, but the signal continues at a low amplitude. Though as much as $\sim$95\% of the GW  energy emission has already occurred, the f-mode continues to the latest times we have simulated as a low hum of progressively increasing frequency\footnote{Sonifications of the signals are available upon request.}. Hence, we see universally for the exploding models a transition from a high-amplitude, lower-frequency stage ($\le$0.3-1.5 seconds, depending upon the progenitor) to a lower-amplitude high-frequency stage ($\ge$1.5 seconds). As Figure \ref{fig:strain} indicates, for the exploding models a very-low frequency memory is superposed that represents a permanent metric strain. There is no such matter memory signal for the black-hole formers (\S\ref{memory}), but the accretion phase continues for them to very late times, abating only slowly as the mantle continues to accrete the mass of the outer mantle until the general-relativistic instability that leads to a black hole ensues. 

It has too often been thought that strain signals such as are depicted in Figures \ref{fig:strain} and \ref{fig:strain_early} are too noisy to be templated cleanly, and this to a degree is true.  There is a lot of stochasticity due to chaotic turbulence. However, the frequency content of these signals tells a different story. In Figures\,\ref{fig:EGW_spec} and \ref{fig:EGW_spec2}, we plot spectrograms of the GW power versus time after bounce for our 3D models and see distinct structures. The most obvious feature is the f/g-mode \citep{murphy:09,2018ApJ...861...10M,torres2019,eggenberger2021,bruel2023} from $\sim$400 Hz early, rising to $\sim$1000-3000 Hz after $\sim$0.8 seconds after bounce. It is in this band that most of the emitted GW power of supernovae resides (see Table \ref{sn_tab}). This is a natural consequence of the fact that the peak in the eigenfunction of the f-mode is in the PNS periphery where the collisions of the accreta with the core are occurring. Hence, the excitation and the fundamental f-mode eigenfunction overlap nearly optimally. Associated with this feature in the earlier phases is a dark band near $\sim$1000-1300 Hz. This has been interpreted as a manifestation of an avoided crossing \citep{2018ApJ...861...10M} between a trapped $\ell = 2$ g-mode and the $\ell = 2$ f-mode. All the spectrograms for all our models show the same modal interaction, though at slightly different frequencies. For instance, at two seconds after bounce, the f-mode frequency is $\sim$1.75 kHz, 1.8 kHz, 2 kHz, and 2.5 kHz for the 9-, 11-, 12.25-, and 23-M$_{\odot}$ models, respectively, reflecting the variation in model PNS masses. Early on power is in the lower frequency component (mostly a trapped g-mode, mixed with the f-mode), and then it jumps to the higher frequency component (mostly the f-mode). This modal repulsion, or ``bumping," is a common feature in asteroseismology \cite{aizenman} and seems generic in core-collapse seismology.

All the models show the early prompt convection phase, with power from $\sim$300 to $\sim$2000 Hz. After this, all the models manifest a ``haze" of emission that extends above the f/g-mode to frequencies up to $\sim$2000 to $\sim$5000 Hz. The duration of this haze is from $\sim$0.25 to $\sim$1.5 seconds and tracks the phase of vigorous accretion (see \S\ref{excitation}). Individual PNS pulsation modes, likely the $\ell = 2; n=1$ p-mode, can also be seen superposed in this ``haze" and extending beyond it to later times.  This is particularly the case for the 12.25-, 15.01-, and 23-M$_{\odot}$ models. It is only for the models with the most vigorous accretion onto the core that this mode is clearly seen at later times.

\begin{figure*}
    \centering
    \includegraphics[width=0.79\textwidth]{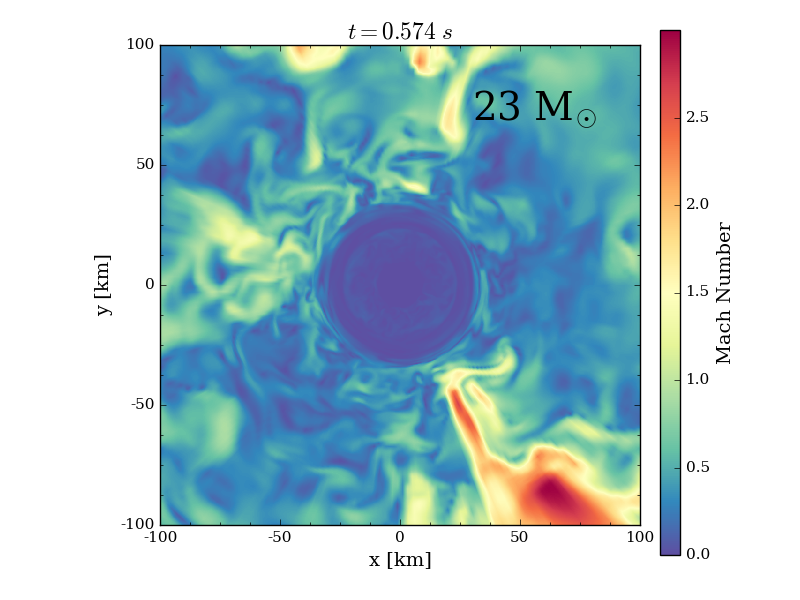}
   \includegraphics[width=0.79\textwidth]{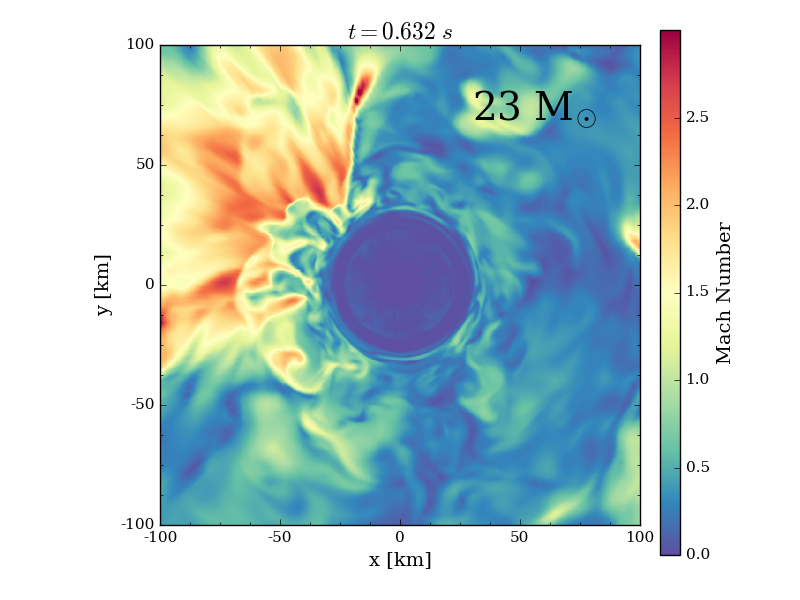}
    \caption{Plots (2D slices of 3D data) of the Mach number distribution in the inner 100 kilometers of the 23-M$_{\odot}$ simulation at 574 and 632 milliseconds after bounce.  Seen are representative supersonic plumes crashing onto the PNS core.  Such plumes not only excite the GW emission, but themselves emit a component of the GW signal. See text in \S\ref{description} for a discussion.}
    \label{fig:infall}
\end{figure*}

The origin of this haze is still a bit unclear, though it is definitely excited by the pummeling accreta (\S\ref{excitation}). The p-mode frequencies of the PNS for radial node numbers from 1 to 10 reside in this space and we could be seeing an overlapping and unresolved superposition of these modes. However, exploding models experience simultaneous explosion and accretion, the latter through infalling funnels that are few in number, can achieve supersonic speeds, and dance over the PNS surface. It is likely that the time-changing quadrupole moment of these funnels as they impinge upon the PNS surface is the source of this power. The timescales of their deceleration are about right, those timescales have a spread which could translate into a broad feature, and at any particular time they represent a low angular order perturbation. Importantly, however, we don't see this haze for the black-hole formers 12.25-M$_{\odot}$ and 14-M$_{\odot}$. It is only for the exploding models that there is a breaking of spherical symmetry that results in simultaneous explosion and supersonic funnel infall. Figure \ref{fig:infall} portrays two snapshots of the Mach-number distribution of the inner 100 km of the residue of the 23-M$_{\odot}$ model, clearly showing such funnel collisions. The accretion onto the proto-black-hole is always subsonic. Though the haze constitutes at most only a few 10's of percent of the total emission, its origin is clearly an interesting topic for future scrutiny.

In Figure\,\ref{fig:heff_spec}, we plot spectrograms for two representative models of the effective strain versus time after bounce. The effective strain is defined as the average of both strain polarizations,
\begin{equation}
h_{\mathrm{eff}} = 0.5 \left(h_{+} + h_{\times}\right)\,.
\end{equation} This figure provides a focus on the low-frequency regions.
We see for the 14-M$_{\odot}$ black-hole former some power near $\sim$100 Hz, which we identify with the spiral-SASI \citep{blondin_shaw} (see \S\ref{black}). Such a mode emerges in our model set only for the proto-black-holes (see \S\ref{black}).  In the red blot in the lower left-hand corner of the left panel may be a signature of the traditional SASI \citep{blondin:03}. We generally see little evidence in the GW signature of this SASI, but always see the spiral SASI when the explosion is aborted
and the stalled shock recedes. In the right panel of Figure \ref{fig:heff_spec}, the red band is a signature of the matter memory associated with the asymmetrical explosion
of the 23-M$_{\odot}$ model. Whether the traditional SASI is seen to the left of this band is unclear, but the early recession of its shock before explosion might be conducive to its brief appearance.

\subsection{Signatures of Failed Explosions - The Prelude to Black Hole Formation}
\label{black}

If a black hole forms by late-time fallback after many, many seconds to hours after the launching of a stalled shock that seemed to herald a successful explosion (but didn't), then the GW signal will be similar to those seen in the context of successful explosions. If, however, the stalled shock is never ``reignited," it will slowly settle to progressively smaller radii and the mantle of the progenitor core will continue to accrete through it onto the PNS.  Eventually, the fattening PNS will experience the general-relativistic instability to a black hole, at which time the GW emission abruptly terminates within less than a millisecond. This latter phase could take many seconds to many minutes to reach\footnote{For the 14-M$_{\odot}$ model, we estimate a black hole formation time (using a maximum baryon mass of 2.477 M$_{\odot}$ from Steiner et al. \citep{2013ApJ...774...17S} at the onset of collapse) of $\sim$500 seconds.}. The GW signature of this modality of black hole formation, representatives of which are our 14- and 12.25-M$_{\odot}$ models, has particular diagnostic features that set these evolutions apart. First, the breaking of symmetry that results in the simultaneous accretion of lower-entropy plumes with the explosion of high-entropy bubbles does not occur. The result is that for this channel of black hole formation the infalling plumes do not dance over the PNS, do have high Mach numbers, and don't excite the higher-frequency ``haze" that we have identified for the exploding models seen in the associated spectrograms (Figures \ref{fig:EGW_spec} and \ref{fig:EGW_spec2}). We do see in Figure \ref{fig:EGW_spec2} power not only in the dominant f-mode, but weakly in an overtone p-mode as well. However, as shown in Table \ref{sn_tab} for the 12.25 M$_{\odot}$ black-hole former, the fraction of the total GW energy radiated in the f-mode is correspondingly higher, as much as $\sim$95\% of the total, than for the exploding models that also generate power in the haze.

\begin{figure*}
    \centering
    \includegraphics[width=0.87\textwidth]{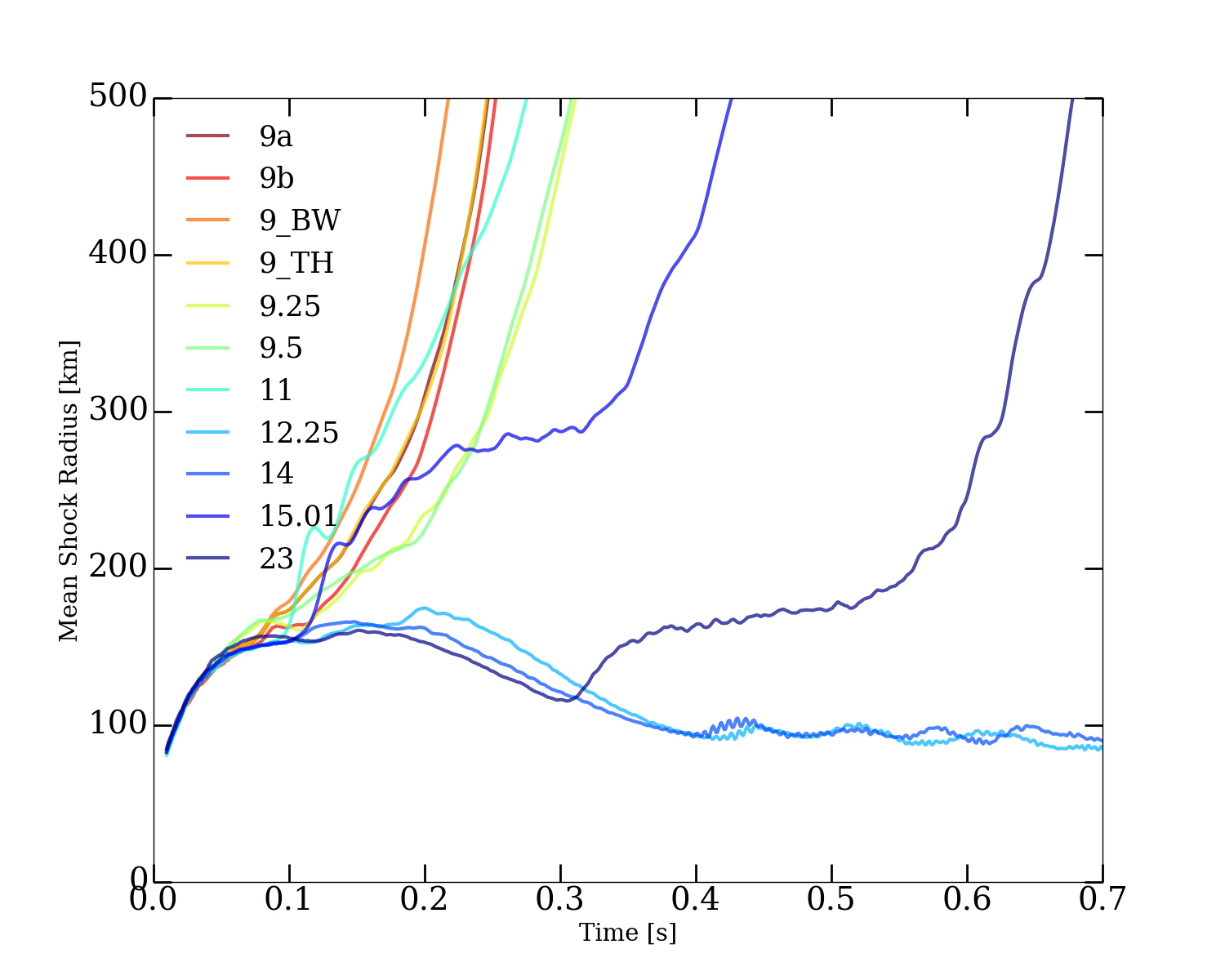}
    \caption{Early post-bounce behavior of the mean shock radii (in km) for all the models as a function of time (in seconds after core bounce). All models explode except for the 12.25- and 14-M$_{\odot}$ progenitors, which show evidence for a spiral SASI after $\sim$400 and $\sim$300 ms post-bounce, respectively, in the form of $\sim$10 ms oscillations in the shock radii. Note that the 12.25- and 14-M$_{\odot}$ (non-exploding) progenitors also shows a secular $\sim$70 ms oscillation in the shock radii.}
    \label{fig:rs1}
\end{figure*}

\begin{figure*}
    \centering
    \includegraphics[width=0.45\textwidth]{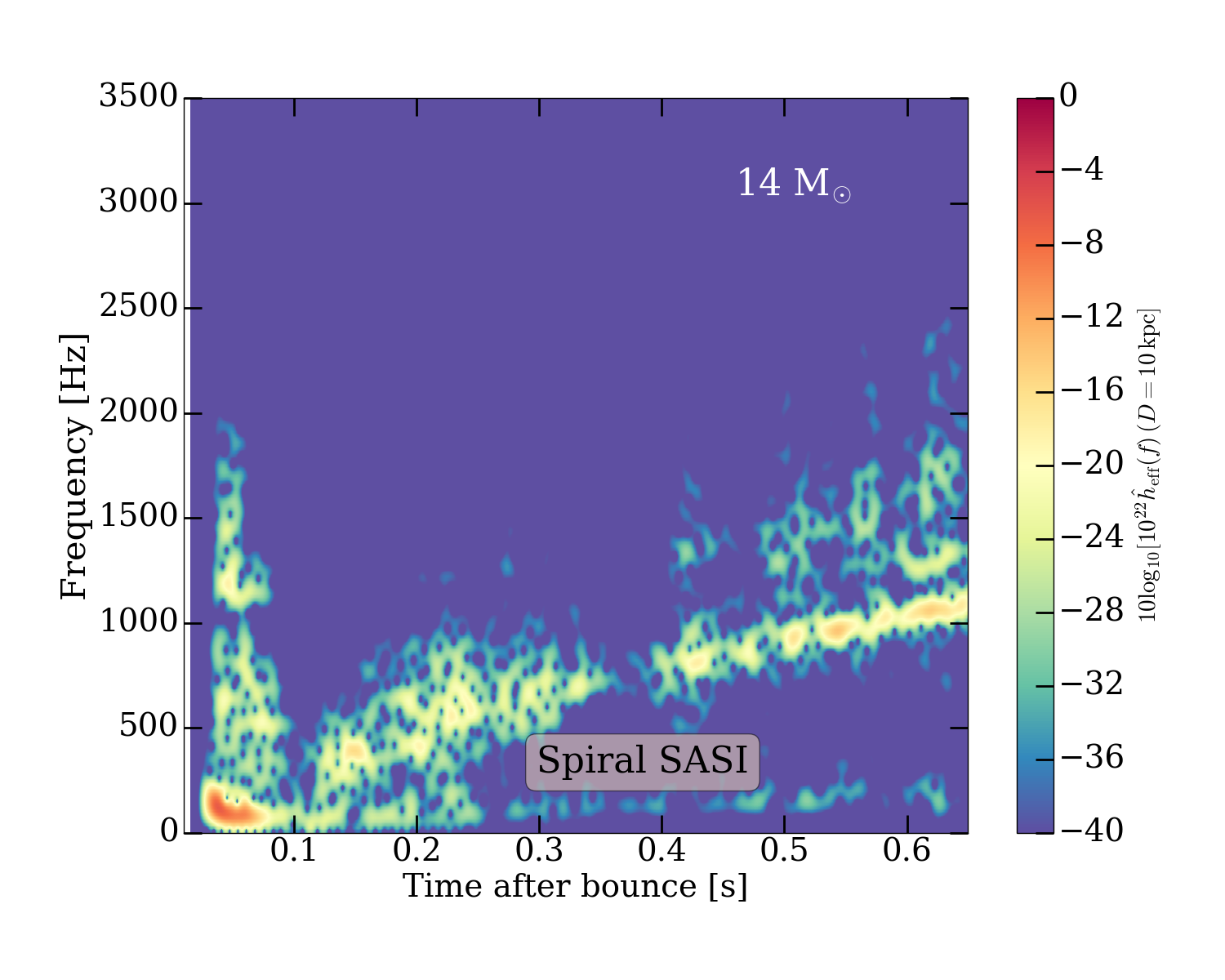}
   \includegraphics[width=0.45\textwidth]{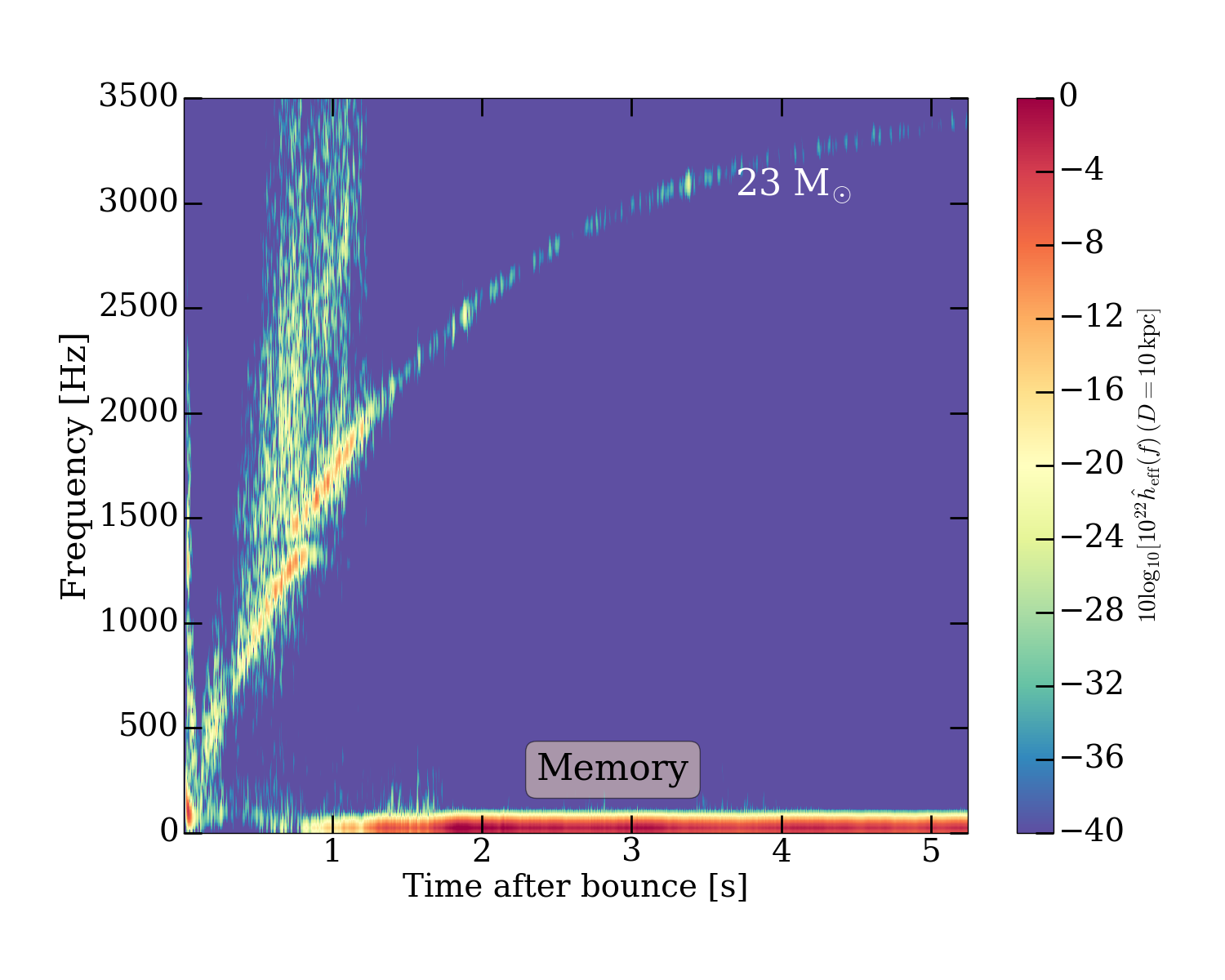}
    \caption{Plots of the effective strains as a function of time after bounce (in seconds) for two representative 3D models (the black-hole former 14-M$_{\odot}$ and late-exploding 23-M$_{\odot}$ model). Note that we see a dark red band at low ($\le$50 Hz) frequencies for the exploding model associated with the asymmetric ejecta at later times (matter ``memory"). The non-exploding model shows a spiral SASI signature at $\sim$100 Hz. This quasi-periodic feature is also seen after $\sim$0.4 seconds in Figure \ref{fig:rs1} for both the 14-M$_{\odot}$ and 12.25 M$_{\odot}$ black-hole formers. Such a feature shows up only for non-exploding models when the shock radius has shrunken below $\sim$100 km. These features are much less prominent in the gravitational-wave energy spectra because they are low-frequency, and, hence, low-power components. There may be a hint of the SASI itself in the lower left-hand corner of the panel to the right. See text in \S\ref{description} for a discussion.}
    \label{fig:heff_spec}
\end{figure*}

\begin{figure*}
    \centering
    \includegraphics[width=0.45\textwidth]{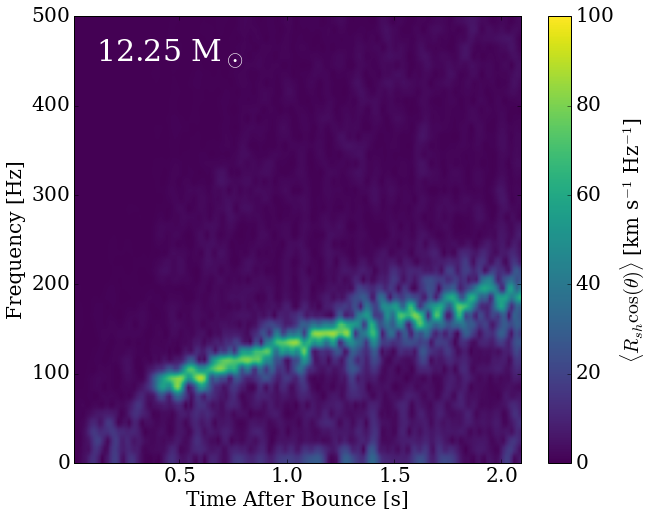}
    \includegraphics[width=0.45\textwidth]{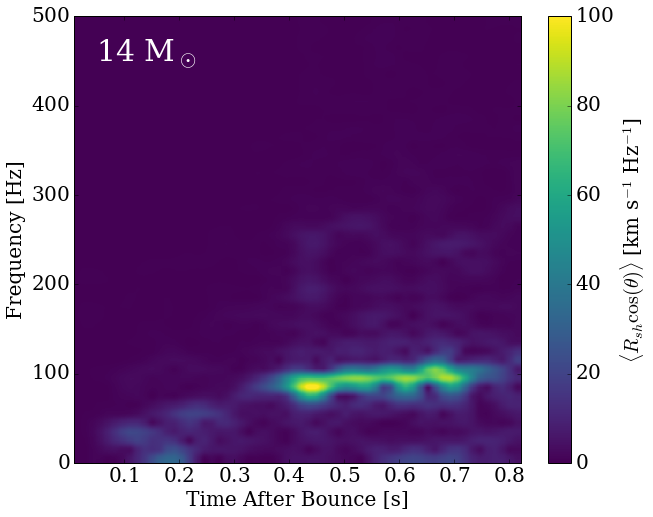}
    \includegraphics[width=0.45\textwidth]{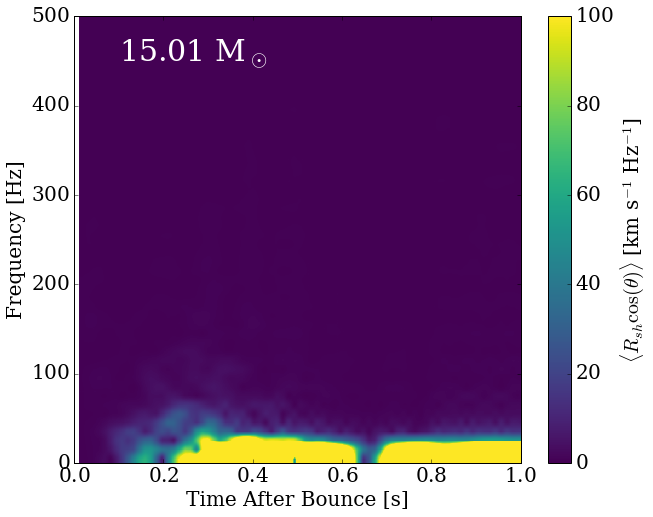}
    \includegraphics[width=0.45\textwidth]{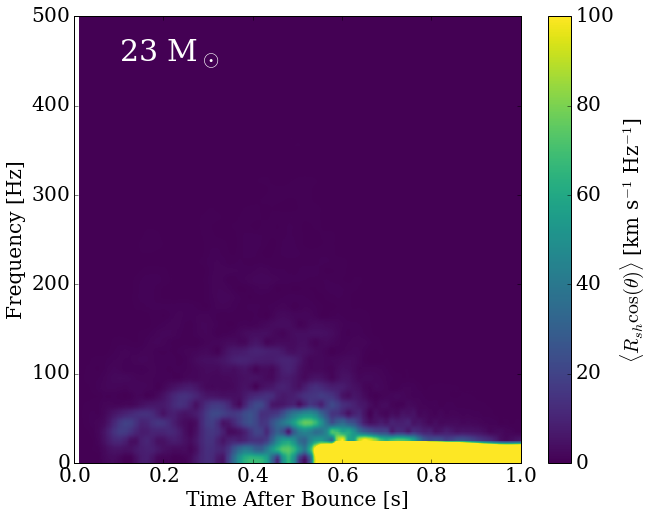}
    \caption{In this figure, we show the spectrogram of the z-component of shock dipole position for four models. Note the focus on lower frequencies. In the non-exploding 12.25- and 14-M$_{\odot}$ models, the spiral-SASI modes are clearly seen and manifest gradually increasing frequencies as the PNS core shrinks. The exploding models (for example, the 15.01 and 23 M$_{\odot}$ models depicted here) don't show this feature, but do manifest the even lower frequency signals due to the matter memory caused by explosion anisotropy (see also Figure \ref{fig:heff_spec}). We note that the 23-M$_{\odot}$ model may show a weak SASI signature (analogous with that found in the 25-M$_{\odot}$ model in reference \protect\citep{vartanyan2019}) that vanishes at $\sim 0.55$ seconds and is perhaps indicative of its delayed explosion time. The onset of explosion corresponds with the rise of the low-frequency signal memory in the bright yellow band.}
    \label{fig:spiral-sasi}
\end{figure*}

This channel of black hole formation also experiences the emergence of what we identify as the spiral-SASI \citep{blondin_shaw}. This is seen in Figure \ref{fig:rs1} in the clear $\sim$100-200 Hz periodicity of the late-time mean shock position of both the 14- and 12.25-M$_{\odot}$ models after $\sim$300$-$400 milliseconds after bounce and very clearly in the spectrogram of the shock dipole depicted in Figure \ref{fig:spiral-sasi}. Generally, this feature emerges after the mean stalled shock radius sinks below $\sim$100 km and is not seen in exploding models. The timescale of the periodicity scales roughly with $\Delta {R_s}/c_s + \Delta{R_s}/v_{acc}$, where $R_s$, $c_s$, and $v_{acc}$ are the shock radius, speed of sound, and post-shock accretion speed \citep{foglizzo:07}. 

Another feature seen most clearly in Figure \ref{fig:rs1} in the context of these black-hole formers is a much longer-timescale modulation of the mean shock position with a period near $\sim$70 ms. Not seen clearly in the GW spectrograms or strain plots (though there may be a hint in the strain plot for the 14-M$_{\odot}$ model), this oscillation may be due to a global pulsation mode associated with the neutrino heating, cooling, and transport of the mantle, but this speculation remains to be verified. Nevertheless, this feature has never before been identified in studies of 3D CCSN and is interesting in itself. Finally, as Figure \ref{fig:strain} suggests for the 14- and 12.25-M$_{\odot}$ models, since those cores that form black holes by this channel do not explode, they are expected to have no net low-frequency matter memory component.

\subsection{Total Gravitational-Wave Energy Radiated}\label{energy}

\begin{figure*}
    \centering
    \includegraphics[width=0.9\textwidth]{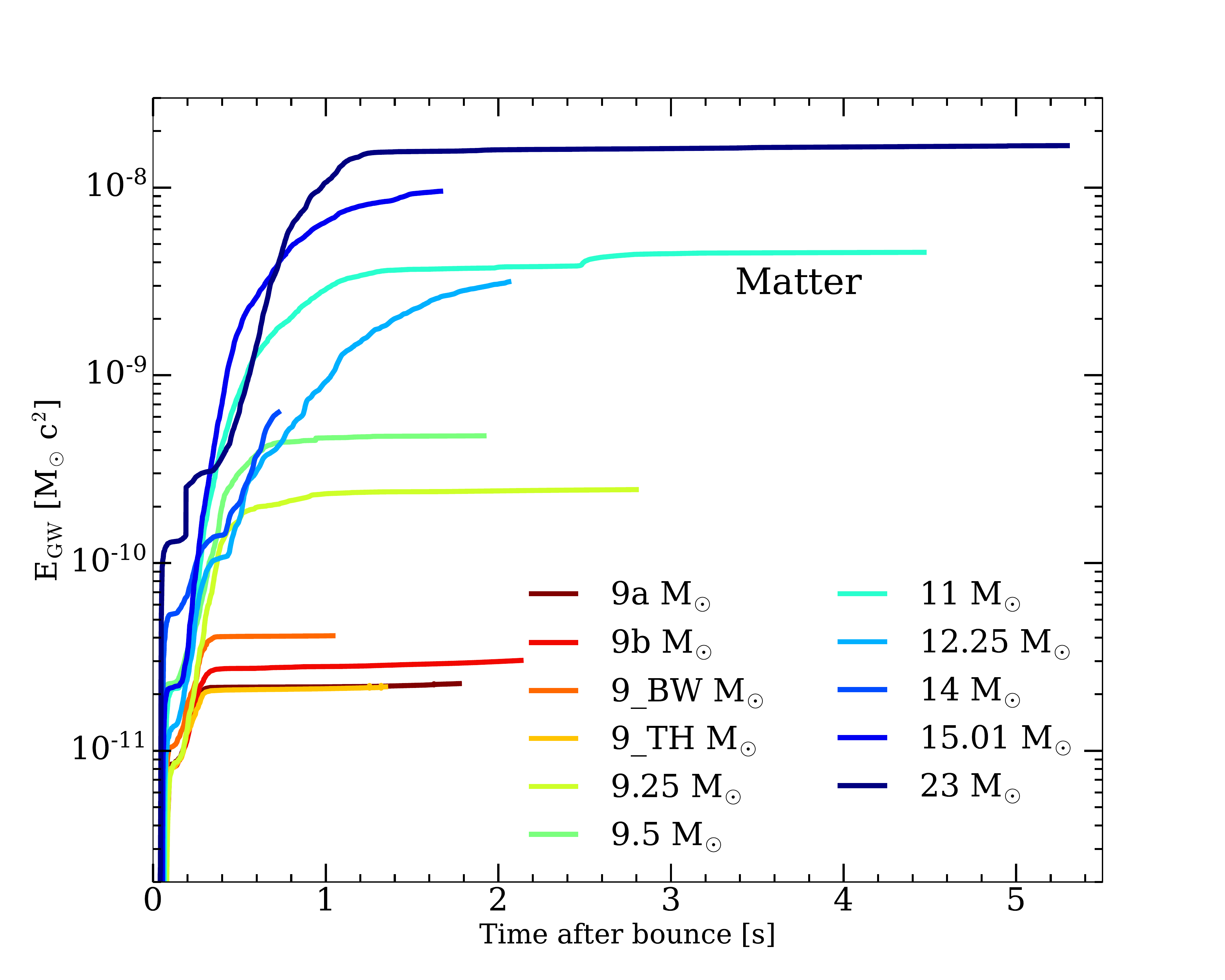}
    \caption{The gravitational-wave energy from matter motions (in M$_{\odot}$\,c$^{2}$) for all the 3D models highlighted in this paper as a function of time after bounce (in seconds). Note that the total GW energy radiated differs by $\sim$three orders of magnitude from the least massive, 9-M$_{\odot}$ to the most massive, 23-M$_{\odot}$ progenitor. This energy grows by three orders of magnitude for the most massive progenitors over the first $\sim$2 seconds of simulation, but is already asymptoting shortly after one second post-bounce. All models show rapid growth in the first $\sim$50 ms, associated with the onset of prompt convection driven by the overturn of the shocked mantle dynamically generated at and after bounce as the shock stalls initially into accretion. After this phase, the neutrino-driven turbulence between the shock and the proto-neutron star core grows in vigor
    over a period of $\sim$100 milliseconds and excites a spectrum of core pulsational f- and some p-modes and  likely generates a GW component due to the impinging of the plumes onto the PNS that all together constitute the bulk of the gravitational radiation issuing from the supernova.  This phase can last from hundreds of milliseconds to $\le$1.5 seconds, depending upon progenitor, after which the strains subside to a hum dominated by the fundamental $\ell = 2$ f-mode and (weakly) overtones. This last phase can last for many seconds. However, the signals from those models destined to leave black holes are still vigorous for a longer period of time, since accretion persists for these models until the black hole forms, after which the signal ceases abruptly. See the text for a discussion.}
    \label{fig:EGW}
\end{figure*}

In Figure\,\ref{fig:EGW}, we plot versus time the integrated radiated GW energy due to matter motions. Model 23-M$_{\odot}$ radiates the most GW energy ($\sim$3.0$\times$10$^{46}$ erg, or $\sim$2$\times$10$^{-8}$ M$_{\odot}$ c${^2}$ after $\sim$5 seconds), while the collection of 9-M$_{\odot}$ models radiates the least.  There are a few important features of this plot.  First, we see that we have captured what is basically the entire GW signal for many of the models (the 15.01-M$_{\odot}$, 14-M$_{\odot}$, and 12.25-M$_{\odot}$ model emissions are still climbing). Within $\sim$1.5 seconds, most exploding models have radiated $\ge$95\% of the total energy to be radiated, and after $\sim$2 seconds they have radiated $\ge$98\%. Table \ref{sn_tab} provides the total energy radiated via the f/g-mode, as well as the fraction of this total radiated in the f-mode after 1.5 seconds. Not shown in Table \ref{sn_tab} is the fact that more than 95\% of the total GW energy radiated after 1.5 seconds is via the f-mode.
As indicated for the 12.25-M$_{\odot}$ model in Table \ref{sn_tab} and Figure \ref{fig:EGW}, due to continuing accretion the black hole formers radiate to later times than the exploding models, and this mostly in the f-mode.

\begin{figure*}
    \centering
    \includegraphics[width=0.9\textwidth]{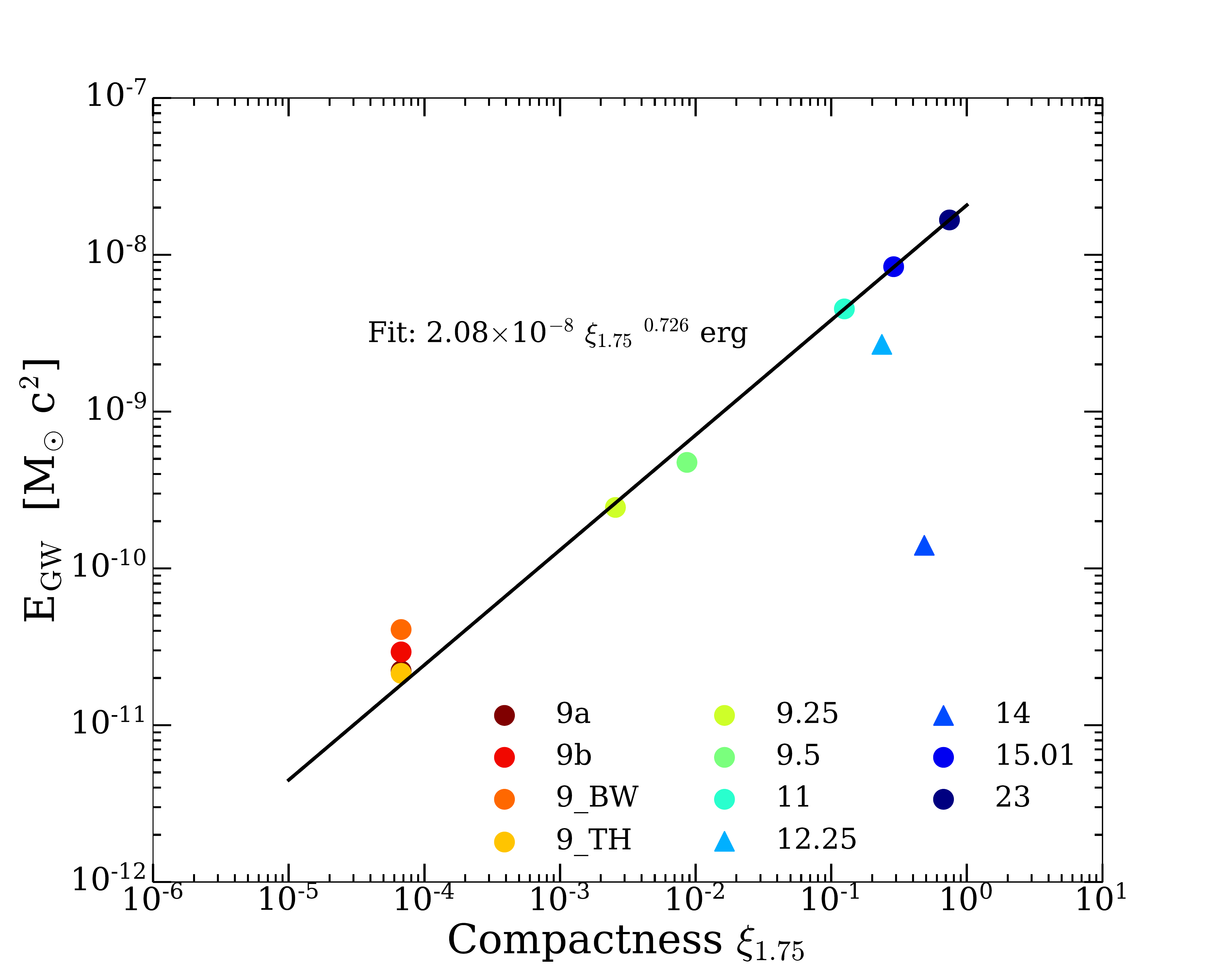}
    \caption{The gravitational-wave energy from matter motions (in M$_{\odot}$c$^{2}$) for all the 3D models in this study versus the model compactness at 1.75 M$_{\odot}$ \citep{2011ApJ...730...70O}. Note the clear linear correlation on the log-log scale, with a power-law slope of $\sim$0.73. The outliers are the non-exploding models (12.25- and 14-M$_{\odot}$, shown with triangle markers) and these models generally take longer to radiate a given amount of GW energy (see Figure\,\ref{fig:EGW}). The black-hole formers don't as a rule manifest turbulence that is as vigorous as those models that eventually explode.  All exploding models show a steep growth in the GW energy associated with the onset of vigorous turbulence. The 23-M$_{\odot}$, which explodes the latest, has the latest growth spurt in gravitational energy. By contrast, non-exploding models show a slower rate of of increase of GW emission, associated with a slower growth rate of turbulence. However, due to the persistently stalled shock and continued accretion, the black-hole formers will generate GWs for a longer time and will saturate their total GW emission more slowly. Finally, a spiral SASI mode emerges for the black-hole formers when the stalled shock wave retreats significantly (below $\sim$100 km). This mode modulates the stalled accretion shock radius and the GW emissions of these black-hole progenitors quasi-periodically.}
    \label{fig:EGW_eta}
\end{figure*}

Figure \ref{fig:EGW} shows that the various phases described in \S\ref{description} are recapitulated via stair steps until finally asymptoting.  Moreover, the continuum of models highlighted in this paper demonstrate collectively that the radiated GW signal energies vary by as much as three orders of magnitude from the lowest-mass to the higher-mass models. This is a consequence of the differences in their initial density profiles (see Figure \ref{fig:rho}), and directly from the resulting mass accretion histories. Even more directly, as Figure \ref{fig:EGW_eta} demonstrates, there is a strong monotonic relation for exploding models between the total GW energy radiated and the compactness \citep{2011ApJ...730...70O} of the initial progenitor ``Chandrasekhar" core\footnote{We define the compactness here as $\xi_M= \frac{M/M_{\odot}}{R(M)/1000\, \mathrm{km}}$, where $M=1.75$ M$_{\odot}$.}. Though compactness does not correlate with explodability \citep{burrows_2020,2021ApJ...916L...5V,2021Natur.589...29B,wang, tsang2022}, it does seem to correlate with residual neutron star mass, radiated neutrino energy, and, as now indicated in Figure \ref{fig:EGW_eta}, the total GW energy radiated. In fact, we derive a power-law with index $\sim$0.73 between the two.

Finally, we note that the collection of 9-M$_{\odot}$ models don't behave exactly the same. This is due to the fact that these models were simulated with slightly different code variants (as we continued to update and upgrade F{\sc{ornax}}); one (the 9a model) had small artificial imposed perturbations in the initial model and three different supercomputers were used. The natural chaos in the flow and in the simulations will pick up on any slight variations and amplify them, with the result that the evolution can be slightly different. Though the qualitative behavior of all four of these 9-M$_{\odot}$ models is the same, they exploded at slightly different times (see Figure \ref{fig:strain}), with the resulting different developments of their GW signals. As a consequence, the radiated energy varies by about a factor of two. In a crude sense, we might view this spread as an imperfect indicator of the likely spread in Nature due to the chaos of turbulence for the ``same" progenitor, but we have certainly not demonstrated this.

\subsection{Avoided Crossing and Trapped g-mode}
\label{avoided}

\begin{figure*}
    \centering
    \includegraphics[width=0.45\textwidth]{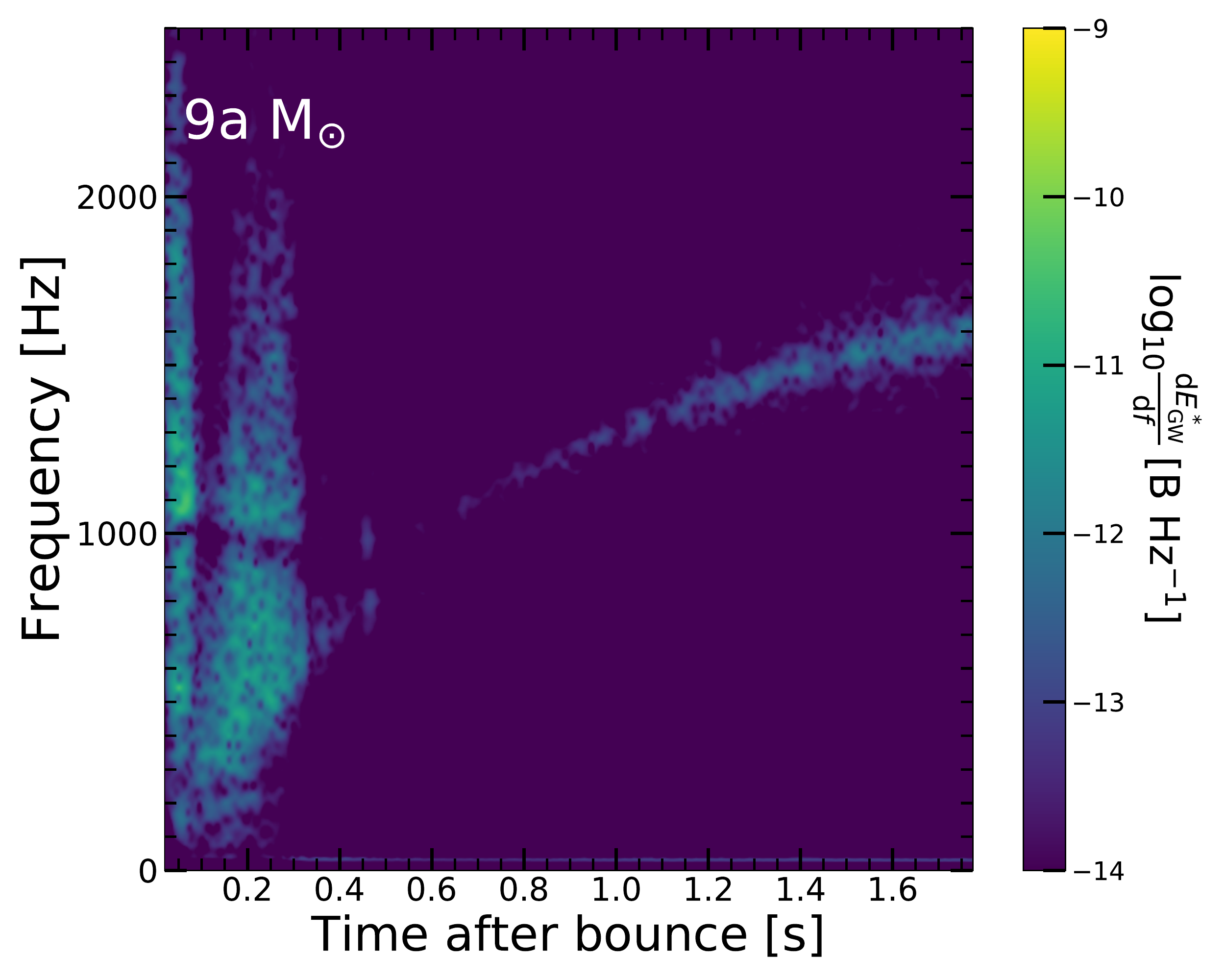}
    \includegraphics[width=0.45\textwidth]{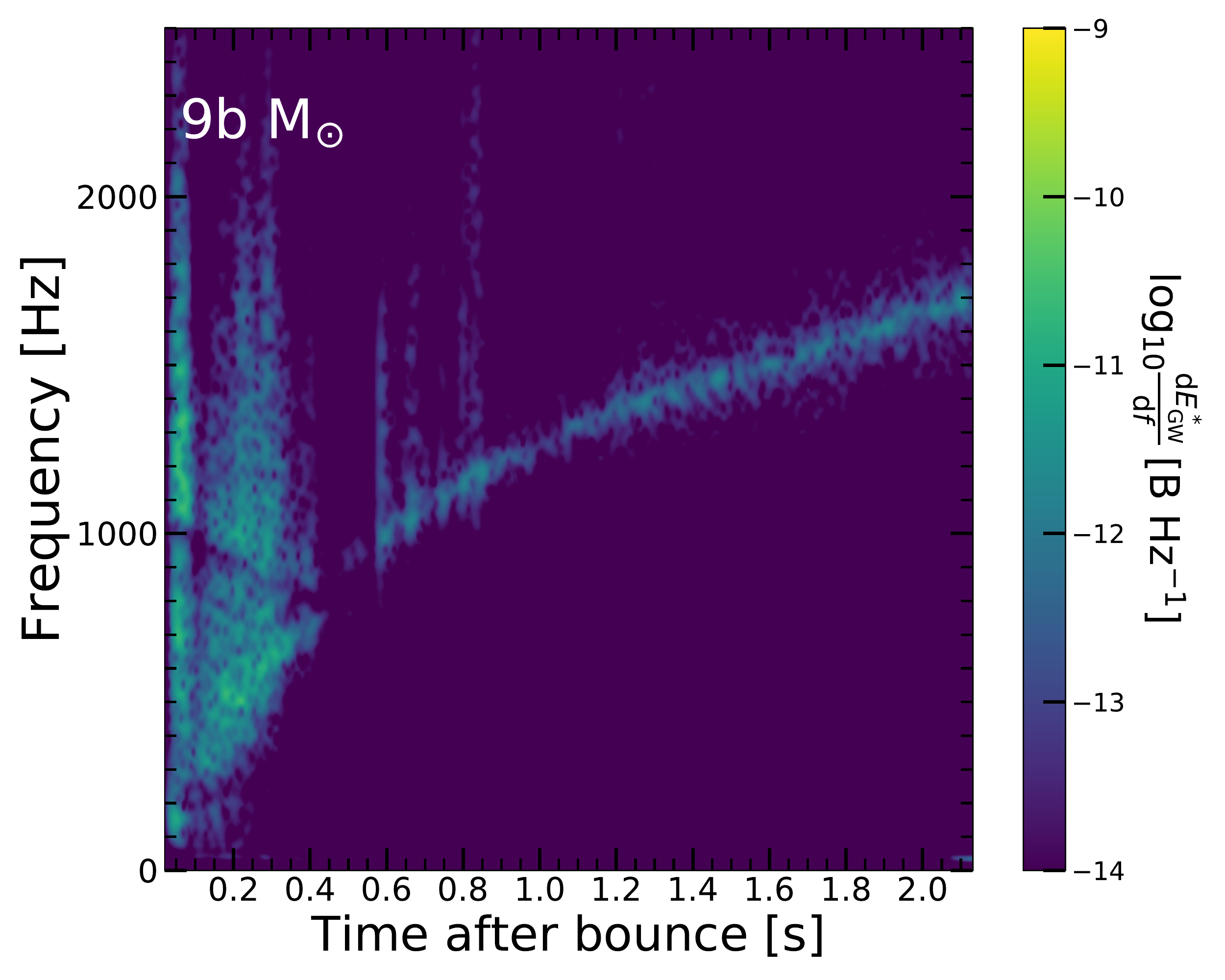} 
    \includegraphics[width=0.45\textwidth]{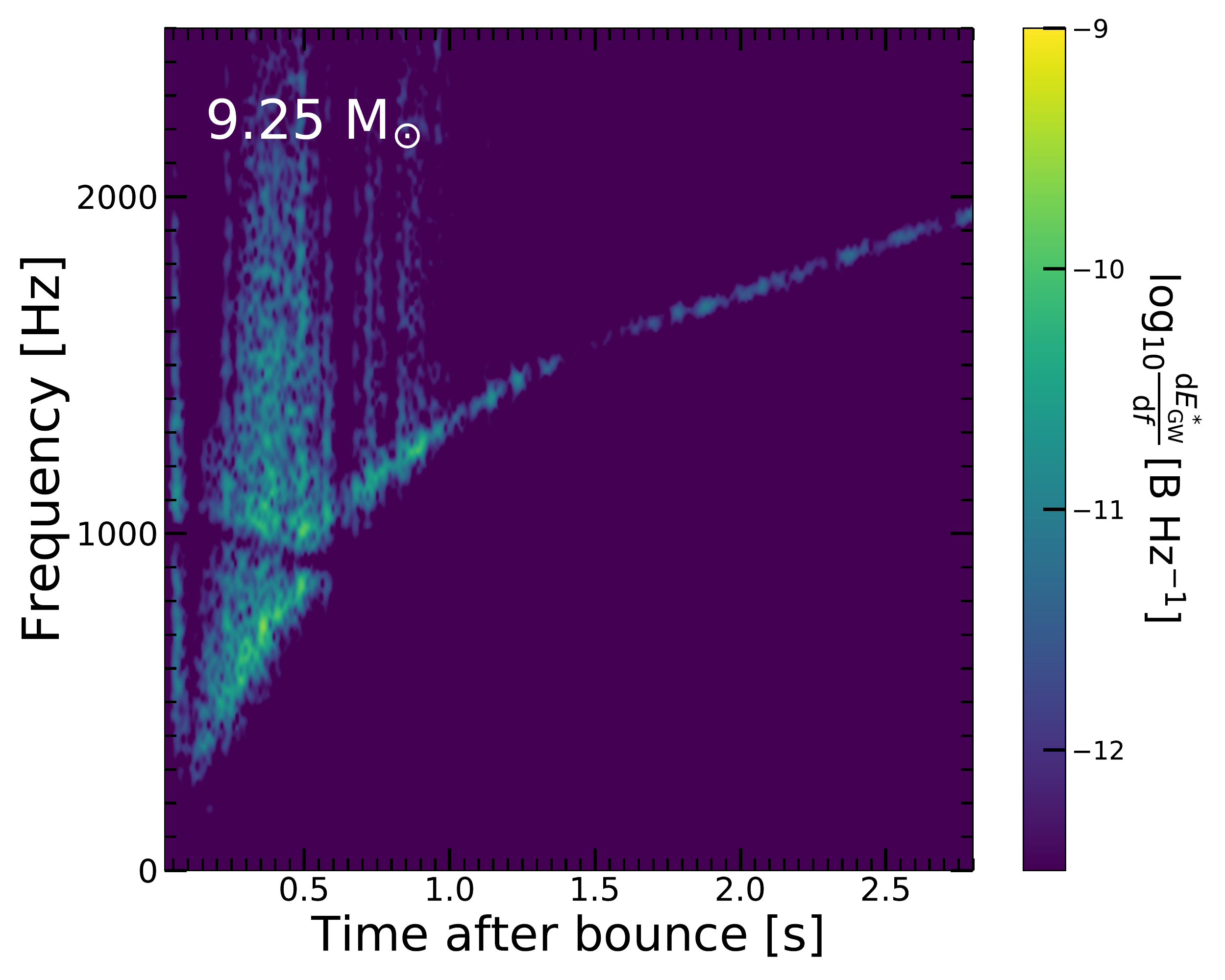}     
    \includegraphics[width=0.45\textwidth]{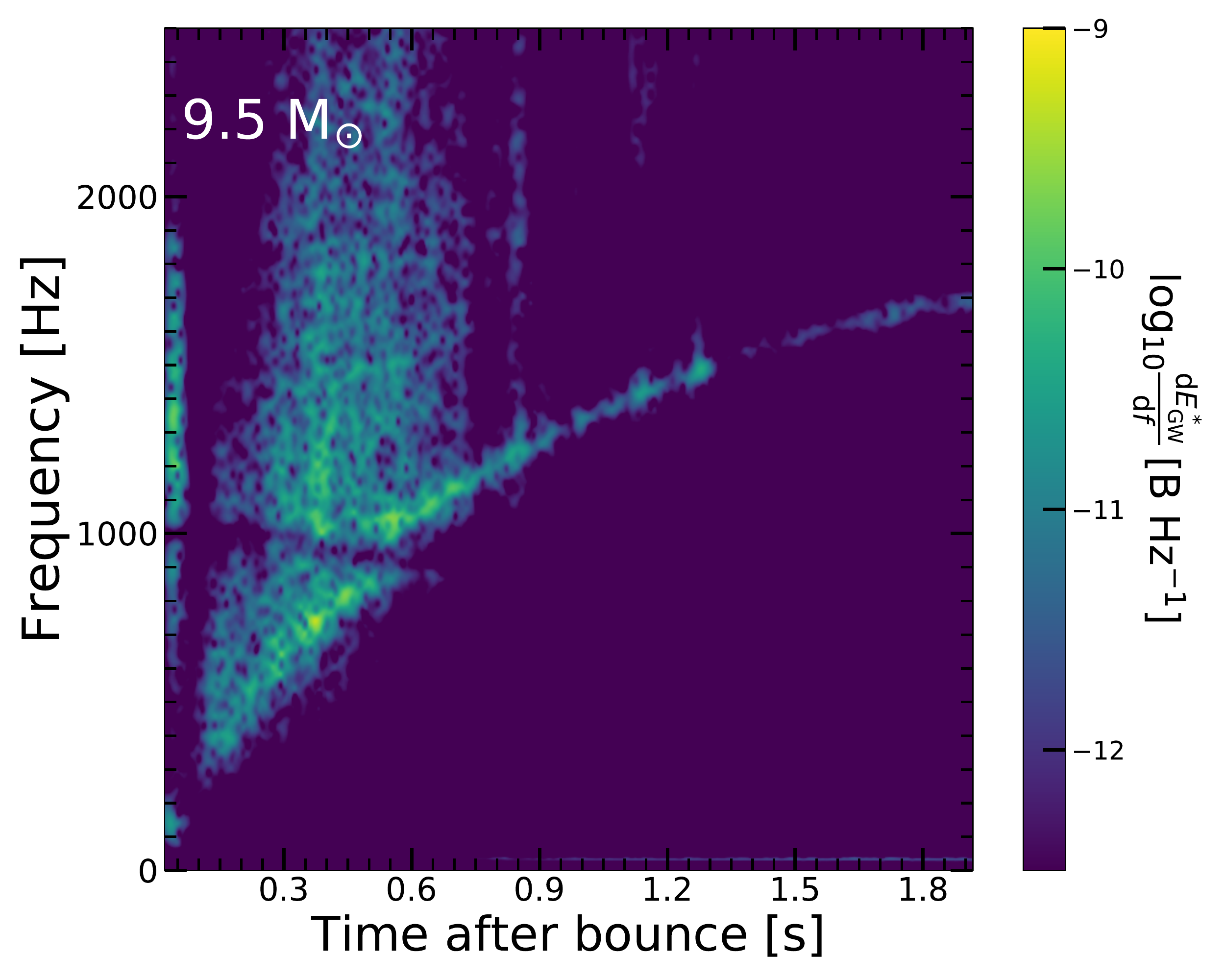} 
    \caption{In this figure and the next, we provide the gravitational-wave energy spectrogram (to a normalization constant of order unity) at 10 kiloparsecs (kpc) as a function of time after bounce (in seconds) for the 3D models highlighted in this paper. Note that the frequency ranges and durations differ from panel to panel. The colormap  provides the spectral density in units of Bethes (10$^{51}$ egs) per Hertz, where we have integrated over a running 40-millisecond bin around each time. We have found that the results do not depend significantly upon the width of this window. Though the models in this first set are for 9-, 9.25-, and 9.5 M$_{\odot}$, the description in this caption also applies to the 11-, 12.25-, 15.01-, and 23-M$_{\odot}$ models in the accompanying figure (Figure \ref{fig:EGW_spec2}). All models show a strong feature out to  $\sim$50 ms and up to $\sim$2000 Hz associated with prompt convection, followed by a short quiescent phase, during which the turbulence between the shock and the proto-neutron star grows. Thereafter, all models illustrate high frequency behavior before $\sim$one second (even earlier for the rapidly exploding lower progenitor mass models), which truncates in power as the turbulent accretion which excites it subsides. This ``emission haze" is either a superposition of core pulsational $\ell = 2$ p-modes with a range of node numbers or the GW emission from the supersonic plumes themselves impinging upon the PNS core, or some combination of both. The fundamental f/g-mode (early), then the f-mode (later), is manifest and significant in all models for the duration of the simulations. We note that the 15.01-M$_{\odot}$ model (next figure), which has a relatively lower ejecta mass, sustains turbulent accretion and high frequency haze emission until the end of our simulation, $\sim$2.0 seconds post-bounce.  We also note that we see higher harmonics/overtones after the first second and after the earlier emission haze abates, most visibly in the 15.01- and 23-M$_{\odot}$ models. These later-time PNS pulsational modes are likely the $\ell = 2, n = 1$ p-mode. The fundamental f-mode frequency grows roughly quadratically with time (\protect\cite{vsg2018}) and is associated with the shrinkage of the PNS as it executes its Kelvin-Helmholtz contraction phase, driven by neutrino losses. Except for the very early phase after bounce, most g-modes are suppressed due both to the presence and growth of PNS convection in the deep core and to their lower frequencies ($\sim$a few hundred Hz). Early on, all models show a dark band near $\sim$1000 Hz which may be a signature of an avoided crossing and interference with a trapped g-mode. See text in \S\ref{description} and \S\ref{avoided} for a discussion.}
    \label{fig:EGW_spec}

\end{figure*}

\begin{figure*}
    \centering
    \includegraphics[width=0.45\textwidth]{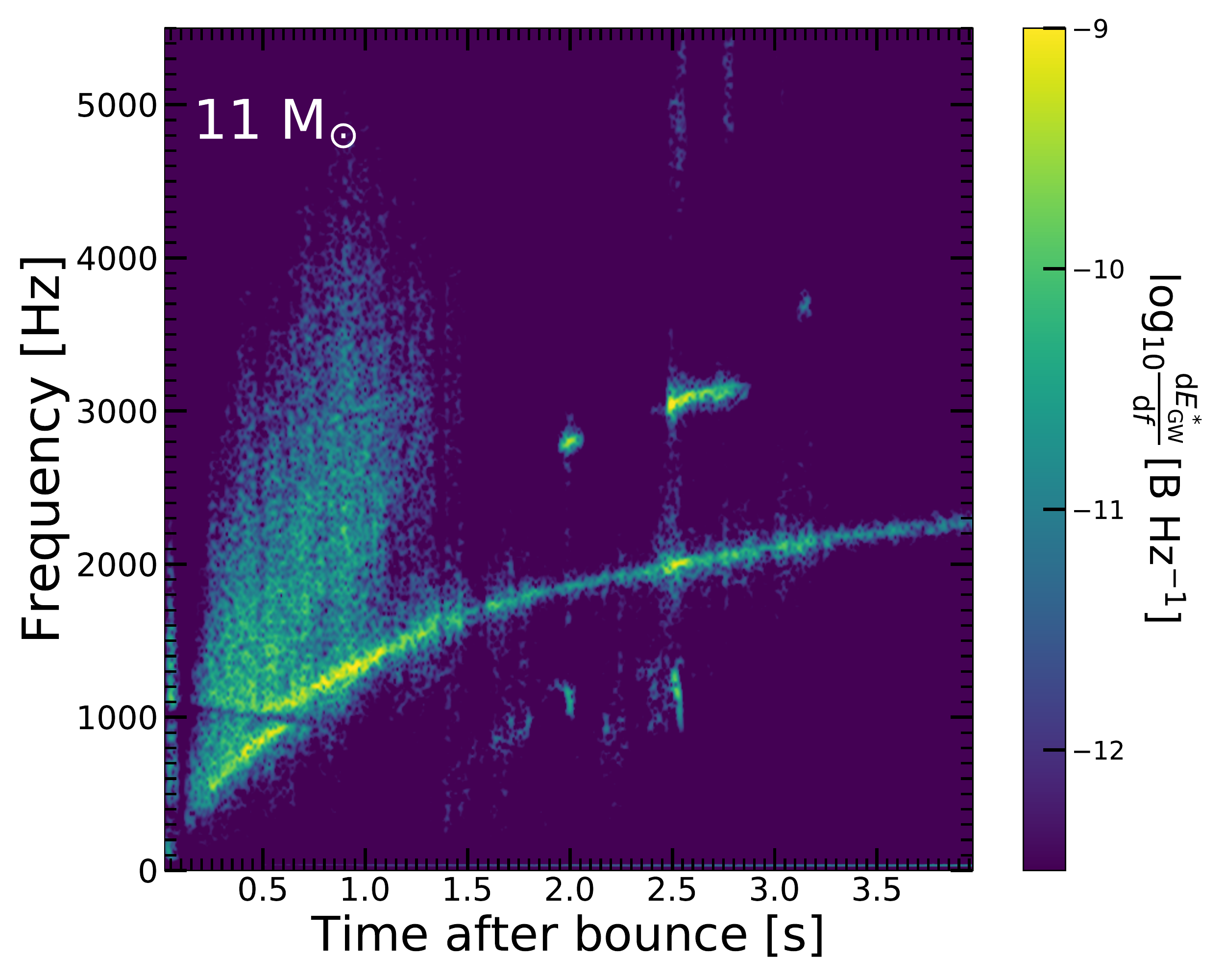}
    \includegraphics[width=0.45\textwidth]{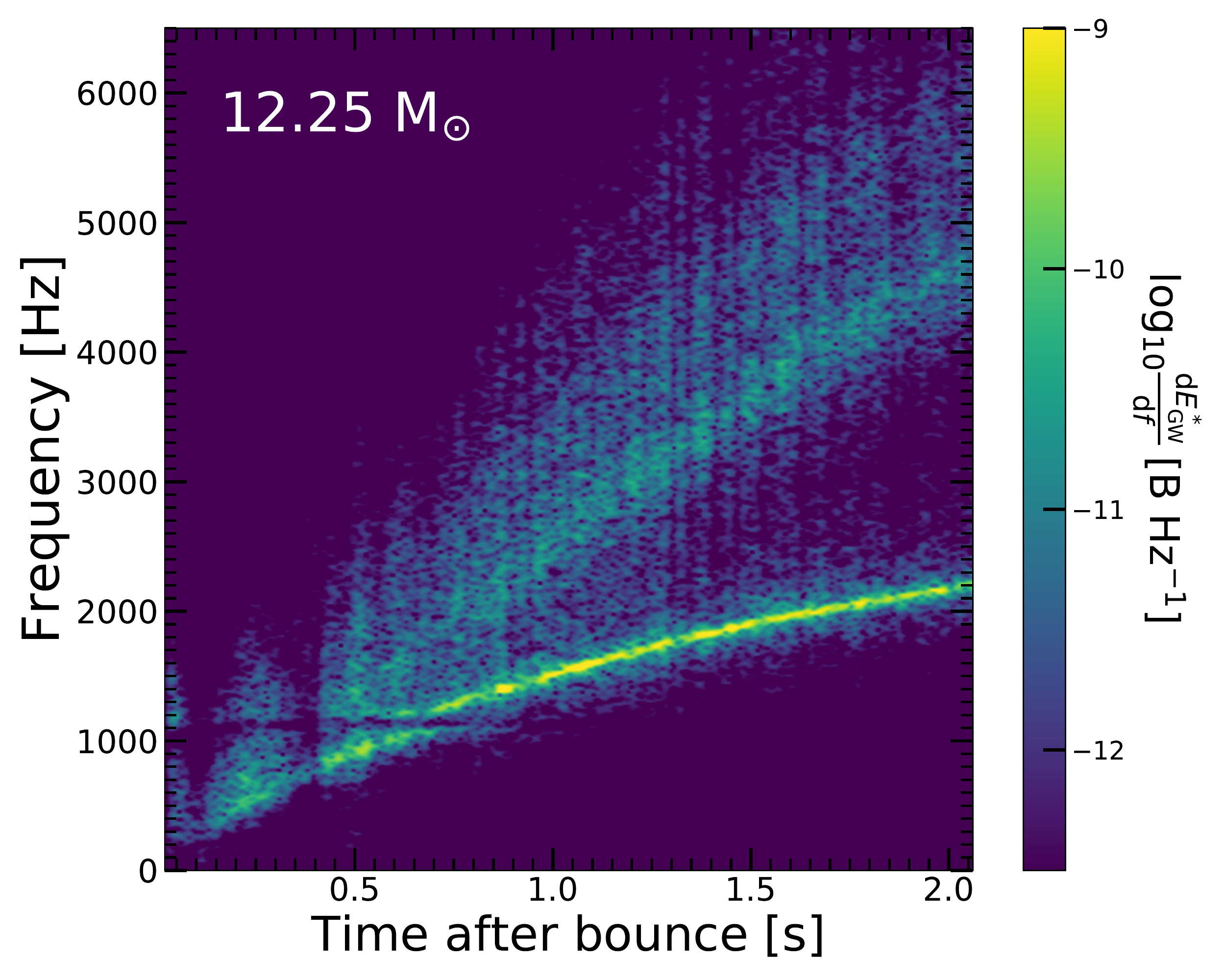}
    \includegraphics[width=0.45\textwidth]{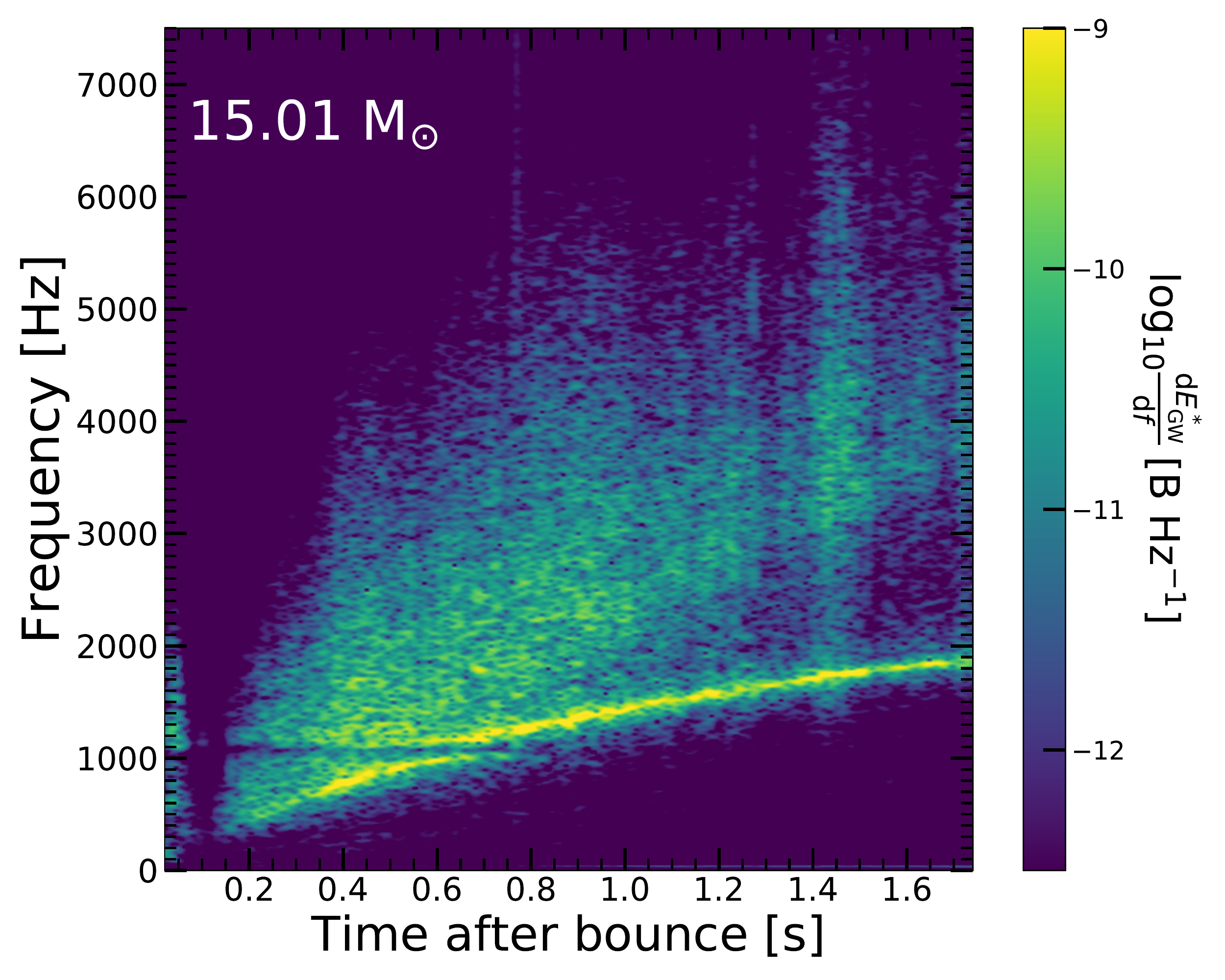}
    \includegraphics[width=0.45\textwidth]{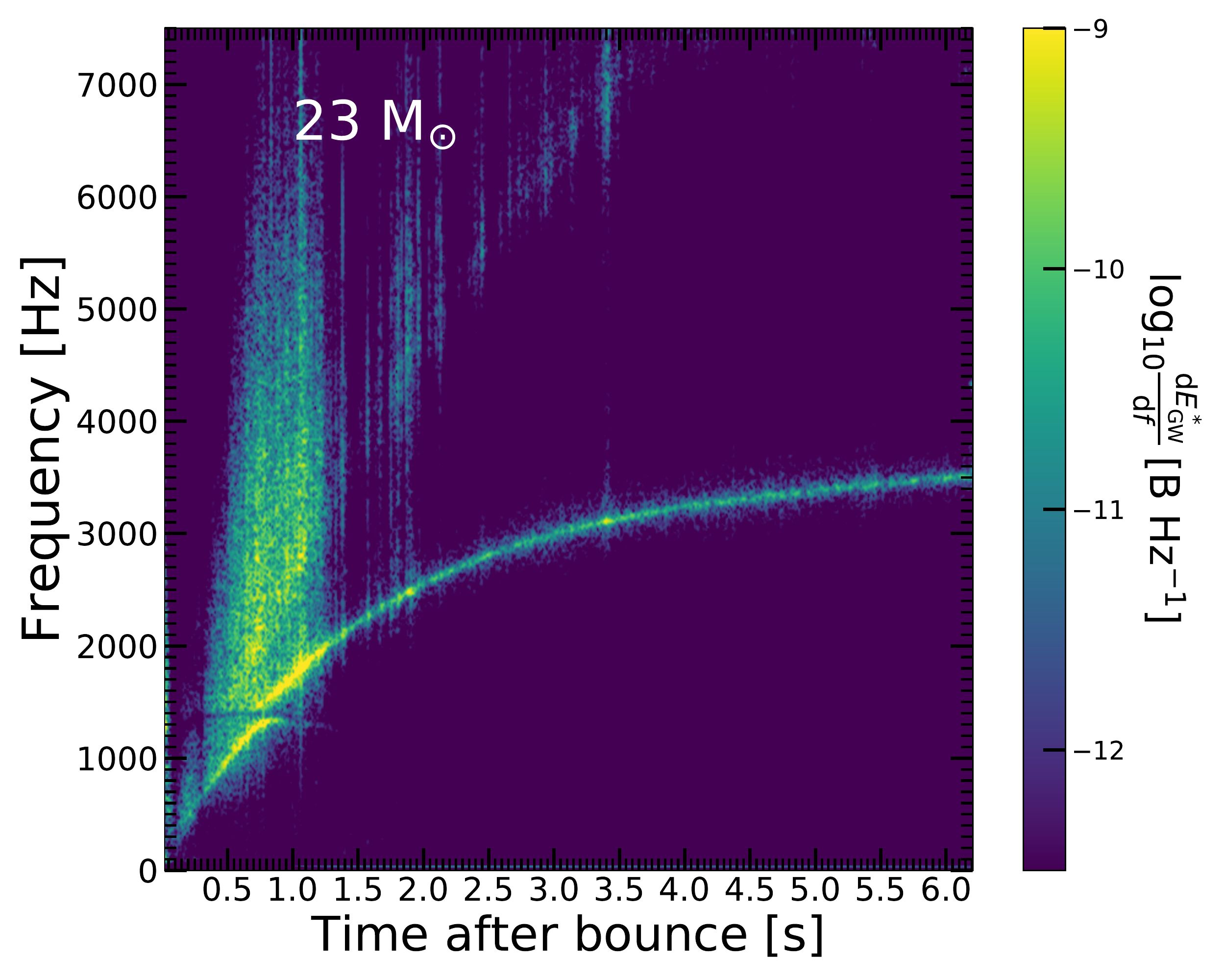}
    \caption{Same as Figure\,\ref{fig:EGW_spec}, but for models 11-, 12.25-, 15.01-, and 23-M$_{\odot}$.}
    \label{fig:EGW_spec2}
\end{figure*}

A distinctive feature seen clearly in the spectrograms for all the models (see Figures \ref{fig:EGW_spec}, \ref{fig:EGW_spec2}) is a dark band near $\sim$1000$-$1300 Hz during the first $\sim$0.3$-$1.0 seconds of the post-bounce evolution.  This is most likely due to an avoided crossing \citep{aizenman} of interfering $\ell = 2$ PNS pulsation modes that are coupled and mixed \citep{2018ApJ...861...10M}. The best current thinking is that the interfering modes are a trapped g-mode and the $\ell = 2$ f-mode \citep{2018ApJ...861...10M,eggenberger2021}, though much work remains to be done to determine the details and the nature of the couplings. Non-linear mode coupling may be involved. There is also evidence for some GW power in the ``bumped" g-mode (that thereafter trends to lower frequencies), seen just after the mode repulsion in Figures \ref{fig:EGW_spec} and \ref{fig:EGW_spec2}, but most clearly in Figure \ref{fig:GW-mdot-corr} below. This is a qualitatively similar feature to that highlighted recently in \citep{jakobus}. Nevertheless, within $\sim$0.3 to $\sim$1.0 seconds (depending upon the progenitor) most power is clearly in the f-mode, where it persists thereafter.

\begin{figure*}
    \centering
    \includegraphics[width=0.4\textwidth]{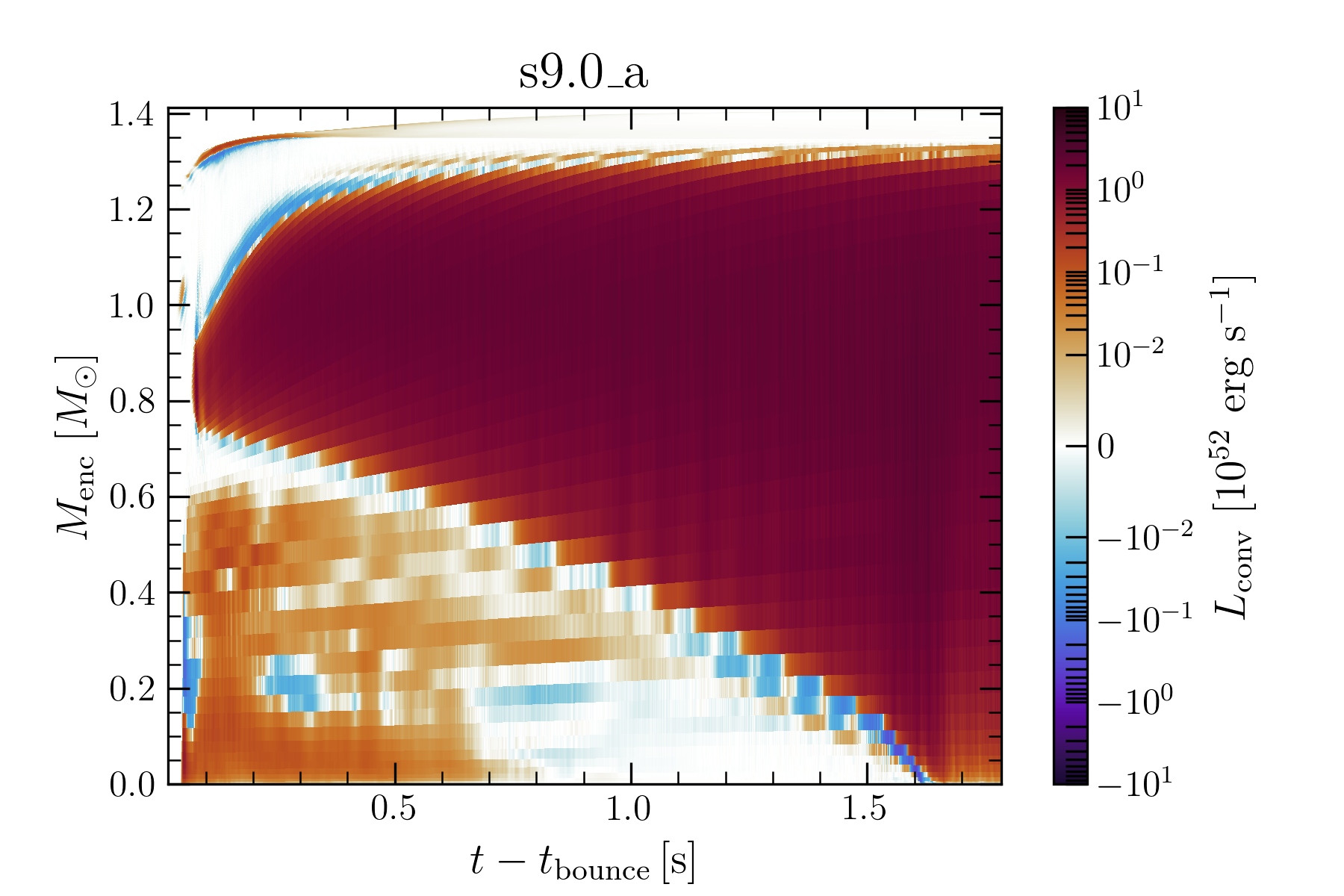}
    \includegraphics[width=0.4\textwidth]{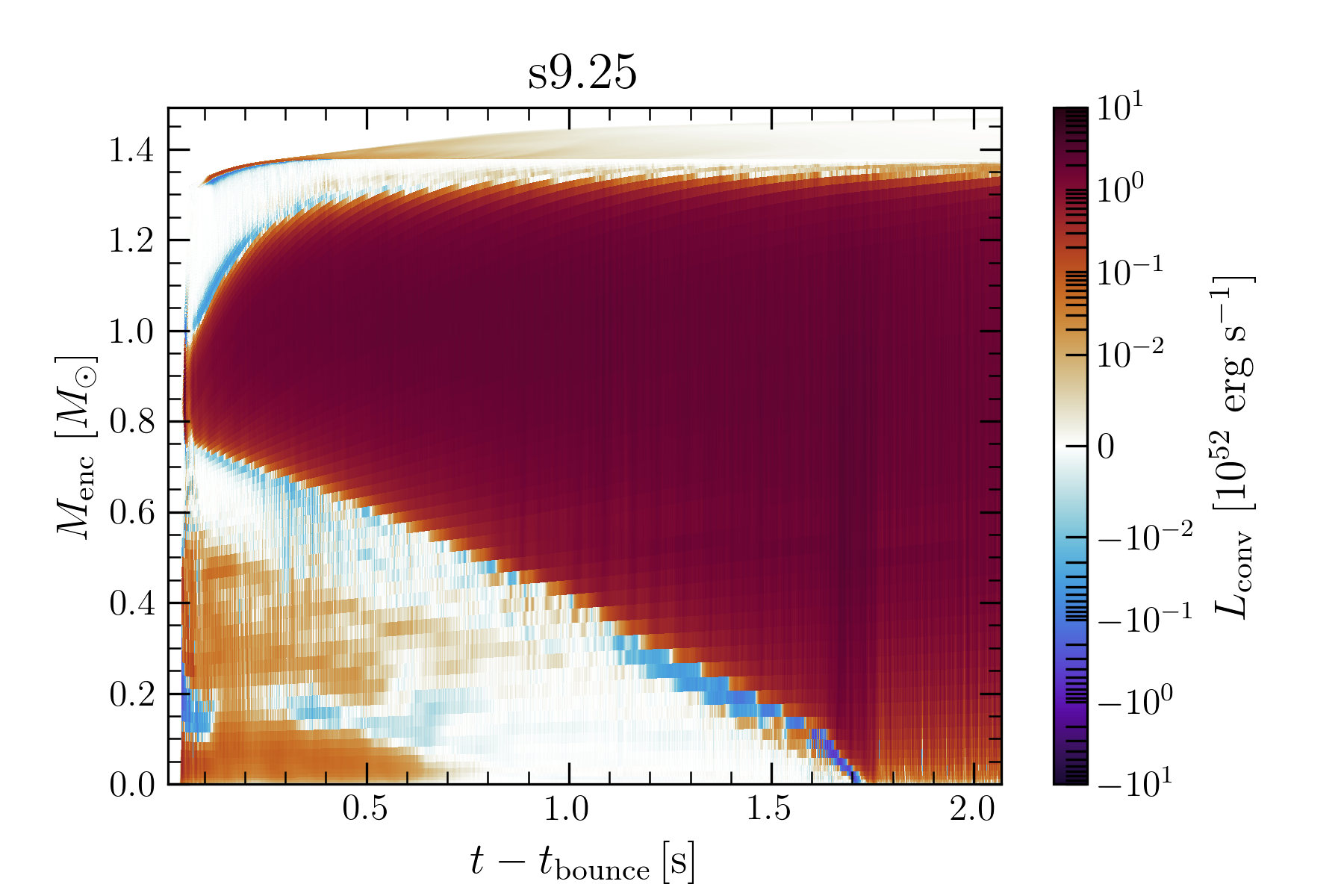}
    \includegraphics[width=0.4\textwidth]{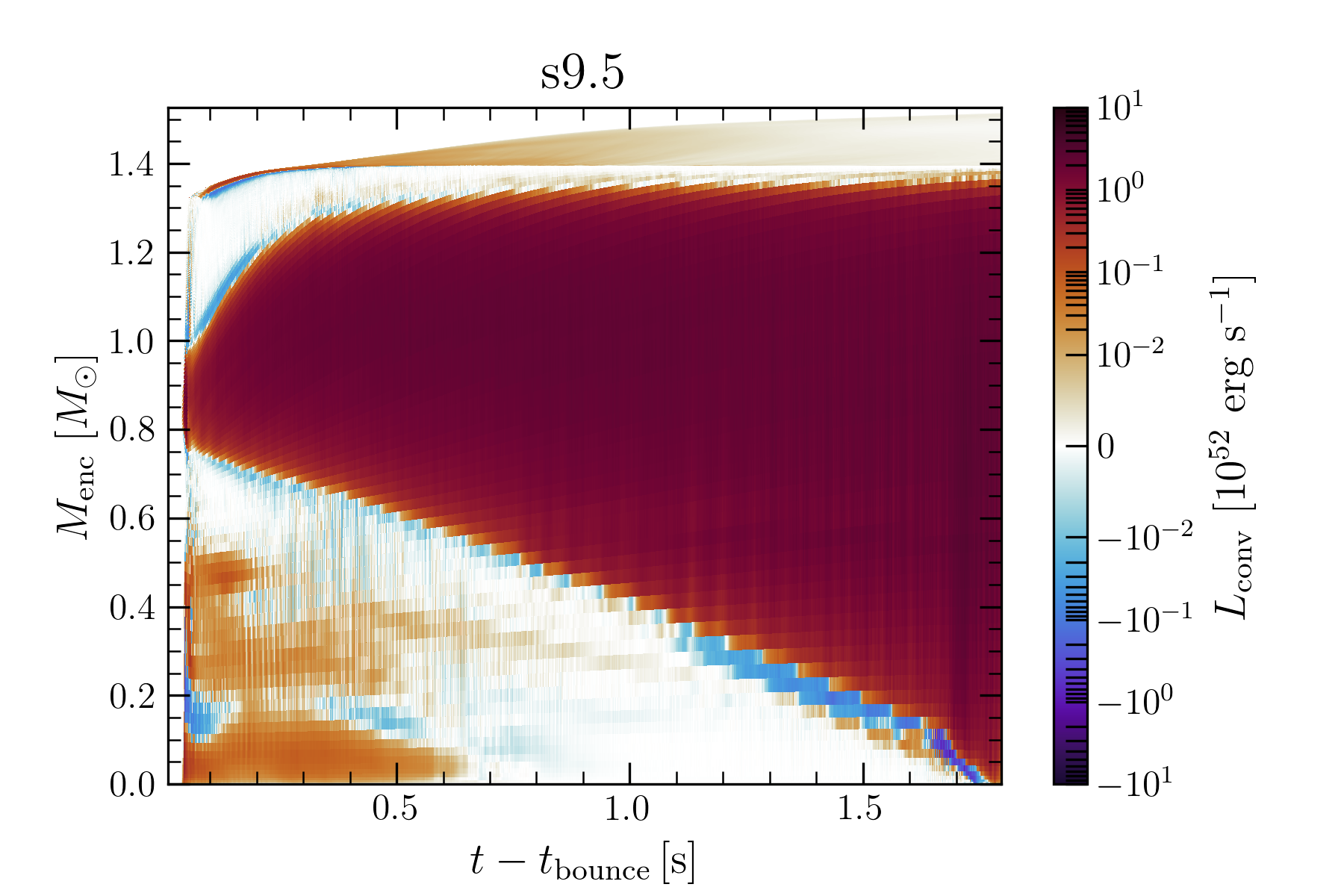}
    \includegraphics[width=0.4\textwidth]{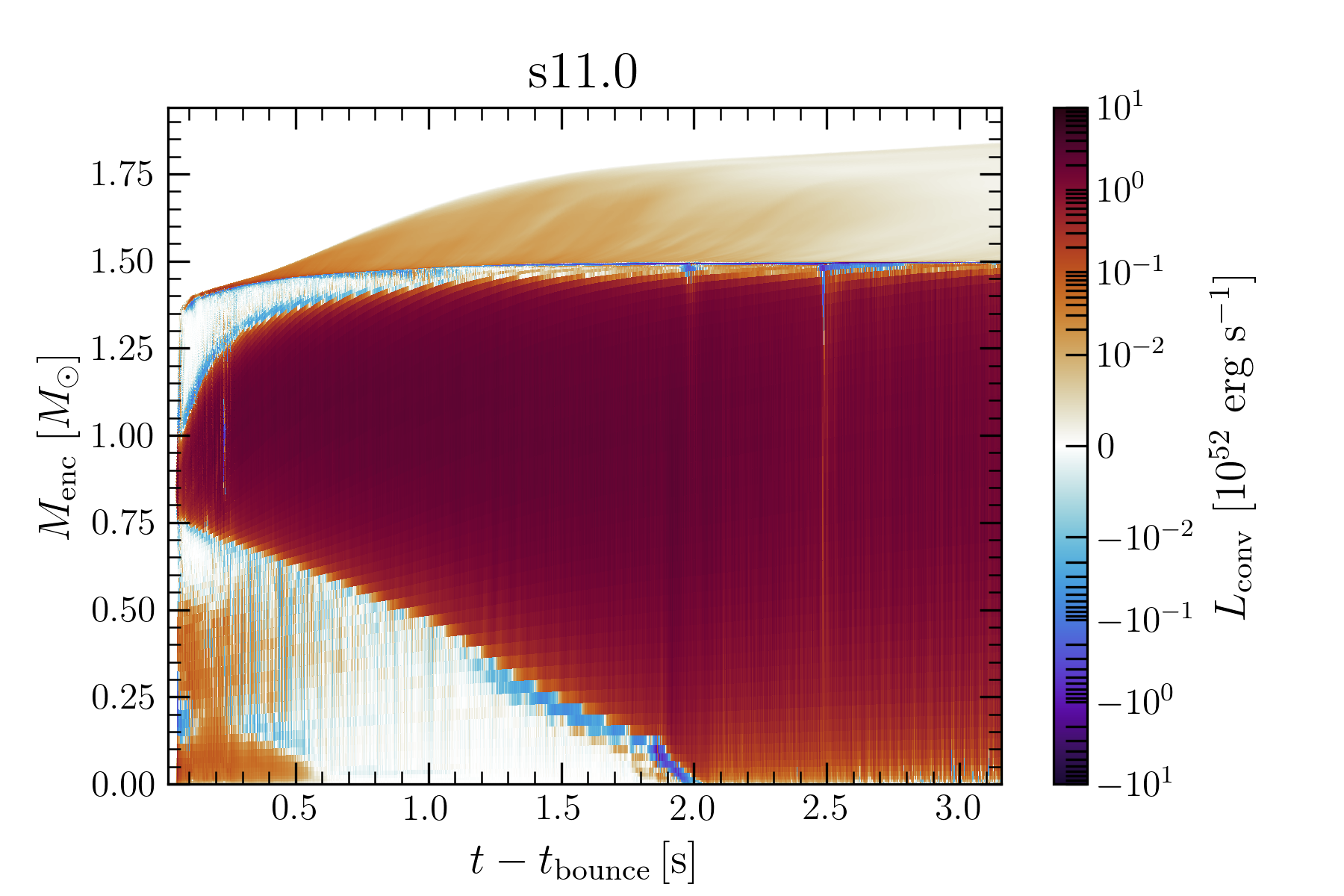}
    \includegraphics[width=0.4\textwidth]{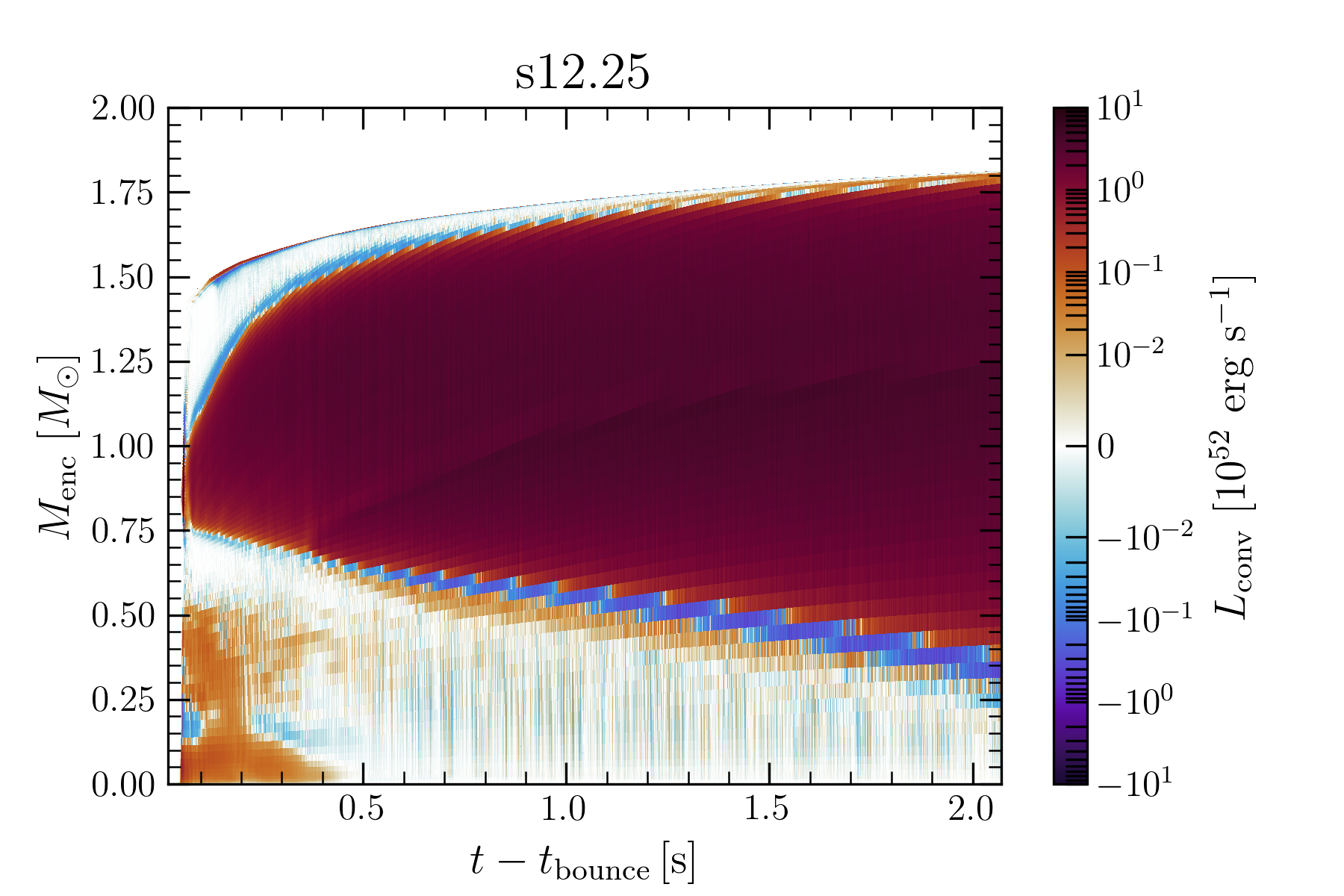}
    \includegraphics[width=0.4\textwidth]{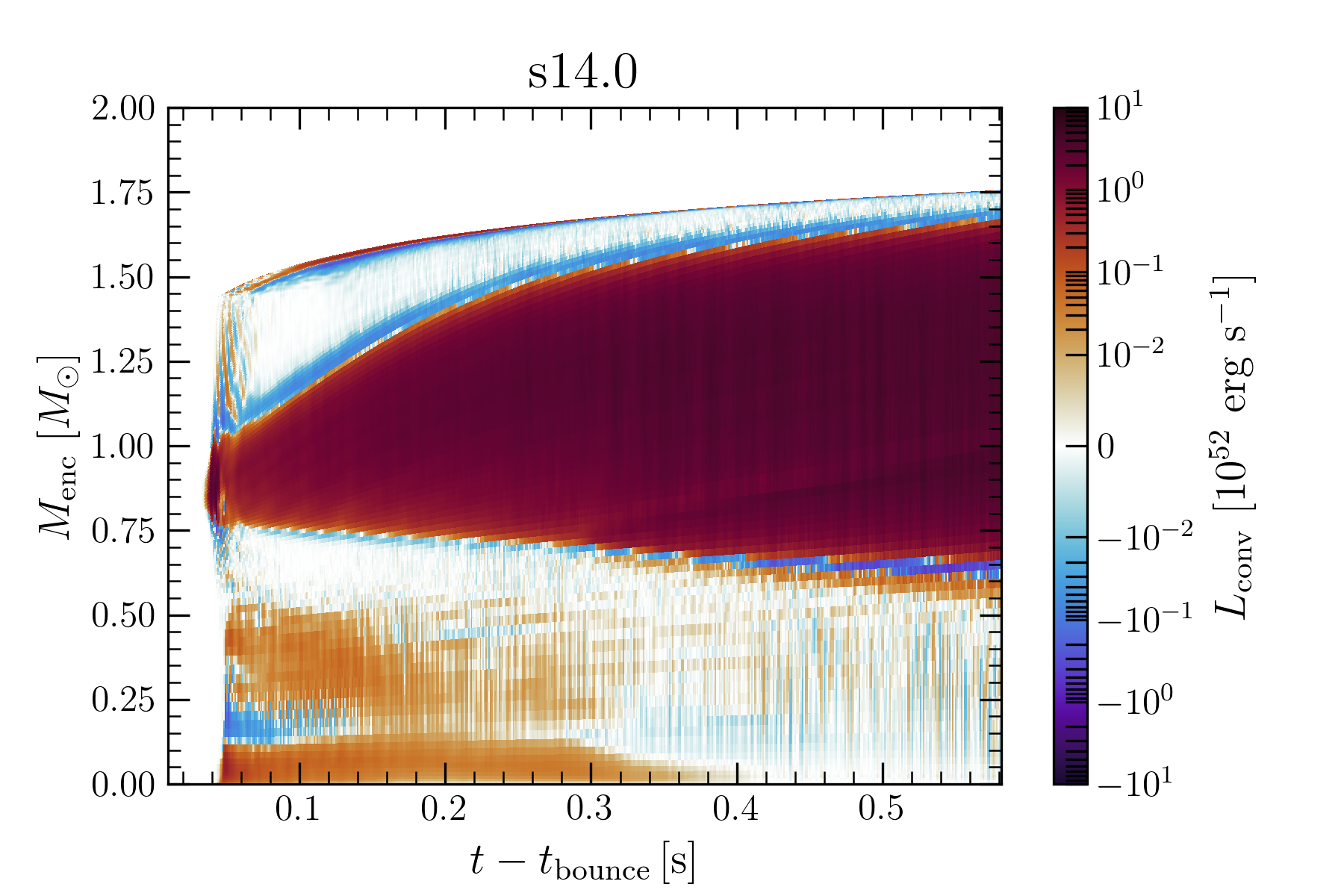}
    \includegraphics[width=0.4\textwidth]{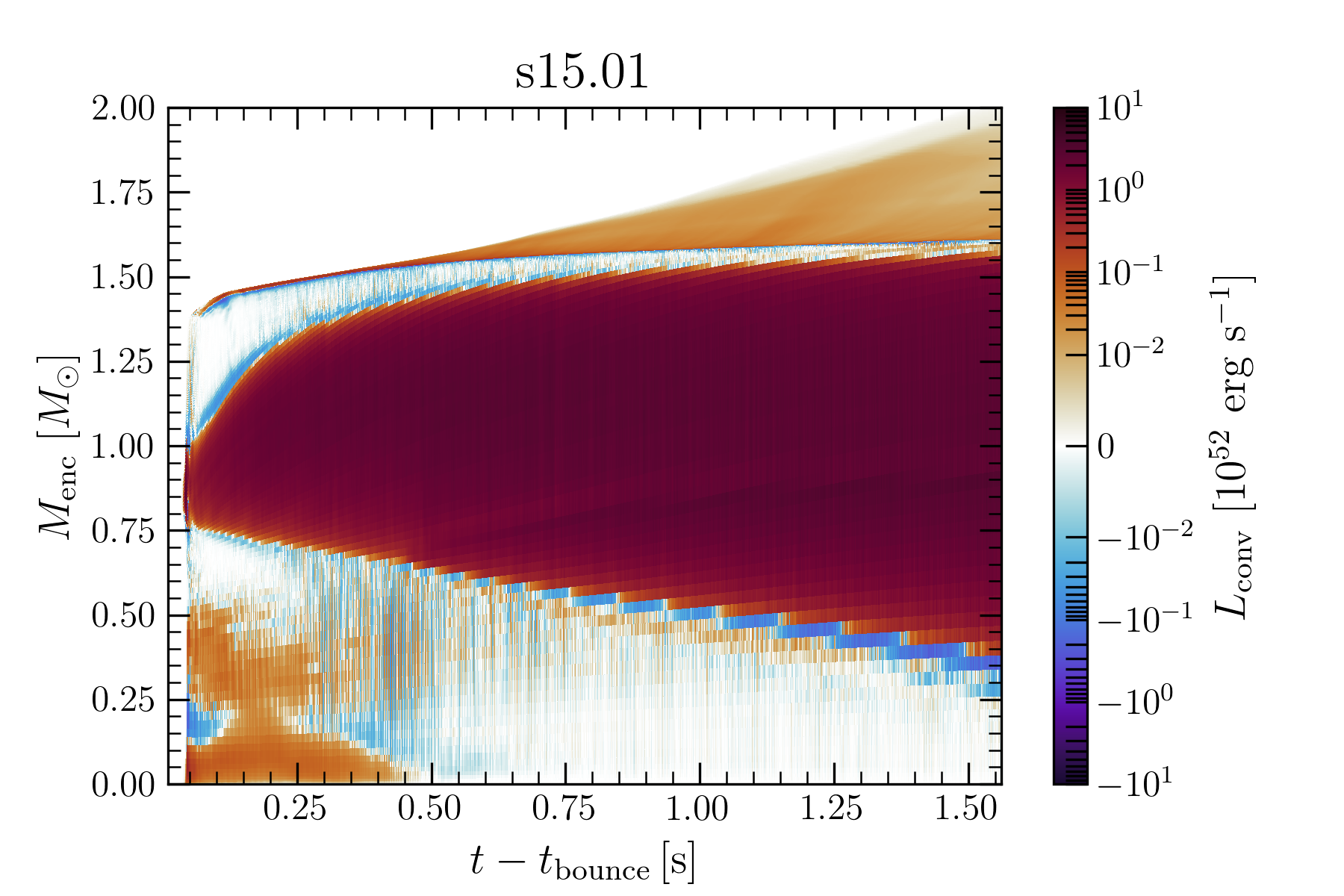}
    \includegraphics[width=0.4\textwidth]{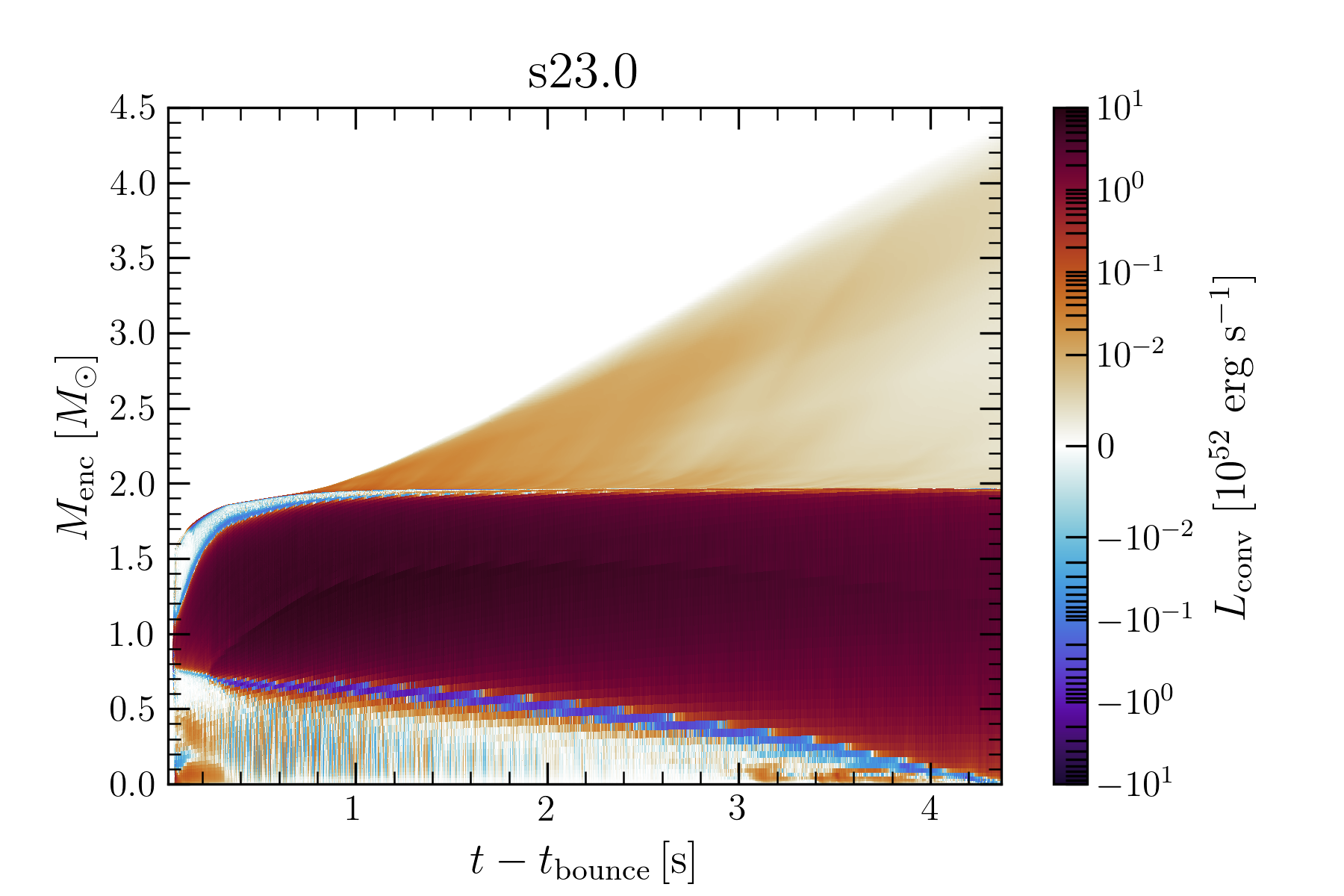}
    \caption{Spacetime diagrams illustrating the convective regions in the PNS for eight representative models, identified by the convective luminosity, as a function of time after bounce. Note that convection reaches the PNS core within $\sim$1.6 to $\sim$4.0 seconds, earlier for the low-mass progenitors and later for the 23-M$_{\odot}$ model. For all models, PNS convection starts in a shell and then grows to encompass much of the core.  The fact that g-modes are evanescent in convective regions bears on the suppression of g-mode excitation in general, but also on the eventual severing of the interaction/coupling between a trapped g-mode interior to it and the PNS surface where the plumes are exciting much of the GW emission.  See text for a discussion (\S\ref{avoided}).}
    \label{fig:pns_cmap}
\end{figure*}

G-modes, however, are generally at low frequencies below $\sim$500 Hz and don't contribute much to the GW signature.  Moreover, the presence of lepton-gradient-driven PNS convection \citep{2006ApJ...645..534D,nagakura_pns} introduces a
region in the PNS for which g-modes are evanescent and non-propagating. Figure \ref{fig:pns_cmap} depicts the evolution of PNS convection for most of the models presented in this work. One sees clearly that the extent of PNS convection starts in a narrow shell, but grows wider with time.  By $\sim$1.6 to $\sim$4.0 seconds
PNS convection has grown to encompass the center and most of the residual PNS and will persist beyond the simulation times of this study \citep{roberts_12c}. During the early post-bounce phase, though most g-modes have frequencies too low to couple with the f- and p-modes, at an early stage before the region of PNS convection has grown too thick it is possible for a g-mode trapped mostly interior to PNS convection to couple with them. With time, the coupling will be broken by the evolving thickness of the convective shell; it is this growth that eventually severs the coupling with the outer regions where the impinging plumes are providing the excitation and that leads to the jump to the pure f-mode. 
However, much work remains to be done to fully demonstrate the details of this coupling and ``bumping" transition.  Nevertheless, the manifest presence of this avoided crossing in the GW spectrograms and in the GW signature of core collapse universally is an interesting direct marker of the presence of PNS convection that deserves further study. 


\subsection{The Angular Anisotropy of the Matter Gravitational-Wave Emissions}
\label{anisotropy}

\begin{figure*}
    \centering
    \includegraphics[width=0.49\textwidth]{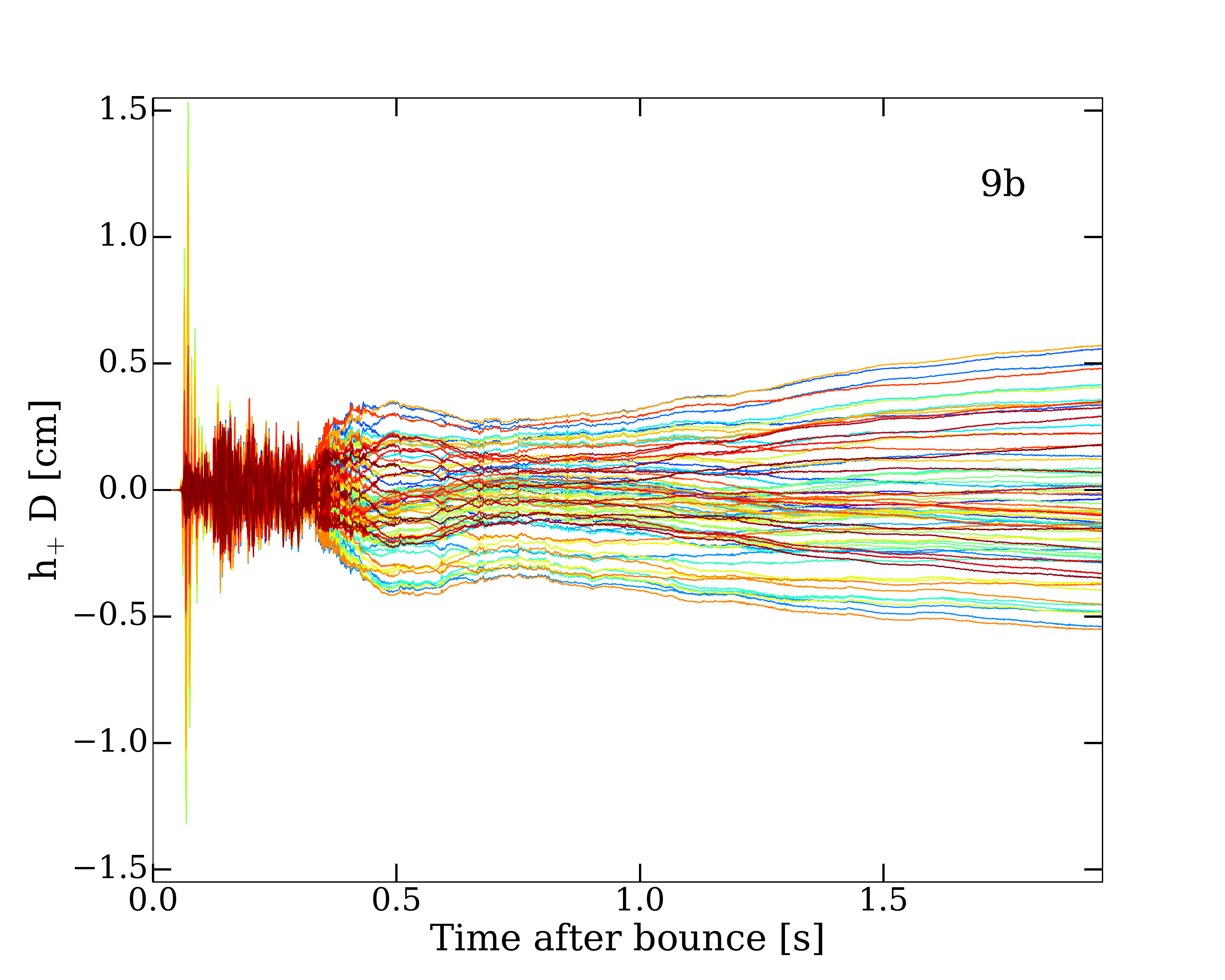}
    \includegraphics[width=0.49\textwidth]{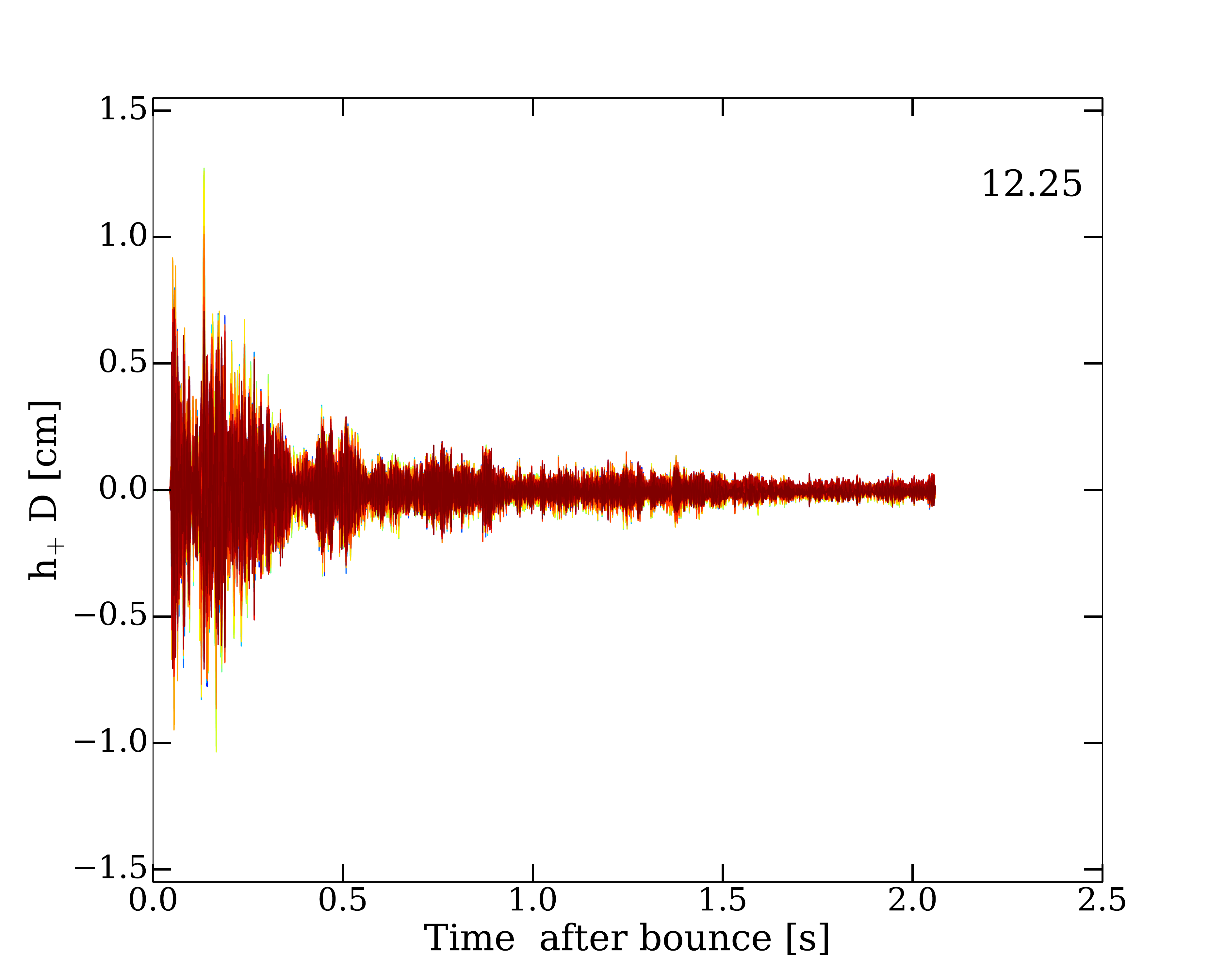}
    \includegraphics[width=0.49\textwidth]{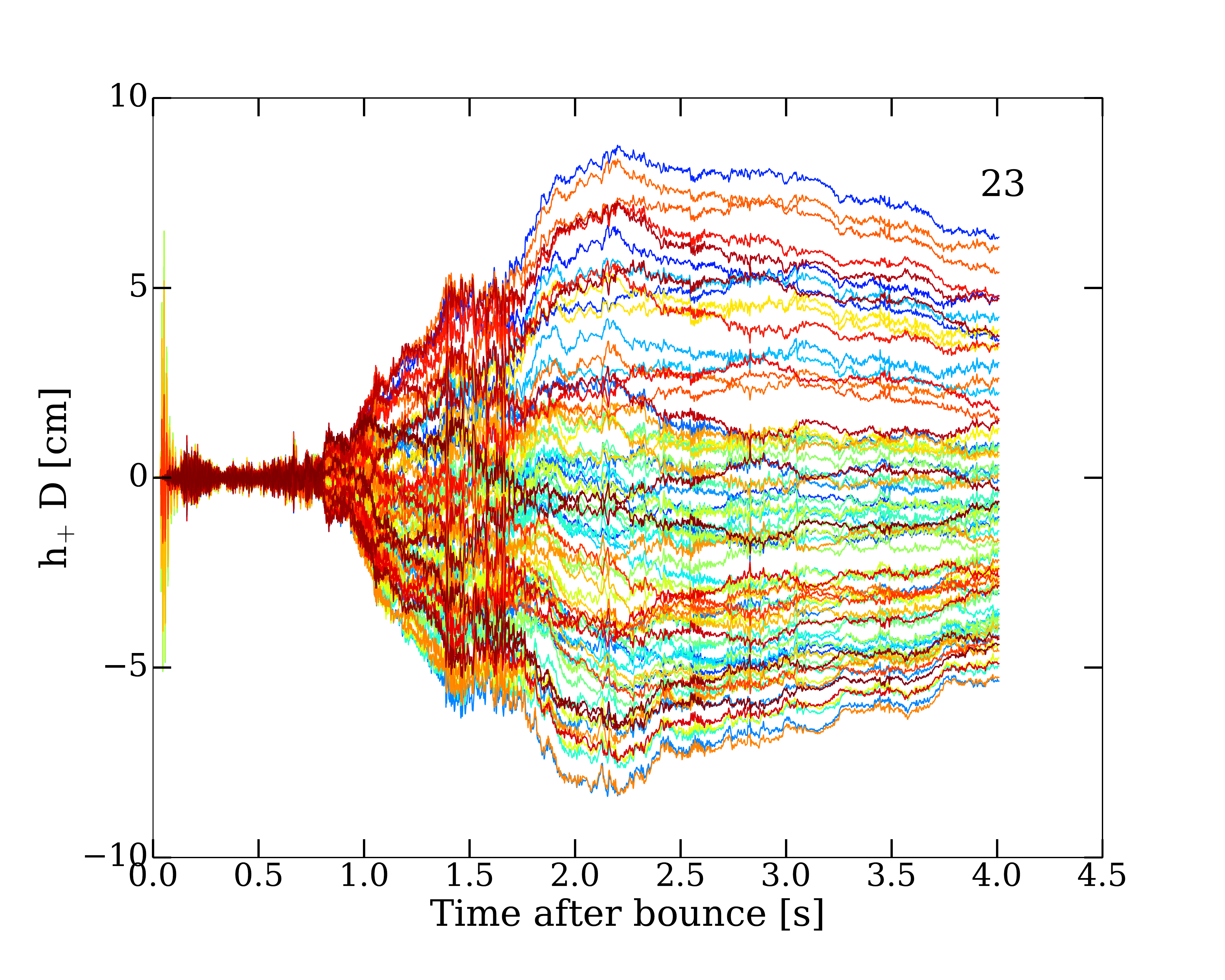}
    \caption{The gravitational-wave strain h$_+$D (cm) as a function of time after bounce (in seconds) for the 9b model, the non-exploding model 12.25-M$_{\odot}$, and model 23-M$_{\odot}$ for various, arbitrary, viewing angles illustrated by color. These plots do not include the neutrino memory (\S\ref{memory}). Note the large anisotropy by viewing angle, which is dominated at late times by the matter memory, which is sourced by the large-scale explosion asymmetry. In the black hole model (12.25-M$_{\odot}$), we see no memory, and little anisotropy, at late times. Though we show only the $+$ polarization, the results are consistent for both polarizations. We see that in principle some measure of the degree and angle of the matter explosion asymmetry can be discerned by the sign and magnitude of this memory.  However, the neutrino memory, if only weakly correlated with the matter memory, will complicate the extraction of such a feature. Nevertheless, the difference in the frequency spectra of the matter and neutrino memories generically, with the former at a higher average frequency, may allow one someday to disentangle the two.}
    \label{fig:matter_anis}
\end{figure*}

In Figure\,\ref{fig:matter_anis}, we plot the matter strain for the 9b, 12.25-, and 23-M$_{\odot}$ models as a function of time after bounce to illustrate the anisotropy with various,  arbitrarily chosen viewing angles. Note that significant anisotropy manifests at late times in the (low-frequency) memory component, which captures the large-scale asymmetry in the explosion ejecta. The non-exploding 12.25-M$_{\odot}$ model, by comparison, has virtually no anisotropy. 
The early high-frequency component is stochastic, whereas the low-frequency late-time memory show secular time-evolution that does not average to zero and indicates a metric shift, reaching values of $\sim$5 cm for the various massive models (showing a general trend with the progenitor mass/explosion asymmetry). We find similar significant anisotropy in the (low-frequency) neutrino memory component (discussed in \S\ref{memory}). Importantly, however, when calculating the total inferred ``isotropically-equivalent" radiated GW energy, which is dominated by the higher-frequency component in and near the LIGO band, as a function of angle we find that it varies by $\sim$10 to $\sim$15\% around an angle-averaged mean.  This implies that, though the higher-frequency emissions are indeed anisotropic, the integrated high-frequency signals are only weakly dependent on angle. 

\subsection{The Neutrino Memory Component}
\label{memory}

In addition to the matter memory, asymmetries in the emission of neutrinos produce another low-frequency memory component, the neutrino memory \citep{epstein1978,burrows1996,emuller,2020ApJ...901..108V,2022PhRvD.105j3008R,2022PhRvD.106d3020M}. In Figure\,\ref{fig:strain_nu}, we plot the GW strain due to anisotropic neutrino emission as a function of time after bounce for the models studied here. The neutrino strain is significantly larger in magnitude than the matter contribution, reaching over 1000 cm for the most massive progenitors. There is generally a hierarchy of strain amplitude with progenitor mass, reflecting the sustained turbulent accretion in more massive progenitors, which results in higher neutrino luminosities and generally more anisotropic explosions. The 11-M$_{\odot}$ model is an exception and fields the highest strain amplitude. In addition, the neutrino memory shows much lower frequency evolution and more secular time-evolution than the matter component, which is fundamentally because it is a cumulative time-integral of the anisotropy-weighted neutrino luminosity (see Appendix \,\ref{appendix} as well as \cite{vartanyan2019}). The difference in the mean frequencies of the neutrino and matter memories may, therefore, provide a means someday to distinguish them observationally.

In Figure\,\ref{fig:EGW_neutrino}, we plot the GW energy due to neutrino emissions as a function of time after bounce for our various 3D models. There are several key and distinguishable features. First, like the GW energy due to matter motions displayed in Figure\,\ref{fig:EGW}, we see growth by over two orders of magnitude (but not three, as in the former) in the neutrino memory. Additionally, we generally see a hierarchy with progenitor mass, however with the 11-M$_{\odot}$ surpassing the 23-M$_{\odot}$ until $\sim$4.5 seconds. Unlike the matter component of the GW energy, the neutrino component shows sustained growth for our longest duration model (the 23-M$_{\odot}$ model). In comparison with Figure\,5 from \cite{vartanyan2020}, this emphasizes again the need to carry simulations out to late times to capture the entire signal. Note that, despite the higher strains seen in the neutrino component of the GW signature and its sustained growth, due to the much smaller frequencies it is still more than two orders of magnitude less energetic than the matter-sourced GW energy. In addition, though both components capture the development of turbulence, the neutrino component does not show a prompt convective phase and begins to develop $\sim$100 to 200 ms later than the matter component. 

As with the matter component, the neutrino component is most pronounced for delayed explosions of models with higher compactness reflecting their more vigorous turbulent accretion history and more anisotropic explosions. 

\subsection{Turbulent Accretion Excites Gravitational-Wave Emission}
\label{excitation}

\begin{figure*}
    \centering
    \includegraphics[width=0.34\textwidth]{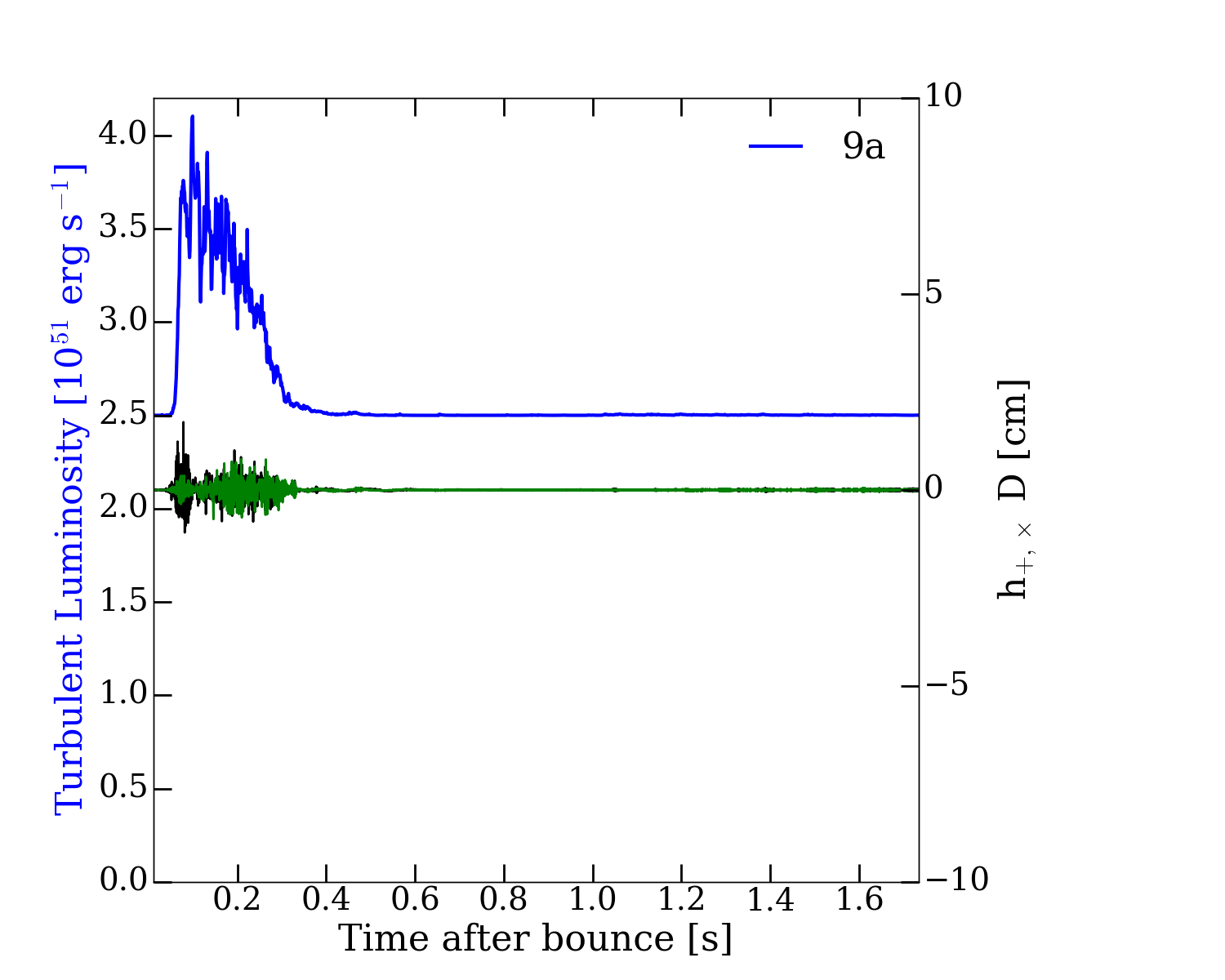}
    \includegraphics[width=0.34\textwidth]{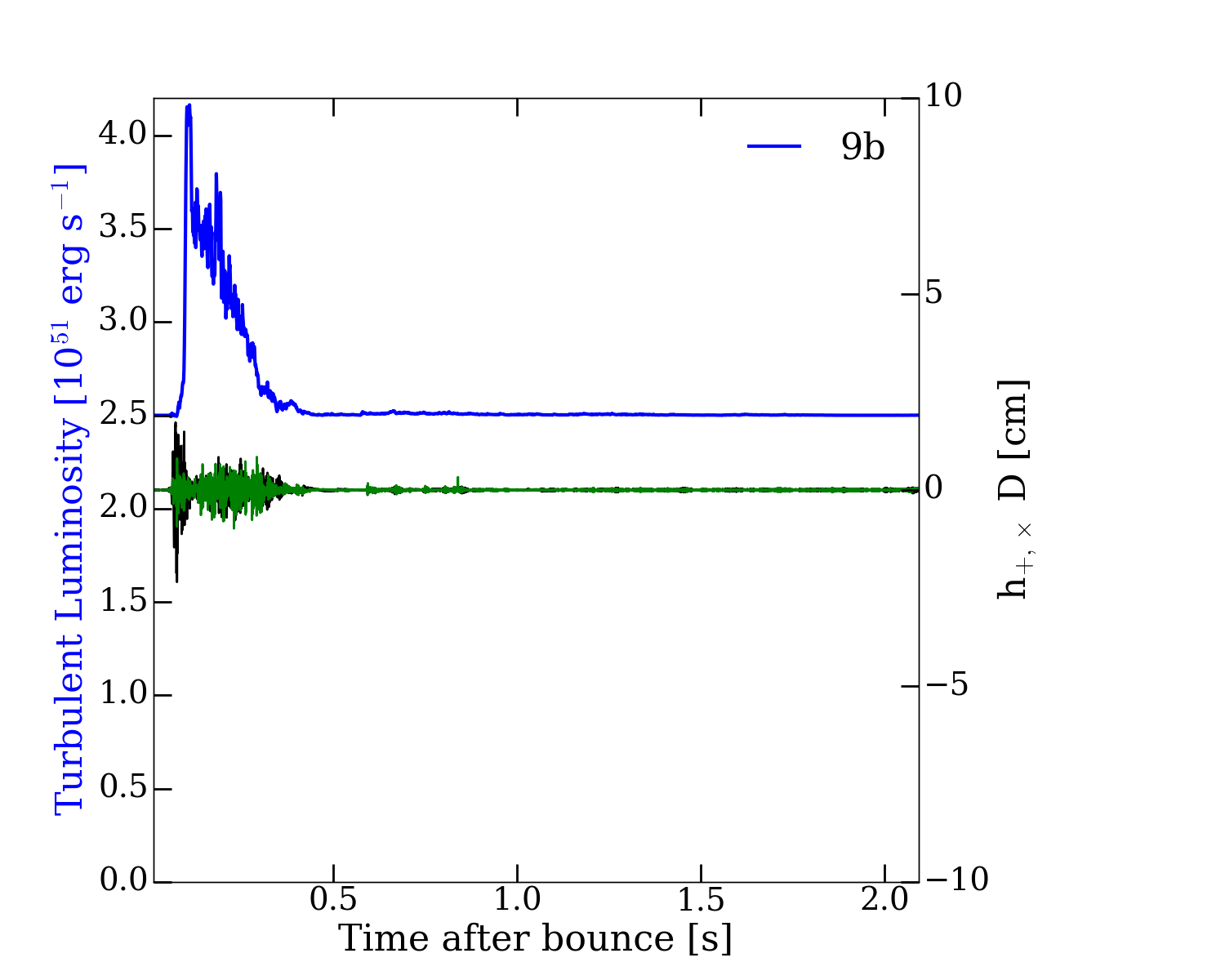}
    \includegraphics[width=0.34\textwidth]{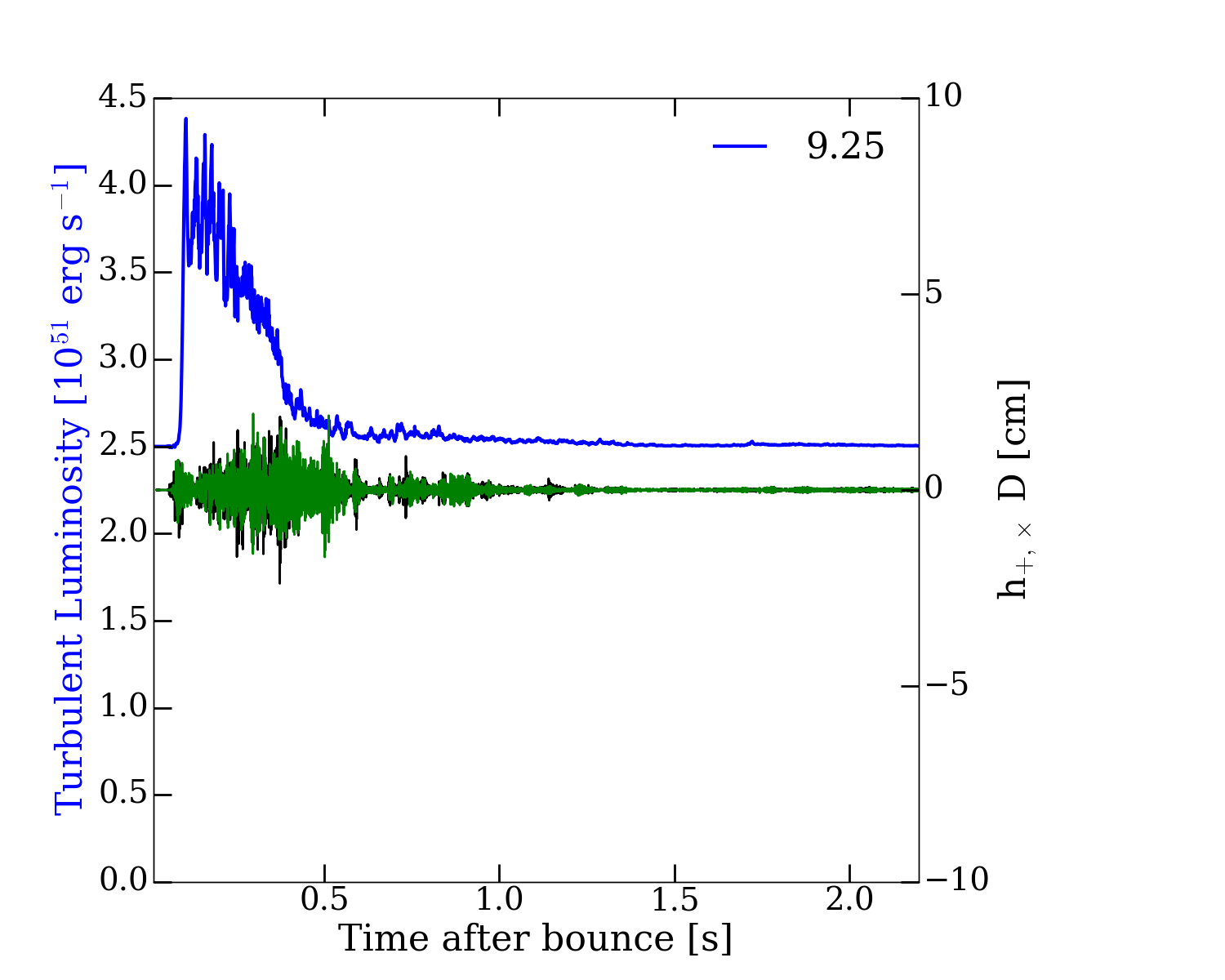}
    \includegraphics[width=0.34\textwidth]{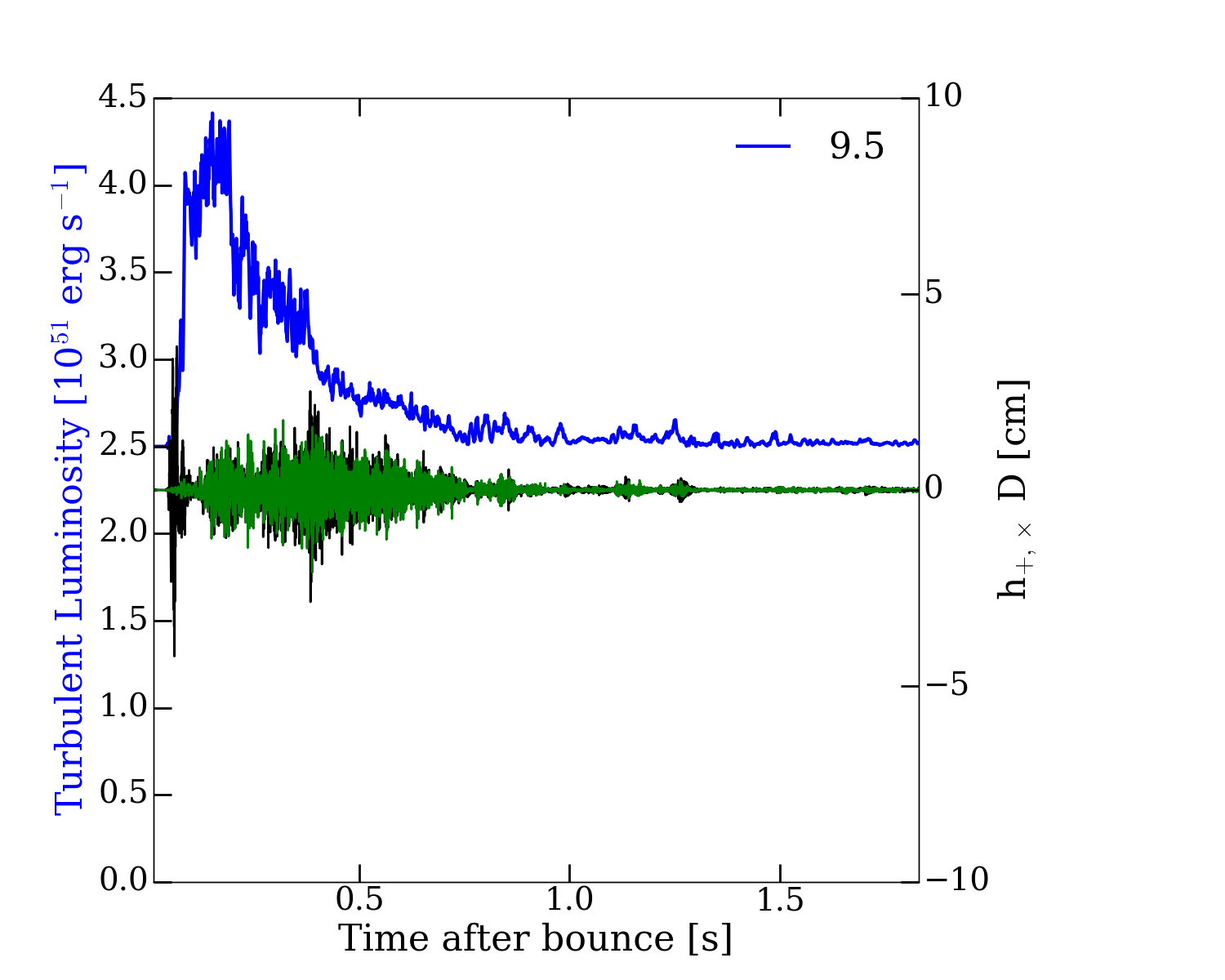}
    \includegraphics[width=0.34\textwidth]{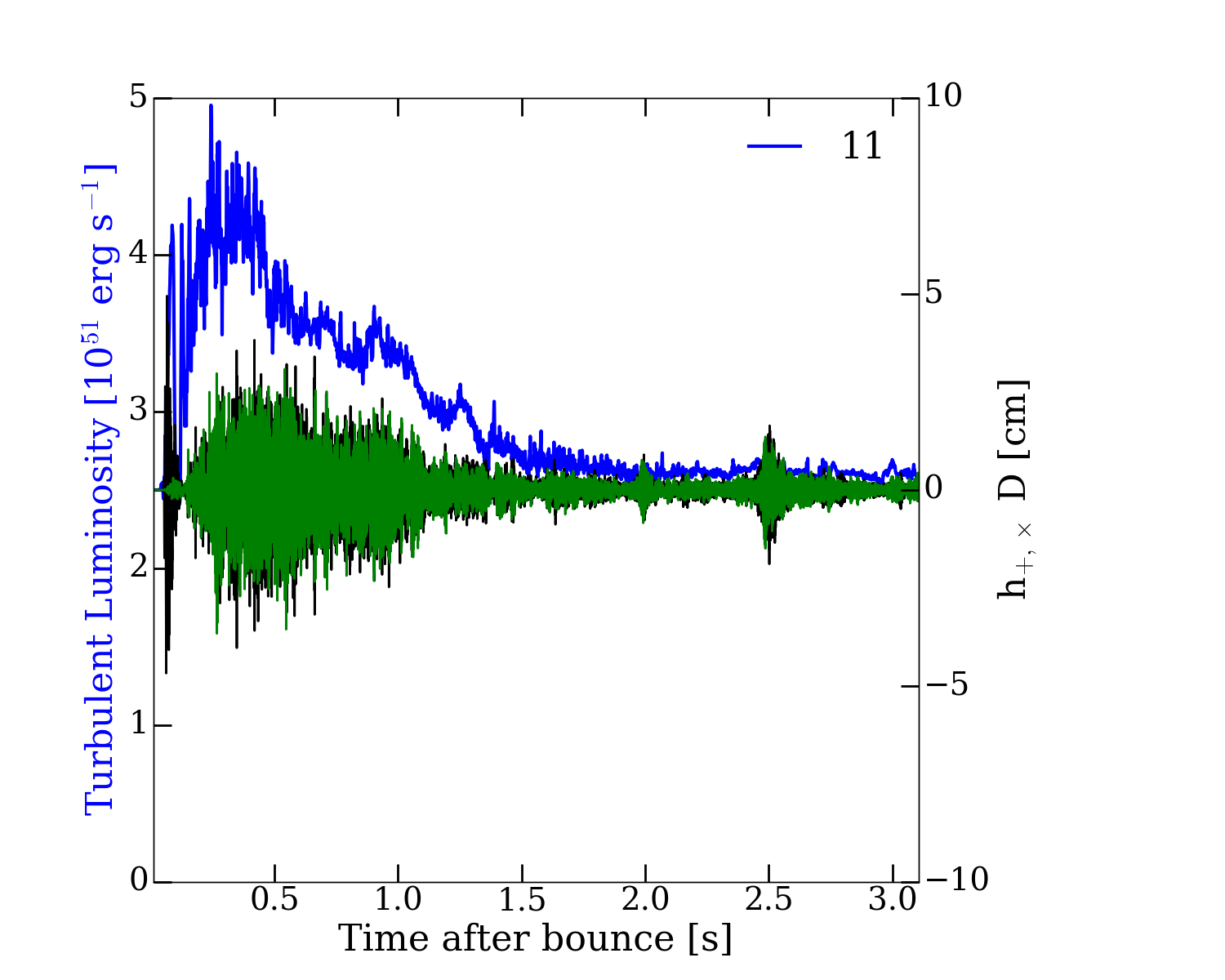}
    \includegraphics[width=0.34\textwidth]{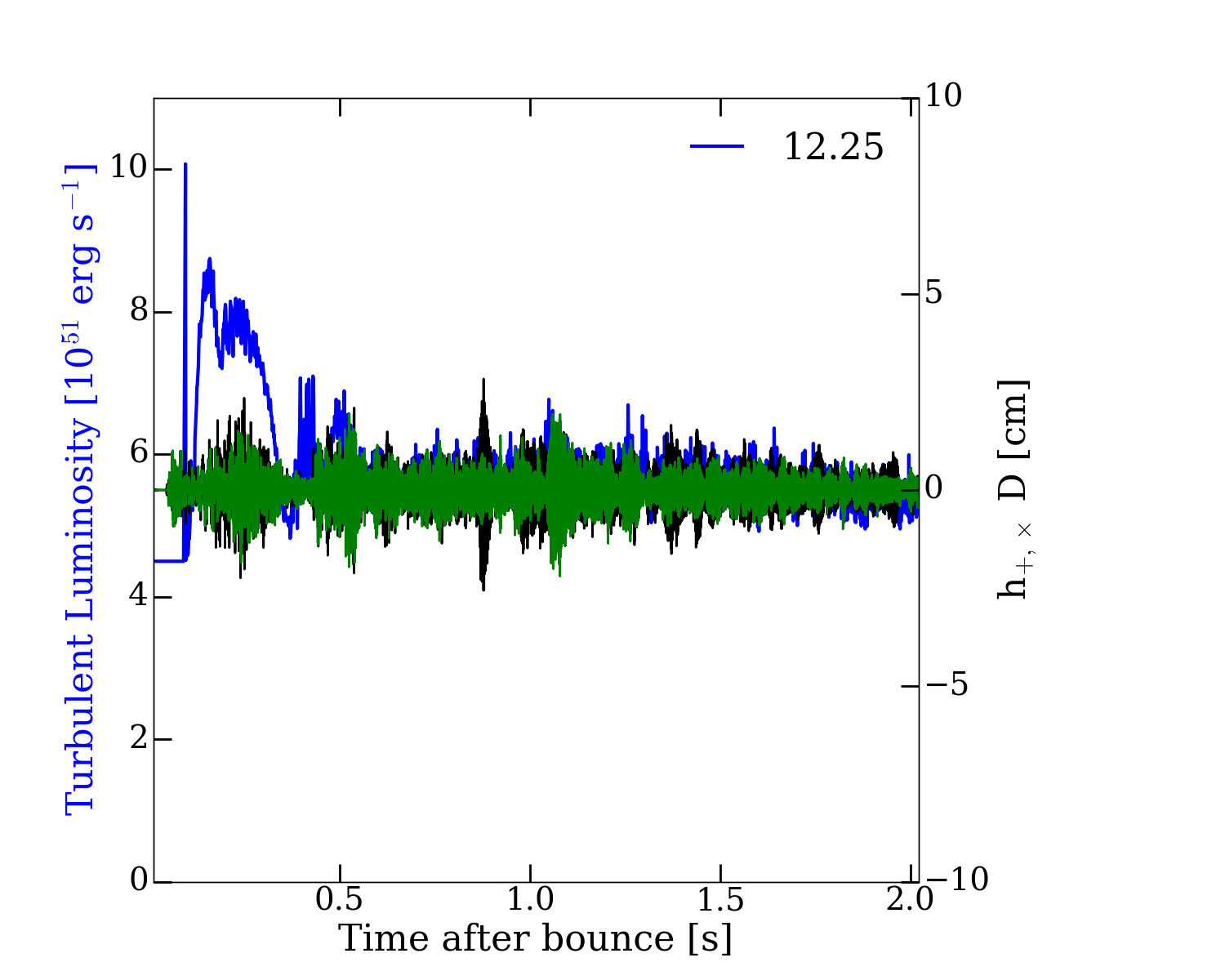}
    \includegraphics[width=0.34\textwidth]{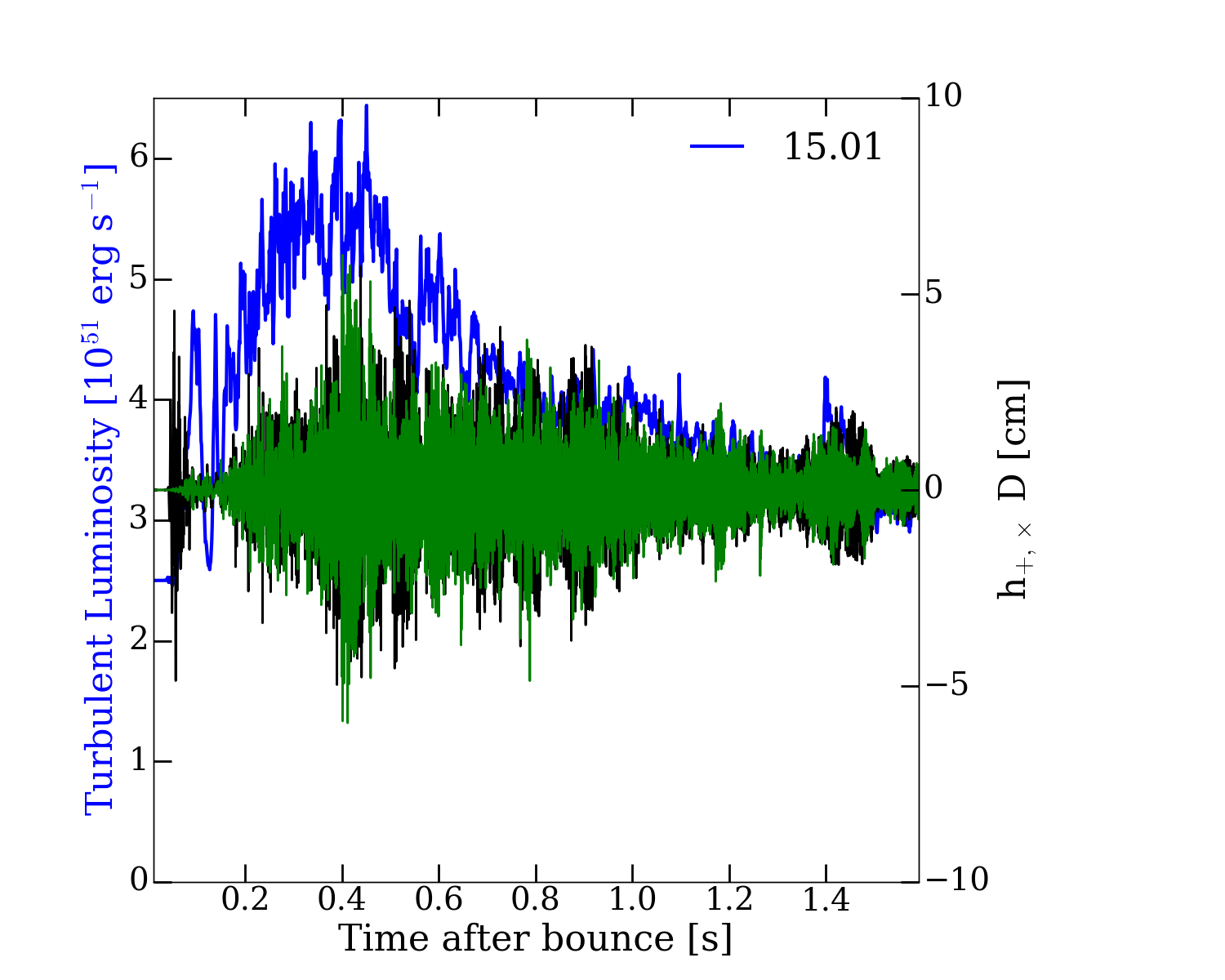}
    \includegraphics[width=0.34\textwidth]{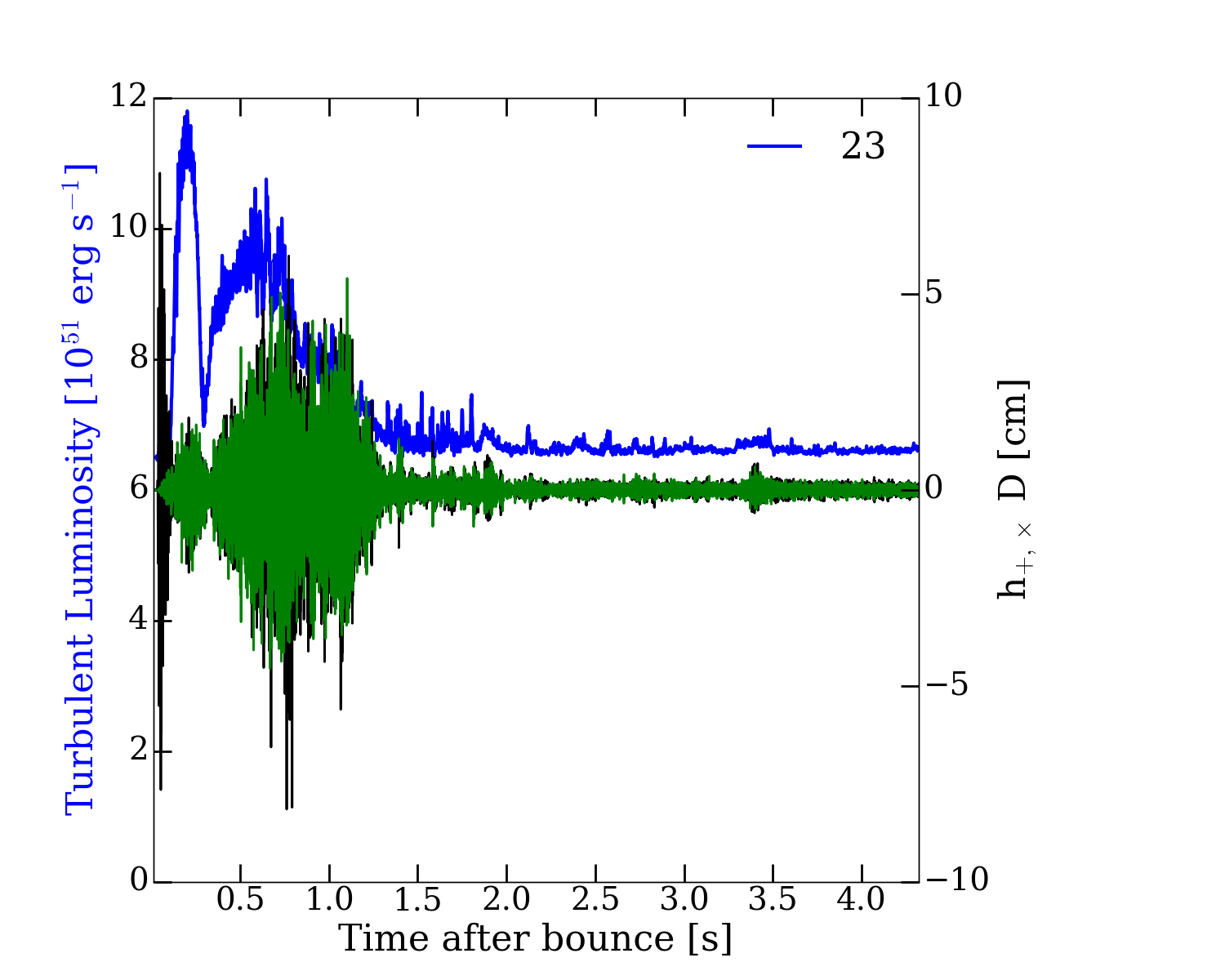}
    \caption{Turbulent luminosity (blue, in 10$^{51}$ erg s$^{-1}$) at 110 km (except for model 12.25-M$_{\odot},$ where it is evaluated at 2.5 times the PNS radius), overplotted with both filtered GW strain polarizations (green and black) multiplied by distance (product in cm) as a function of time after bounce (in seconds). We emphasize the clear correlation shown here $-$ the outer turbulent luminosity closely follows the strain. It reproduces both the global time behavior, rising and following at coincident times with the strain, and the episodic `packets' of accretion (for instance, reproducing the bump in the 15.01-M$_{\odot}$ strain at $\sim$1.35 s and for the 23-M$_{\odot}$ at $\sim$3.5 s). From this figure, it is clear that it is the outer downward turbulent convective flux onto the PNS that drives the gravitational-wave signature after the prompt convection phase.}
    \label{fig:lconv_corr}
\end{figure*}

The major excitation mechanism of GW from CCSN and black hole formers is the pounding accretion onto the PNS core \citep{murphy:09,muller2017b,2013ApJ...766...43M,kuroda2017,2018ApJ...861...10M,radice2019}.  As shown in \S\ref{description}, much of the GW power comes out at the frequencies associated with the pulsational modes of the PNS. To demonstrate the correlation of the gravitational strain with the matter accretion, we first remove the matter memory by applying a high-pass Butterworth filter below 15 Hz to the strains. Then, in Figure\,\ref{fig:lconv_corr} we plot the turbulent hydrodynamic luminosity evolution interior to the shock with this filtered strain timeline. 

The turbulent hydrodynamic flux is defined following \cite{2022ApJ...926..111W} as
\begin{equation}
F_{conv} = \Bigl \langle\left(\frac{1}{2}\rho v_{turb}^2 + u + p \right)v_{turb}^r\Bigr \rangle_{d\Omega}\,,
\end{equation} including the turbulent kinetic energy, the internal energy $u$, and the pressure $p$. The turbulent velocity is defined here as the radial component of the turbulent velocity,

\begin{equation}\label{eq:turb}
    v_{conv}^r = \langle\left(v^r - v_{ave}^r\right)^2\rangle_{d\Omega}^{1/2}\,,
\end{equation} where $v_{ave}^r$ is the density-weighted angle average of the radial velocity. This is calculated at 110 km for all models except the non-exploding 12.25- and 14-M$_{\odot}$ progenitors, whose shocks early on sink below this radius; for these models we calculate the turbulent hydrodynamic flux at 2.5 times the PNS radius (defined here as the density cutoff at 10$^{10}$ g cm$^{-3}$).

The strong correlation throughout their evolution (even in detail) between the turbulent hydrodynamic flux impinging onto the core and the GW strains demonstrates that turbulent accretion through the shock and onto the PNS core is the major agency of GW excitation and emission in CCSN. We note that this correlation was demonstrated
even though the flux was angle-averaged and the strain was for emission along the x-axis. No attempt was made to break
the flux into $\ell = 2; m=[-2, -1, 0, 1, 2]$ components, yet the correlation is clear.

To provide another perspective, in Figure\, \ref{fig:GW-mdot-corr}, we show the relation between the GW energy spectrogram and the accretion rate ``power" in frequency components with $f>25$ Hz (orange lines) for the 9.5- and 23-M$_{\odot}$ models. As the panels demonstrate, after the explosion and the period of heavy infall subsides (during which the effects of individual accretion events overlap), the accretion rate power shows clear correlations with excursions in the GW energy spectrogram. Spikes and gaps on the spectrogram coincide directly with the peaks and troughs on the curve, meaning that the GWs are excited by, or at least correlated with, the short-period variations in accretion rate onto the core. Such mass accretion rate variations directly tie both episodic fallback and outflow events with the GW emission. 

We note that the colormap used for these plots reveals, particularly for the 23-M$_{\odot}$ model, some power in the g-mode ``bumped" by the f-mode and the repulsion between the two modes around $\sim$0.8 seconds (see \S\ref{avoided}). Though weak, for the 23-M$_{\odot}$ model this signature continues almost to $\sim$2.0 seconds, and perhaps beyond.

\subsubsection{Possible Secondary Role of Proto-Neutron-Star Convection}
\label{pns_convection}

\begin{figure*}
    \centering
    \includegraphics[width=0.37\textwidth]{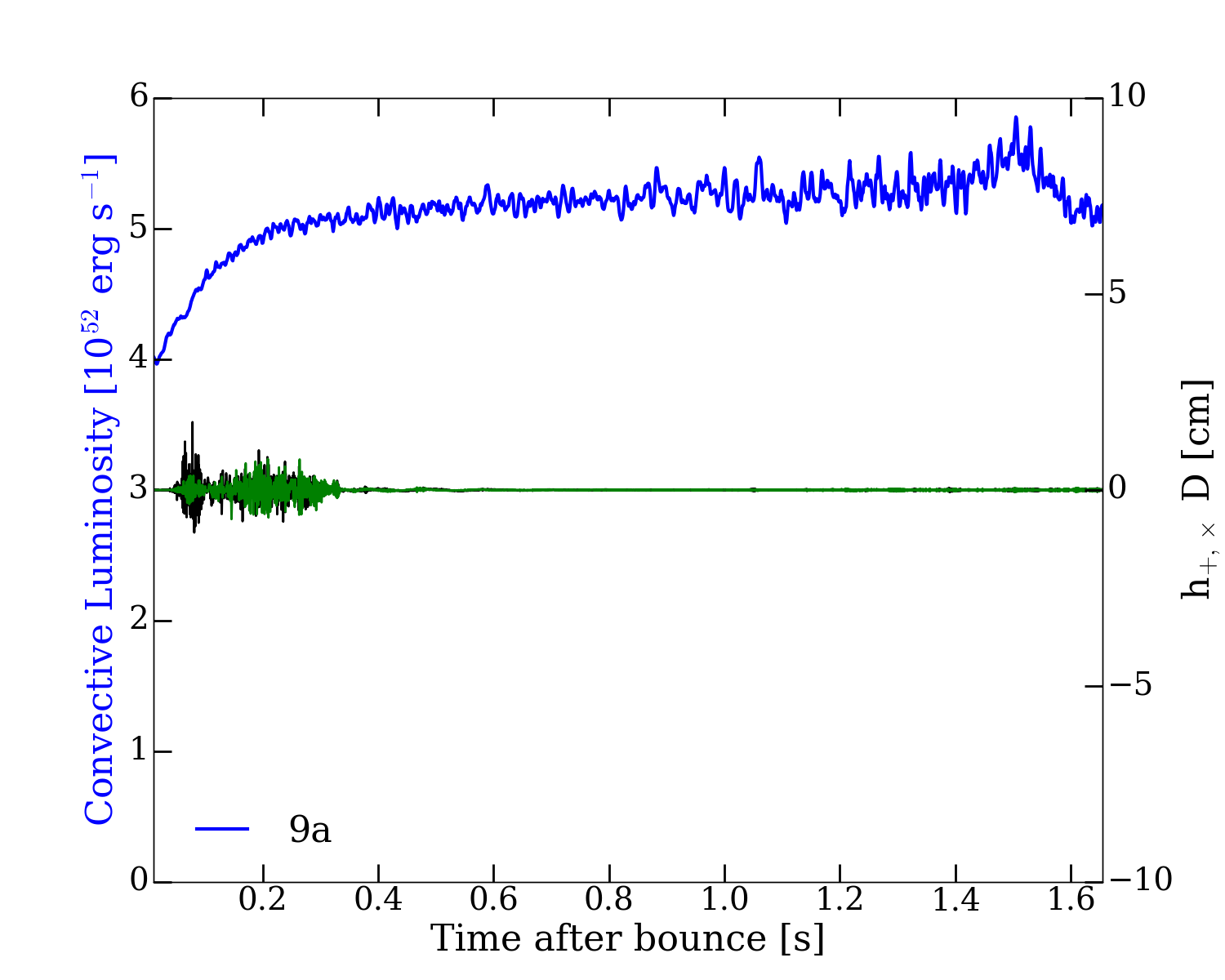}
    \includegraphics[width=0.37\textwidth]{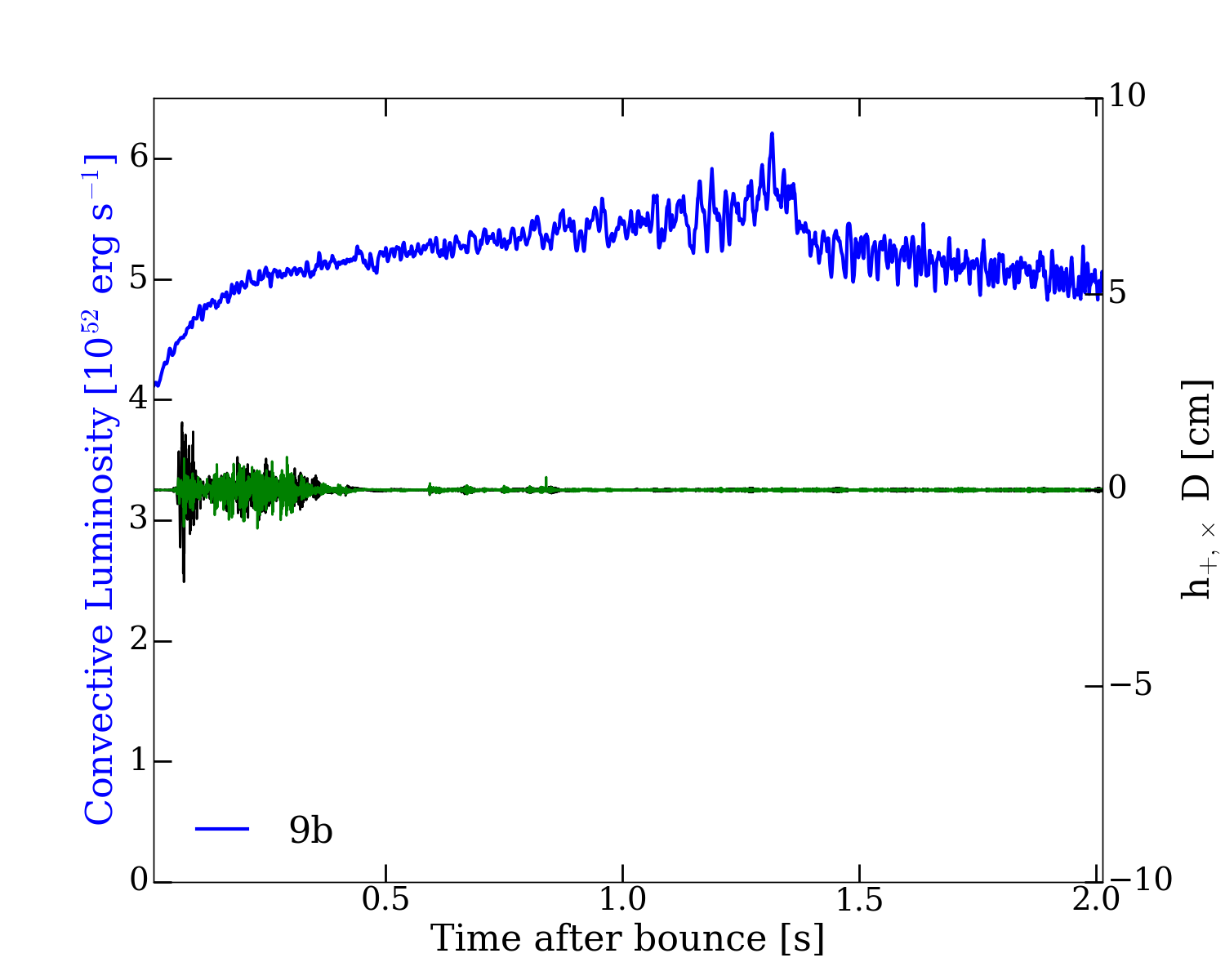}
    \includegraphics[width=0.37\textwidth]{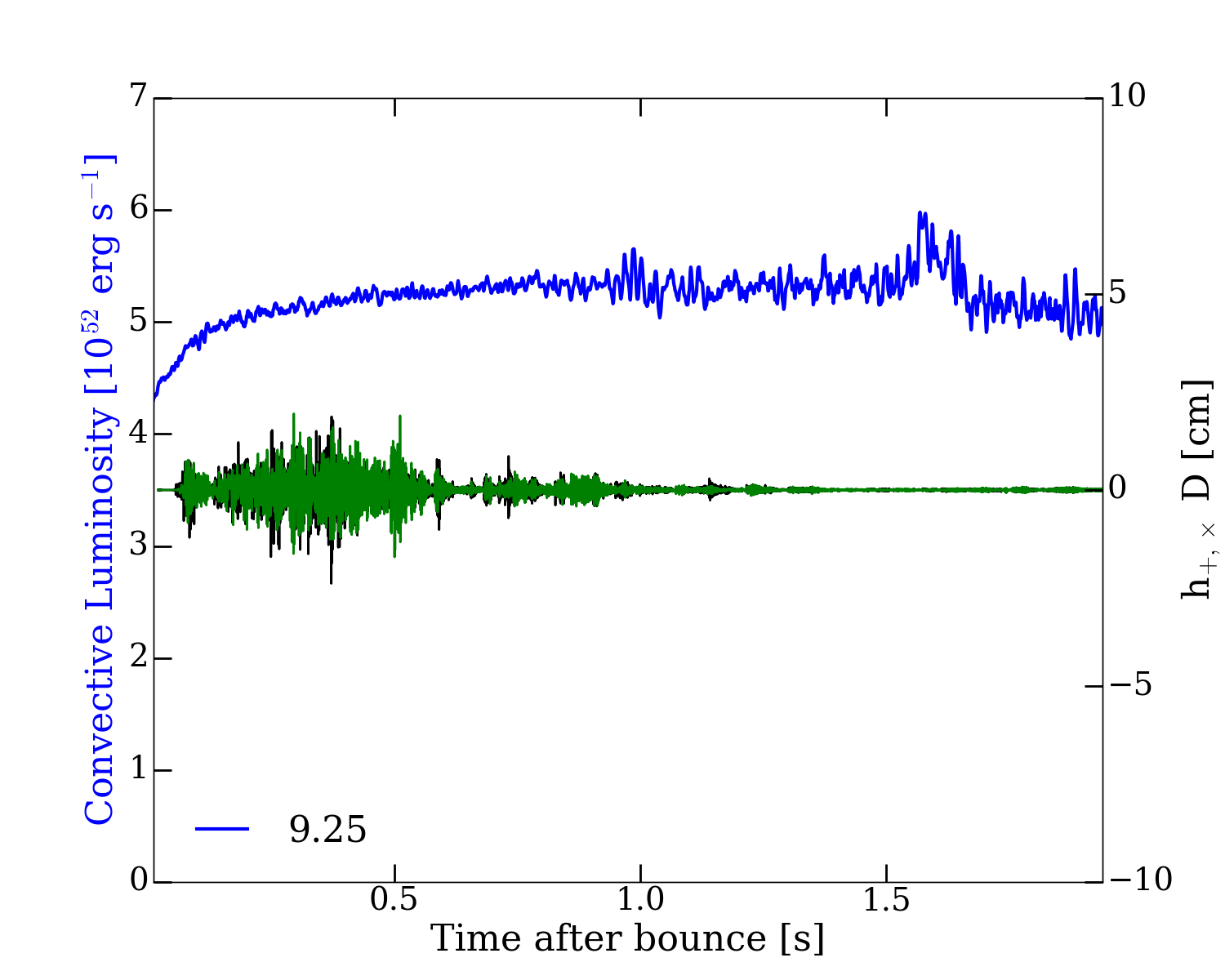}
    \includegraphics[width=0.37\textwidth]{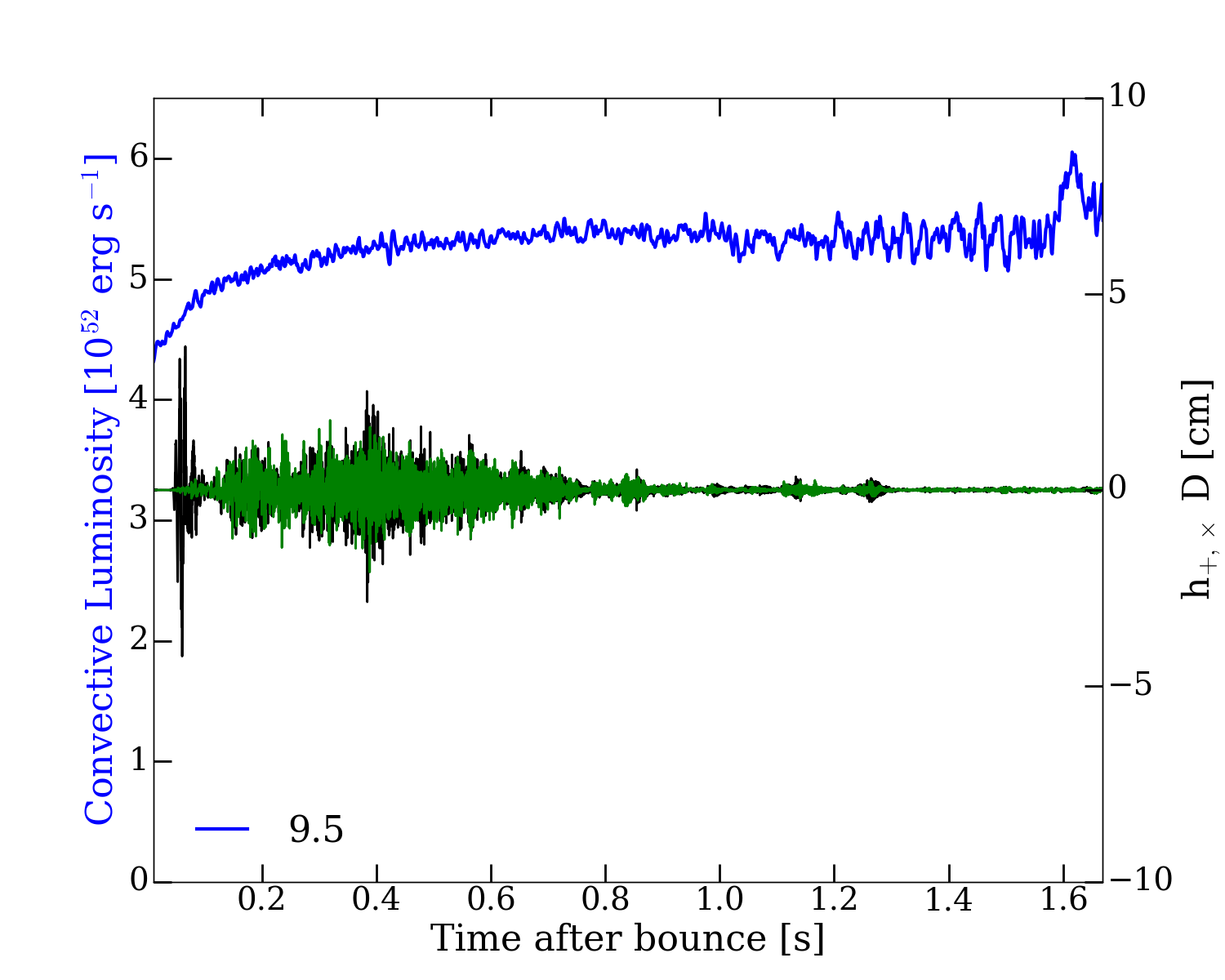}
    \caption{Convective hydrodynamic luminosity within the PNS (blue, in 10$^{51}$ erg s$^{-1}$) at its peak value in the inner PNS, overplotted with both gravitational-wave strain polarizations (black and green) multiplied by distance (product in cm) as a function of time after bounce (in seconds) for models 9a, 9b, 9.25, and 9.5. Figure \,\ref{fig:conv_p2} depicts the same quantities, but for models 11-, 12.25-, 15.01-, and 23-M$_{\odot}$. Note that PNS convection first increases in strength over time and then stays approximately constant  (with some minor excursions) or stays high with a secular evolution for the duration of the simulations. Moreover, PNS convection grows in extent over time to encompass more and more of the core, expanding deeper in radius and reaching the center of the PNS after from $\sim$1.7 seconds (the 9-M$_{\odot}$ model) to $\sim$4 seconds (the 23-M$_{\odot}$ model, see Figure \ref{fig:conv_p2}), all the while maintaining its vigor. This behavior shows little or no correlation with the GW strain, which decreases in magnitude with time from an earlier peak and nearly flattens at late time (dominated then by a humming fundamental f-mode (see Figure \ref{fig:EGW_spec}). In particular, at late times, when the strain has subsided substantially, the inner PNS convective luminosity is always still strong, indicating that PNS convection is not the dominant excitation mechanism of CCSN GW emission.}
    \label{fig:conv_p}
\end{figure*}

\begin{figure*}
    \centering
    \includegraphics[width=0.37\textwidth]{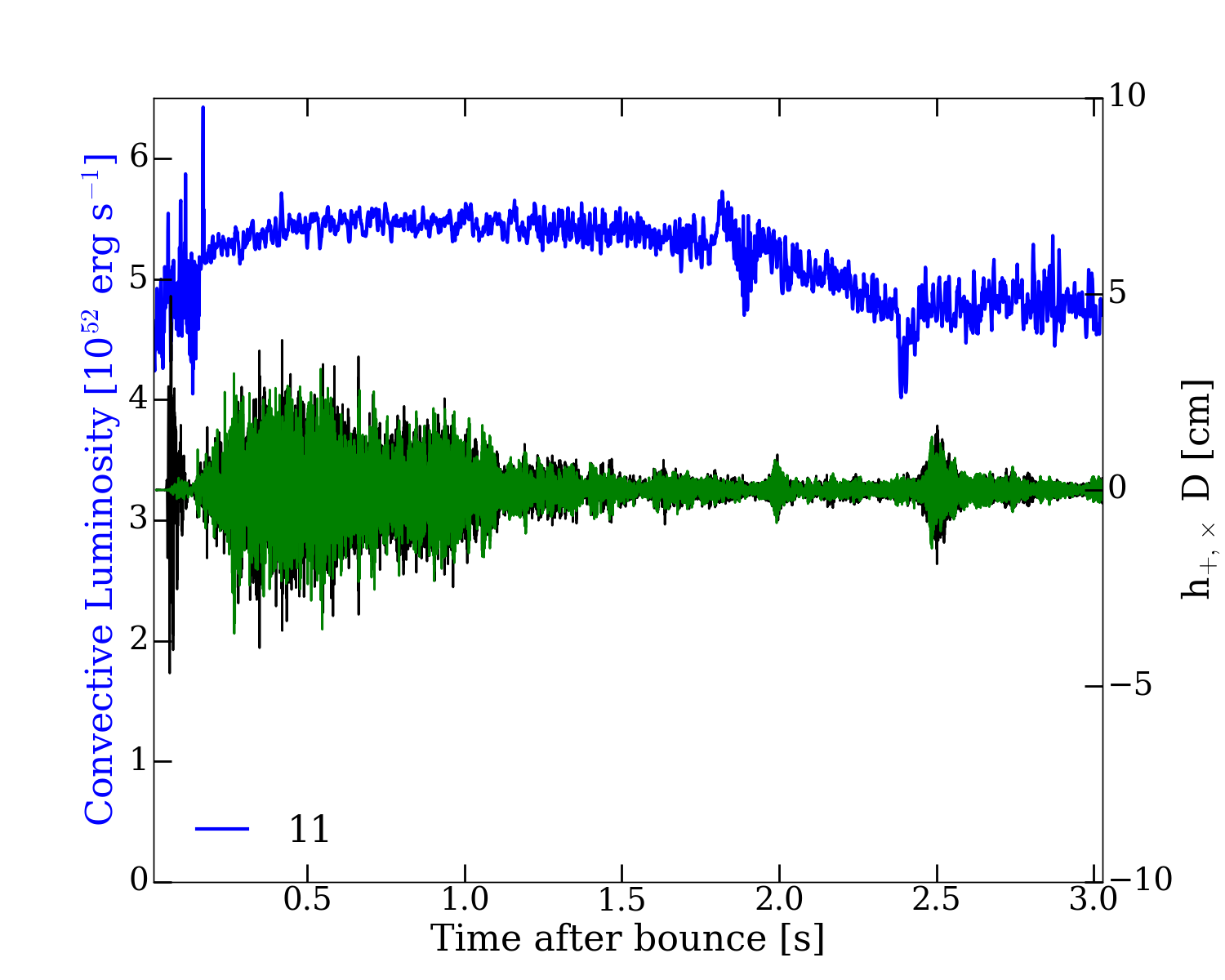}
    \includegraphics[width=0.37\textwidth]{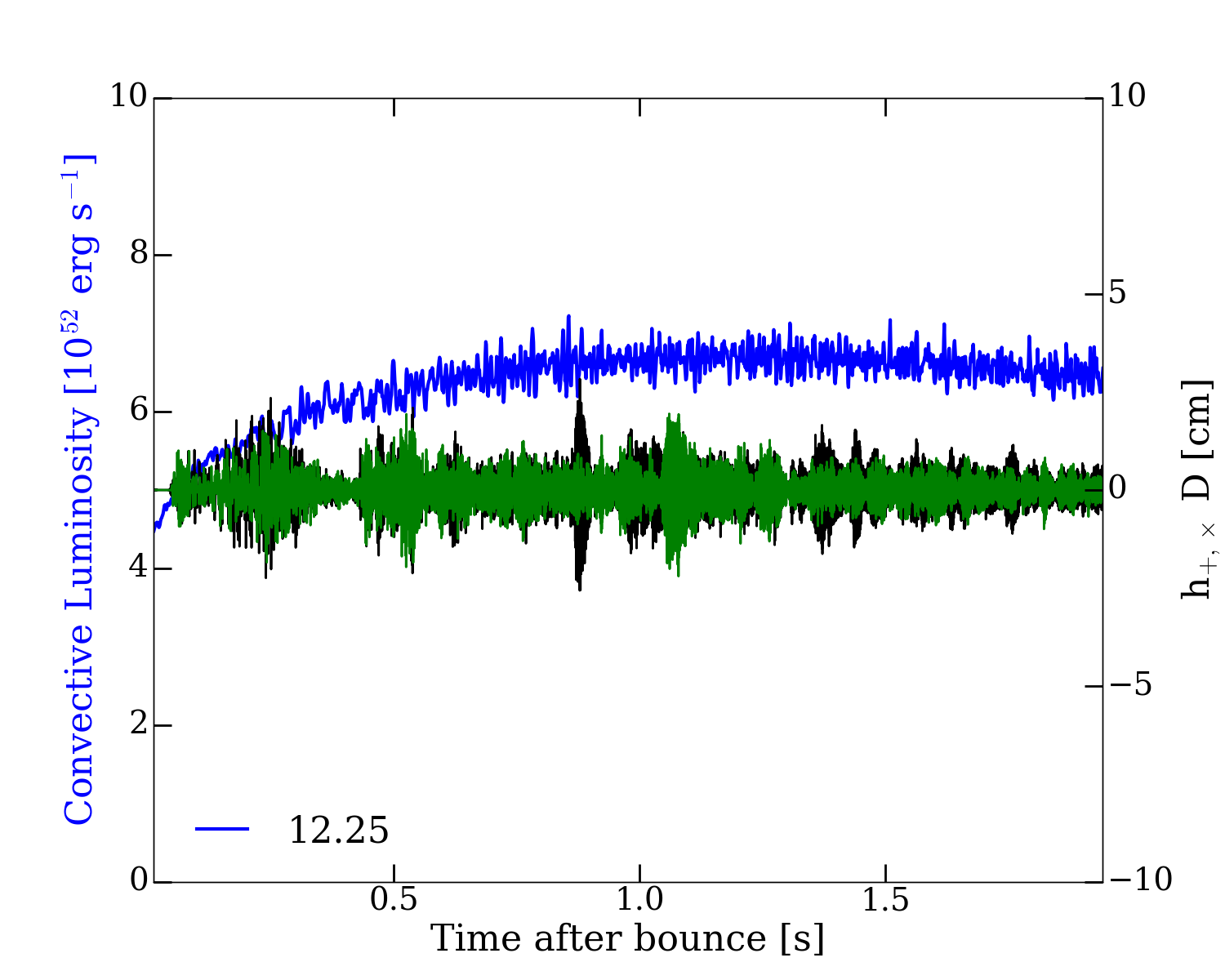}
    \includegraphics[width=0.37\textwidth]{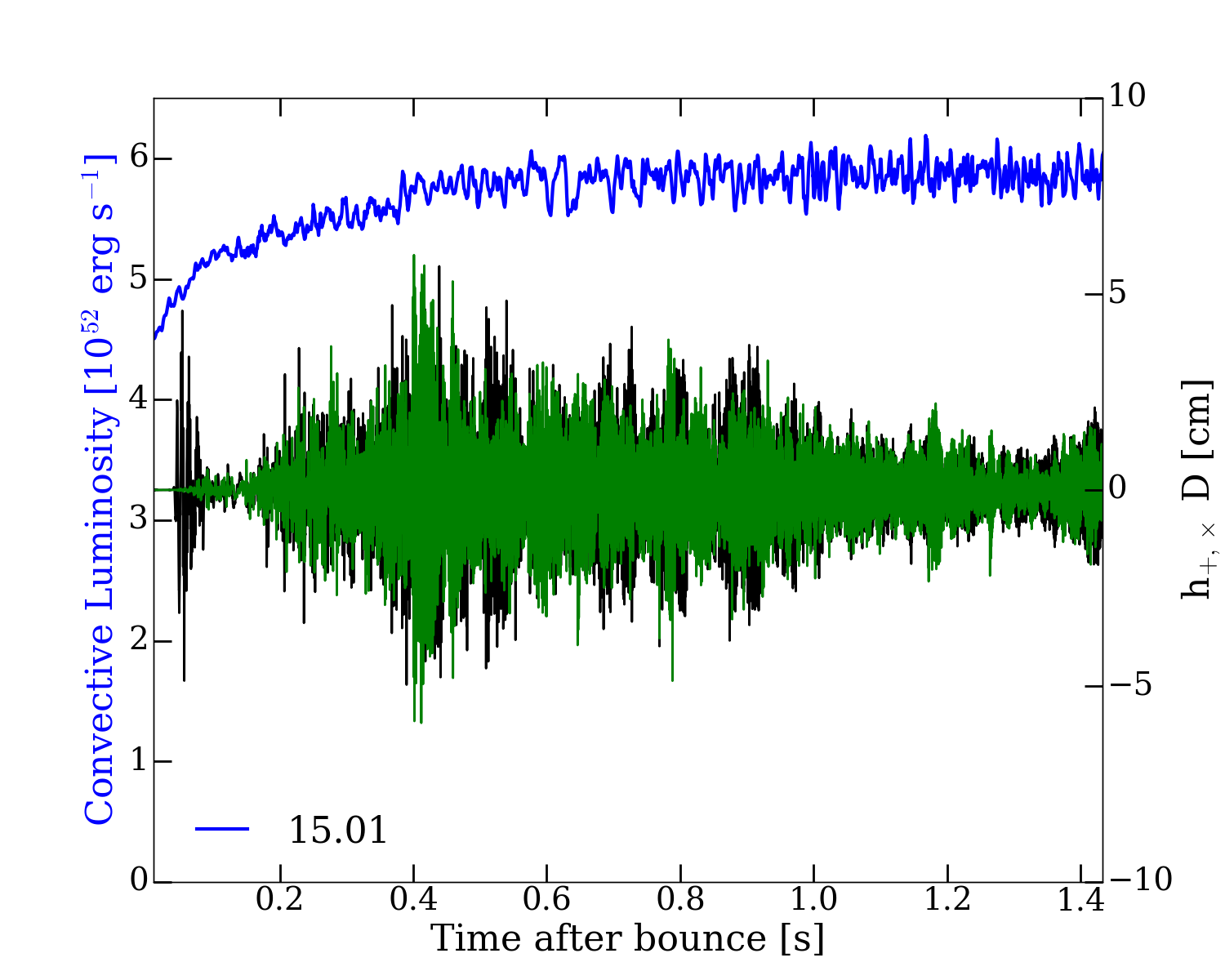}
    \includegraphics[width=0.37\textwidth]{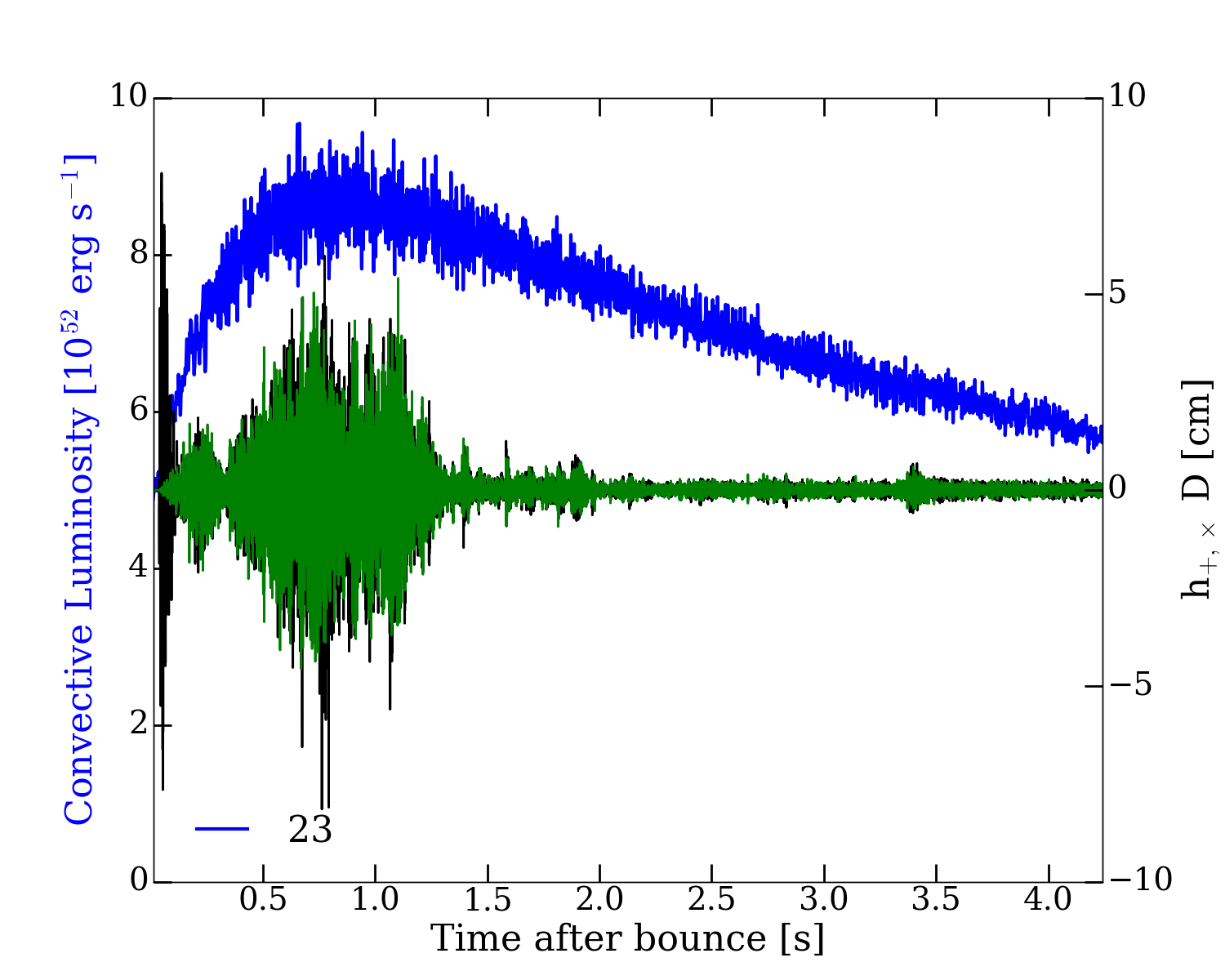}
    \caption{Same as Figure\,\ref{fig:conv_p}, but for models 11-, 12.25-, 15.01-, and 23-M$_{\odot}$.}
    \label{fig:conv_p2}
\end{figure*}

Some have suggested that inner PNS convection itself excites much of the GW emission from CCSN \citep{andresen2017}. In Figures\,\ref{fig:conv_p} and \ref{fig:conv_p2}, we overplot the angle-averaged convective hydrodynamic luminosity at its peak inside the PNS (see also Figures \ref{fig:pns_cmap}) against the Butterworth filtered strain. We follow Eq.\,\ref{eq:turb}, but limit the radius to that of the maximum convective luminosity within the PNS (positive outwards), which lies near $\sim$15 km for the various models. As Figures \ref{fig:conv_p} and \ref{fig:conv_p2} show there seems to be no correlation between the two. This is particularly clear at late times when the PNS convective flux is still large while the GW emissions have all but subsided. If PNS convection were the agency of excitation at all phases, GW emission would not have subsided to such a degree at later times (from 0.3 to 1.5 seconds after bounce). This comparison demonstrates the importance of simulating to late times to capture the entire GW signal.

However, as Figures \ref{fig:EGW_spec} and \ref{fig:EGW_spec2} themselves show, the f-mode persists to the latest times and manifests an episodically modulated (see Figure \ref{fig:GW-mdot-corr}), though continuous, signal.  We have yet to identify the major excitation mechanism for this component at late times. Anisotropic winds and neutrino emissions from the residual core could be causes, but inner PNS convection may be also a factor here. It is also the case that the mode should ring down over a period given by the dominant damping mechanism. Such mechanisms include sound generation, the back reaction of anisotropic winds, neutrino emission coupling and viscosity, non-linear parent-daughter mode coupling, and numerical dissipation. We reiterate, however, that the f-mode continues to ring and produce a weak GW signal for the duration of all our simulations. Clearly, this topic deserves more detailed scrutiny in the future. Nevertheless, this later phase amounts to only a few percent of the total energy emitted.  


\section{Conclusions}
\label{conclusions}

In this paper, we have presented and analyzed the GW signatures of an extensive suite of detailed initially non-rotating 3D core-collapse supernova simulations spanning a wide range of massive-star progenitors. For the first time, most of the published simulations were carried out to late enough times to capture more than 99\% of the total GW emission from such events. Moreover, we have endeavored to dump the relevant quadrupole data at a rate sufficient to effectively eliminate Nyquist sampling errors. We see that the f/g-mode and f-mode oscillation modes of the PNS core carry away most of the GW power and that generically there are avoided crossings and modal interactions likely associated with the evolution, extent, and character of lepton-driven PNS convection. The f-mode frequency inexorably rises as the proto-neutron star core shrinks during its Kelvin-Helmholtz contraction phase, driven by neutrino loses, and its power and frequency behavior are central features of the GW emissions from the core-collapse event. Other modes are also seen in the GW spectra, in particular a $\ell = 2; n = 1$ p-mode and, perhaps directly, a trapped g-mode, though most g-modes are not excited. Whether other p-modes are in evidence is to be determined.

We demonstrate that the GW emission is powered mostly by accretion plumes onto the PNS that excite its modal oscillations and also produce a ``haze" of higher frequency emission also correlated with the phase of violent accretion, after which the signal subsides to be dominated by the chirp of the f-mode signal at low power that nevertheless continues beyond the duration of even these simulations, albeit weakly.  The duration of the major phase of emission varies with exploding progenitor and is generally shorter for the lower-mass progenitors ($\sim$0.3-0.5 seconds) and longer for the higher-mass progenitors ($\sim$1.5 seconds). We find a strong correlation between the total GW energy radiated and the compactness of the progenitor whose mantle explodes as a supernova.  Furthermore, we find that the total GW energy emissions can vary by as much as three orders of magnitude from star to star. Hence, there is a severe progenitor dependence that must be factored into any discussion of detectability. For the black-hole forming models, since accretion is not reversed at any time or at any solid angle, their GW signal lasts until the black hole forms,
tapering off only slowly until then.  In addition, they do not manifest the high-frequency haze seen for the exploding models. For these black-hole formers, we also witness the emergence of a spiral shock motion that modulates the GW emission at a frequency near $\sim$100 Hz that slowly increases as the stalled shock sinks.



\begin{figure*}
    \centering
    \includegraphics[width=0.87\textwidth]{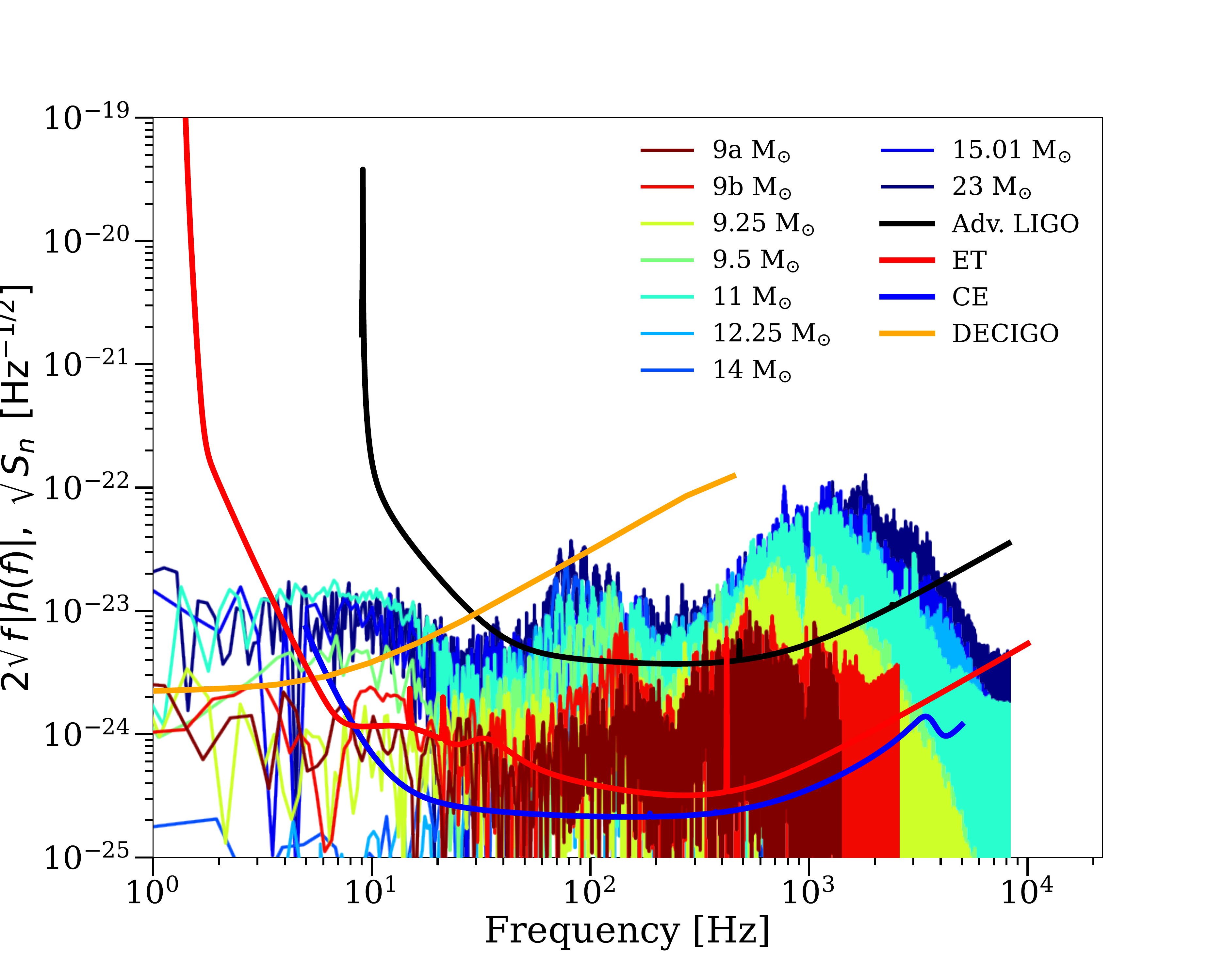}
    \caption{Gravitational-wave sensitivity curve for all the models as a function of time (in seconds after core bounce), compared with the CCSN GW signals at 10 kpc, including both matter (high frequency, dominant above several hundred Hz) and neutrino contributions to GW emissions (low frequency, dominant from sub-Hz to several tens of Hz). Note that DECIGO, the Einstein Telescope, and the Cosmic Explorer can probe down to the lowest-mass progenitors across three decades in frequency ($\sim$10 to $\sim$5000 Hz), whereas aLIGO is sensitive between $\sim$30 to $\sim$3000 Hz only to the most massive progenitors. Our lower limit in frequency is set by the duration of the CCSN simulation. Since our longest simulation is out to ~6.2 seconds, simulations are generally unable to constrain the spectrogram data between 0.1$-$1 Hz. Our upper limit in frequency is set by the Nyquist limit in Table\,\ref{sn_tab}.}
    \label{fig:sens}
\end{figure*}

In Figure\,\ref{fig:sens}, we plot the sensitivity curves with the amplitude spectral densities at 10 kiloparsecs of our 3D models. More massive models generally leave larger footprints, with the 11 M$_{\odot}$ studied being the exception. Current and next-generation detectors can observe, for galactic events, a signature spanning orders of magnitude, from subHz to $\sim$3000 Hz. While Advanced LIGO/Virgo/Kagra (LVK, \citep{aligo, virgo, kagra}) can detect $\sim$30$-$3000 Hz signals for the more massive progenitors, upcoming detectors, including the Einstein Telescope \citep{ET_1,ET_3}, the Cosmic Explorer \citep{ce}, BBO \citep{BBO}, and Decigo \citep{decigo1,decigo2} (at lower frequencies, see also \citep{sedda}) should be able to detect galactic events for all progenitor masses studied here through almost three orders of magnitude in both frequency and total energy radiated. Note that neutrino GWs dominate at lower frequencies (from sub-Hz to 10s of Hz) and matter GWs dominate at higher frequencies (at several hundreds of Hz). However, a detailed retrieval analysis, informed by the best signal-processing approaches, has yet to be performed and is an important topic for future work.

Though we have endeavored here to provide a comprehensive look at the gravitational-wave signatures of core collapse, there remains much yet to understand. Topics unaddressed here are the nuclear equation-of-state dependencies, the role of rapid rotation, the possible signatures of strong magnetic fields \cite{jakobus, 2022arXiv221200200P}, and the results for other progenitor massive stars and from other stellar evolution codes.
Importantly, the analysis of a detected GW signal would be significantly aided if done in concert with a corresponding analysis of the simultaneous neutrino signal. Optimal methodologies with which to extract physical information from such an analysis have yet to be designed. Nevertheless, we have in the GW signal of core-collapse supernovae a direct and real-time window into the supernova mechanism and PNS evolution.  Therefore, such a methodology would likely pay rich scientific dividends when astronomy is finally presented with the opportunity to employ it.   

\section*{Data Availability}

The numerical data associated with this article and sonifications of the GW strains will be shared upon reasonable request to the corresponding author. The GW strains as well as the quadrupole data are available publicly at \url{https://dvartany.github.io/data/} and \url{https://www.astro.princeton.edu/~burrows/gw.3d.new/}.

\section*{Acknowledgments}

We thank Jeremy Goodman, Eliot Quataert, David Radice, Viktoriya Morozova, Hiroki Nagakura, and Benny Tsang for insights and advice during the germination and execution of this project. DV acknowledges support from the NASA Hubble Fellowship Program  grant HST-HF2-51520. We acknowledge support from the U.~S.\ Department of Energy Office of Science and the Office of Advanced Scientific Computing Research via the Scientific Discovery through Advanced Computing (SciDAC4) program and Grant DE-SC0018297 (subaward 00009650), support from the U.~S.\ National Science Foundation (NSF) under Grants AST-1714267 and PHY-1804048 (the latter via the Max-Planck/Princeton Center (MPPC) for Plasma Physics), and support from NASA under award JWST-GO-01947.011-A.   A generous award of computer time was provided by the INCITE program, using resources of the Argonne Leadership Computing Facility, a DOE Office of Science User Facility supported under Contract DE-AC02-06CH11357. We also acknowledge access to the Frontera cluster (under awards AST20020 and AST21003); this research is part of the Frontera computing project at the Texas Advanced Computing Center \citep{Stanzione2020} under NSF award OAC-1818253. In addition, one earlier simulation was performed on Blue Waters under the sustained-petascale computing project, which was supported by the National Science Foundation (awards OCI-0725070 and ACI-1238993) and the state of Illinois. Blue Waters was a joint effort of the University of Illinois at Urbana--Champaign and its National Center for Supercomputing Applications. Finally, the authors acknowledge computational resources provided by the high-performance computer center at Princeton University, which is jointly supported by the Princeton Institute for Computational Science and Engineering (PICSciE) and the Princeton University Office of Information Technology, and our continuing allocation at the National Energy Research Scientific Computing Center (NERSC), which is supported by the Office of Science of the U.~S.\ Department of Energy under contract DE-AC03-76SF00098.



\bigskip
 
\appendix \section{General Equations} \label{appendix}
We follow the method in (\cite{finn}, see also \cite{oohara97,andresen2017, vsg2018,vartanyan2019}) to calculate the gravitational-wave strain tensor. We calculate the first time derivative of the mass quadrupole with the following formula:
\begin{equation}
q_{ij} = \frac{d}{dt}Q_{ij}=\int d^3x\, (v_ix_j+v_jx_i-\frac{2}{3}v_rr),
\end{equation}
where $v$ is the velocity, $x$ is the Cartesian coordinate, $Q_{ij}$ is the transverse-traceless quadrupole tensor, and $r$ is the radius. The transverse-traceless GW strain tensor $h^{TT}_{ij}$ is calculated by taking the numerical time derivative of $q_{ij}$, i.e.,
\begin{equation}
h^{TT}_{ij}=\frac{2G}{c^4D}\frac{dq_{ij}}{dt}\, ,
\end{equation}
where $D$ is the distance to the source.
Hereafter, we drop the superscript ``TT''.
We also calculate and dump the quadrupole $Q_{ij}=\int d^3x\, \rho(x_ix_j-\frac{1}{3}r^2\delta_ij)$, and its numerical derivatives are consistent with the values calculated by the above equation. However, taking numerical derivatives can be viewed as convolving the signal with a window function, thus it will introduce a bit of extra noise at high frequencies.

The ``plus'' and ``cross'' polarized strains along ($\theta,\phi$) direction are given by
\begin{equation}
h_+=\frac{G}{c^4D}\left(\frac{dq_{\theta\theta}}{dt}-\frac{dq_{\phi\phi}}{dt}\right)
\end{equation}

and

\begin{equation}
h_\times=\frac{2G}{c^4D}\frac{dq_{\theta\phi}}{dt}
\end{equation}
where (from \citep{oohara97})
\begin{equation}
\begin{split}
q_{\theta\theta} =& \,\,(q_{xx}\cos^2\phi+q_{yy}\sin^2\phi+2\,q_{xy}\sin\phi\cos\phi)\cos^2\theta\\
&+q_{zz}\sin^2\theta-2\,(q_{xz}\cos\phi+q_{yz}\sin\phi)\sin\theta\cos\theta
\end{split}
\end{equation}
\begin{equation}
\begin{split}
q_{\phi\phi} =& \,\,q_{xx}\sin^2\phi+q_{yy}\cos^2\phi-2\,q_{xy}\sin\phi\cos\phi\\
\end{split}
\end{equation}
\begin{equation}
\begin{split}
q_{\theta\phi}=& \,\,(q_{yy}-q_{xx})\cos\theta\sin\phi\cos\phi+
q_{xy}\cos\theta\,(\cos^2\phi-\sin^2\phi)\\
&+q_{xz}\sin\theta\sin\phi-q_{yz}\sin\theta\cos\phi\, .
\end{split}
\end{equation}

The total energy emitted in GWs is 
\begin{equation}
\begin{split}
E_{GW} &= \int_{0}^{t}\sum_{ij}\frac{G}{5c^5}\left(\frac{d^3Q_{ij}}{dt^3}\right)^2\\
\end{split}
\end{equation}

To calculate GWs from neutrino asymmetries, we follow the prescription of (\cite{emuller}, see also \cite{vartanyan2020}). We include angle-dependence of the observer through the viewing angles $\alpha \in [-\pi,\pi]$ and $\beta \in [0,\pi]$. The time-dependent neutrino emission anisotropy parameter for each polarization is defined as 

\begin{equation}\label{eq:0}
\alpha_S (t,\alpha,\beta) = \frac{1}{\Lambda(t)} \int_{4\pi} d\Omega'\, W_S(\Omega',\alpha,\beta) \frac{d\Lambda}{d\Omega'}(\Omega',t)\,,
\end{equation} where the subscript is $S \in \{+,\times\}$ and the GW strain from neutrinos is defined as: 

\begin{equation} \label{eq:1}
h_S(t,\alpha,\beta) = \frac{2G}{c^4D} \int_0^t dt'\, \Lambda(t')\,\alpha_S(t',\alpha,\beta)\,,
\end{equation} where $\Lambda (t)$ is the angle-integrated neutrino luminosity as a function of time, $D$ is the distance to the source, and W$_\mathrm{S}$ ($\mathrm{\Omega}',\alpha,\beta)$ is the geometric weight for the anisotropy parameter given by  

\begin{equation}
    W_{\mathrm{S}} = \frac{D_{\mathrm{S}} (\theta',\phi',\alpha,\beta)}{N(\theta',\phi',\alpha,\beta)}\,,
\end{equation} where (from \citep{emuller})

\begin{subequations}
\begin{align}
\begin{split}
D_+ &= [1 + (\cos(\phi')\cos(\alpha) + \sin(\phi')\sin(\alpha))\sin(\theta')\sin(\beta)+\cos(\theta')\\&\cos(\beta)]\{[(\cos(\phi')\cos(\alpha)+\sin(\phi')\sin(\alpha))\sin(\theta')\cos(\beta)\\&-\cos(\theta')\sin(\beta)]^2-\sin^2(\theta')(\sin(\phi')\cos(\alpha)-\cos(\phi')\sin(\alpha))^2\}
\end{split}\\
\begin{split}
D_\times &= [1 + (\cos(\phi')\cos(\alpha) + \sin(\phi')\sin(\alpha))\sin(\theta')\sin(\beta)+\cos(\theta')\\&\cos(\beta)]2[(\cos(\phi')\cos(\alpha)+\sin(\phi')\sin(\alpha))\sin(\theta')\cos(\beta)\\&-\cos(\theta')\sin(\beta)]\sin(\theta')(\sin(\phi')\cos(\alpha)-\cos(\phi')\sin(\alpha))^2
\end{split}\\
\begin{split}
N &= [(\cos(\phi')\cos(\alpha)+\sin(\phi')\sin(\alpha))\sin(\theta')\cos(\beta)-\cos(\theta')\\&\sin(\beta)]^2+\sin^2(\theta')(\sin(\phi')\cos(\alpha)-\cos(\phi')\sin(\alpha))^2\,.
\end{split}
\end{align}
\end{subequations}

\clearpage

\bibliographystyle{apsrev}
\bibliography{References}

\begin{thebibliography}{104}
\expandafter\ifx\csname natexlab\endcsname\relax\def\natexlab#1{#1}\fi
\expandafter\ifx\csname bibnamefont\endcsname\relax
  \def\bibnamefont#1{#1}\fi
\expandafter\ifx\csname bibfnamefont\endcsname\relax
  \def\bibfnamefont#1{#1}\fi
\expandafter\ifx\csname citenamefont\endcsname\relax
  \def\citenamefont#1{#1}\fi
\expandafter\ifx\csname url\endcsname\relax
  \def\url#1{\texttt{#1}}\fi
\expandafter\ifx\csname urlprefix\endcsname\relax\def\urlprefix{URL }\fi
\providecommand{\bibinfo}[2]{#2}
\providecommand{\eprint}[2][]{\url{#2}}

\bibitem[{\citenamefont{{Colgate} and {White}}(1966)}]{CoWh66}
\bibinfo{author}{\bibfnamefont{S.~A.} \bibnamefont{{Colgate}}}
  \bibnamefont{and} \bibinfo{author}{\bibfnamefont{R.~H.}
  \bibnamefont{{White}}}, \bibinfo{journal}{\apj}
  \textbf{\bibinfo{volume}{143}}, \bibinfo{pages}{626} (\bibinfo{year}{1966}).

\bibitem[{\citenamefont{{Wilson}}(1985)}]{Wilson1985}
\bibinfo{author}{\bibfnamefont{J.~R.} \bibnamefont{{Wilson}}}, in
  \emph{\bibinfo{booktitle}{Numerical Astrophysics}}, edited by
  \bibinfo{editor}{\bibfnamefont{J.~M.} \bibnamefont{{Centrella}}},
  \bibinfo{editor}{\bibfnamefont{J.~M.} \bibnamefont{{Leblanc}}},
  \bibnamefont{and} \bibinfo{editor}{\bibfnamefont{R.~L.}
  \bibnamefont{{Bowers}}} (\bibinfo{year}{1985}), p. \bibinfo{pages}{422}.

\bibitem[{\citenamefont{{Janka}}(2012)}]{janka2012}
\bibinfo{author}{\bibfnamefont{H.-T.} \bibnamefont{{Janka}}},
  \bibinfo{journal}{Annual Review of Nuclear and Particle Science}
  \textbf{\bibinfo{volume}{62}}, \bibinfo{pages}{407} (\bibinfo{year}{2012}),
  \eprint{1206.2503}.

\bibitem[{\citenamefont{{Burrows}}(2013)}]{burrows2013}
\bibinfo{author}{\bibfnamefont{A.}~\bibnamefont{{Burrows}}},
  \bibinfo{journal}{Reviews of Modern Physics} \textbf{\bibinfo{volume}{85}},
  \bibinfo{pages}{245} (\bibinfo{year}{2013}), \eprint{1210.4921}.

\bibitem[{\citenamefont{{Burrows} and {Vartanyan}}(2021)}]{2021Natur.589...29B}
\bibinfo{author}{\bibfnamefont{A.}~\bibnamefont{{Burrows}}} \bibnamefont{and}
  \bibinfo{author}{\bibfnamefont{D.}~\bibnamefont{{Vartanyan}}},
  \bibinfo{journal}{\nat} \textbf{\bibinfo{volume}{589}}, \bibinfo{pages}{29}
  (\bibinfo{year}{2021}), \eprint{2009.14157}.

\bibitem[{\citenamefont{{Takiwaki} et~al.}(2012)\citenamefont{{Takiwaki},
  {Kotake}, and {Suwa}}}]{takiwaki:12}
\bibinfo{author}{\bibfnamefont{T.}~\bibnamefont{{Takiwaki}}},
  \bibinfo{author}{\bibfnamefont{K.}~\bibnamefont{{Kotake}}}, \bibnamefont{and}
  \bibinfo{author}{\bibfnamefont{Y.}~\bibnamefont{{Suwa}}},
  \bibinfo{journal}{\apj} \textbf{\bibinfo{volume}{749}}, \bibinfo{eid}{98}
  (\bibinfo{year}{2012}), \eprint{1108.3989}.

\bibitem[{\citenamefont{{Hanke} et~al.}(2013)\citenamefont{{Hanke},
  {M{\"u}ller}, {Wongwathanarat}, {Marek}, and {Janka}}}]{hanke_13}
\bibinfo{author}{\bibfnamefont{F.}~\bibnamefont{{Hanke}}},
  \bibinfo{author}{\bibfnamefont{B.}~\bibnamefont{{M{\"u}ller}}},
  \bibinfo{author}{\bibfnamefont{A.}~\bibnamefont{{Wongwathanarat}}},
  \bibinfo{author}{\bibfnamefont{A.}~\bibnamefont{{Marek}}}, \bibnamefont{and}
  \bibinfo{author}{\bibfnamefont{H.-T.} \bibnamefont{{Janka}}},
  \bibinfo{journal}{\apj} \textbf{\bibinfo{volume}{770}}, \bibinfo{eid}{66}
  (\bibinfo{year}{2013}).

\bibitem[{\citenamefont{{Lentz} et~al.}(2015)\citenamefont{{Lentz}, {Bruenn},
  {Hix}, {Mezzacappa}, {Messer}, {Endeve}, {Blondin}, {Harris}, {Marronetti},
  and {Yakunin}}}]{lentz:15}
\bibinfo{author}{\bibfnamefont{E.~J.} \bibnamefont{{Lentz}}},
  \bibinfo{author}{\bibfnamefont{S.~W.} \bibnamefont{{Bruenn}}},
  \bibinfo{author}{\bibfnamefont{W.~R.} \bibnamefont{{Hix}}},
  \bibinfo{author}{\bibfnamefont{A.}~\bibnamefont{{Mezzacappa}}},
  \bibinfo{author}{\bibfnamefont{O.~E.~B.} \bibnamefont{{Messer}}},
  \bibinfo{author}{\bibfnamefont{E.}~\bibnamefont{{Endeve}}},
  \bibinfo{author}{\bibfnamefont{J.~M.} \bibnamefont{{Blondin}}},
  \bibinfo{author}{\bibfnamefont{J.~A.} \bibnamefont{{Harris}}},
  \bibinfo{author}{\bibfnamefont{P.}~\bibnamefont{{Marronetti}}},
  \bibnamefont{and} \bibinfo{author}{\bibfnamefont{K.~N.}
  \bibnamefont{{Yakunin}}}, \bibinfo{journal}{\apjl}
  \textbf{\bibinfo{volume}{807}}, \bibinfo{eid}{L31} (\bibinfo{year}{2015}),
  \eprint{1505.05110}.

\bibitem[{\citenamefont{{Burrows} et~al.}(2018)\citenamefont{{Burrows},
  {Vartanyan}, {Dolence}, {Skinner}, and {Radice}}}]{burrows:16}
\bibinfo{author}{\bibfnamefont{A.}~\bibnamefont{{Burrows}}},
  \bibinfo{author}{\bibfnamefont{D.}~\bibnamefont{{Vartanyan}}},
  \bibinfo{author}{\bibfnamefont{J.~C.} \bibnamefont{{Dolence}}},
  \bibinfo{author}{\bibfnamefont{M.~A.} \bibnamefont{{Skinner}}},
  \bibnamefont{and} \bibinfo{author}{\bibfnamefont{D.}~\bibnamefont{{Radice}}},
  \bibinfo{journal}{\ssr} \textbf{\bibinfo{volume}{214}}, \bibinfo{eid}{33}
  (\bibinfo{year}{2018}).

\bibitem[{\citenamefont{{Melson}
  et~al.}(2015{\natexlab{a}})\citenamefont{{Melson}, {Janka}, and
  {Marek}}}]{melson:15a}
\bibinfo{author}{\bibfnamefont{T.}~\bibnamefont{{Melson}}},
  \bibinfo{author}{\bibfnamefont{H.-T.} \bibnamefont{{Janka}}},
  \bibnamefont{and} \bibinfo{author}{\bibfnamefont{A.}~\bibnamefont{{Marek}}},
  \bibinfo{journal}{\apjl} \textbf{\bibinfo{volume}{801}}, \bibinfo{eid}{L24}
  (\bibinfo{year}{2015}{\natexlab{a}}), \eprint{1501.01961}.

\bibitem[{\citenamefont{{Melson}
  et~al.}(2015{\natexlab{b}})\citenamefont{{Melson}, {Janka}, {Bollig},
  {Hanke}, {Marek}, and {M{\"u}ller}}}]{melson:15b}
\bibinfo{author}{\bibfnamefont{T.}~\bibnamefont{{Melson}}},
  \bibinfo{author}{\bibfnamefont{H.-T.} \bibnamefont{{Janka}}},
  \bibinfo{author}{\bibfnamefont{R.}~\bibnamefont{{Bollig}}},
  \bibinfo{author}{\bibfnamefont{F.}~\bibnamefont{{Hanke}}},
  \bibinfo{author}{\bibfnamefont{A.}~\bibnamefont{{Marek}}}, \bibnamefont{and}
  \bibinfo{author}{\bibfnamefont{B.}~\bibnamefont{{M{\"u}ller}}},
  \bibinfo{journal}{\apjl} \textbf{\bibinfo{volume}{808}}, \bibinfo{eid}{L42}
  (\bibinfo{year}{2015}{\natexlab{b}}), \eprint{1504.07631}.

\bibitem[{\citenamefont{{Takiwaki} et~al.}(2016)\citenamefont{{Takiwaki},
  {Kotake}, and {Suwa}}}]{TaKoSu16}
\bibinfo{author}{\bibfnamefont{T.}~\bibnamefont{{Takiwaki}}},
  \bibinfo{author}{\bibfnamefont{K.}~\bibnamefont{{Kotake}}}, \bibnamefont{and}
  \bibinfo{author}{\bibfnamefont{Y.}~\bibnamefont{{Suwa}}},
  \bibinfo{journal}{\mnras} \textbf{\bibinfo{volume}{461}},
  \bibinfo{pages}{L112} (\bibinfo{year}{2016}), \eprint{1602.06759}.

\bibitem[{\citenamefont{{Roberts} et~al.}(2016)\citenamefont{{Roberts}, {Ott},
  {Haas}, {O'Connor}, {Diener}, and {Schnetter}}}]{roberts:16}
\bibinfo{author}{\bibfnamefont{L.~F.} \bibnamefont{{Roberts}}},
  \bibinfo{author}{\bibfnamefont{C.~D.} \bibnamefont{{Ott}}},
  \bibinfo{author}{\bibfnamefont{R.}~\bibnamefont{{Haas}}},
  \bibinfo{author}{\bibfnamefont{E.~P.} \bibnamefont{{O'Connor}}},
  \bibinfo{author}{\bibfnamefont{P.}~\bibnamefont{{Diener}}}, \bibnamefont{and}
  \bibinfo{author}{\bibfnamefont{E.}~\bibnamefont{{Schnetter}}},
  \bibinfo{journal}{\apj} \textbf{\bibinfo{volume}{831}}, \bibinfo{eid}{98}
  (\bibinfo{year}{2016}), \eprint{1604.07848}.

\bibitem[{\citenamefont{{Ott} et~al.}(2018)\citenamefont{{Ott}, {Roberts}, {da
  Silva Schneider}, {Fedrow}, {Haas}, and {Schnetter}}}]{ott2018_rel}
\bibinfo{author}{\bibfnamefont{C.~D.} \bibnamefont{{Ott}}},
  \bibinfo{author}{\bibfnamefont{L.~F.} \bibnamefont{{Roberts}}},
  \bibinfo{author}{\bibfnamefont{A.}~\bibnamefont{{da Silva Schneider}}},
  \bibinfo{author}{\bibfnamefont{J.~M.} \bibnamefont{{Fedrow}}},
  \bibinfo{author}{\bibfnamefont{R.}~\bibnamefont{{Haas}}}, \bibnamefont{and}
  \bibinfo{author}{\bibfnamefont{E.}~\bibnamefont{{Schnetter}}},
  \bibinfo{journal}{\apjl} \textbf{\bibinfo{volume}{855}}, \bibinfo{eid}{L3}
  (\bibinfo{year}{2018}).

\bibitem[{\citenamefont{{M{\"u}ller} et~al.}(2017)\citenamefont{{M{\"u}ller},
  {Melson}, {Heger}, and {Janka}}}]{muller2017}
\bibinfo{author}{\bibfnamefont{B.}~\bibnamefont{{M{\"u}ller}}},
  \bibinfo{author}{\bibfnamefont{T.}~\bibnamefont{{Melson}}},
  \bibinfo{author}{\bibfnamefont{A.}~\bibnamefont{{Heger}}}, \bibnamefont{and}
  \bibinfo{author}{\bibfnamefont{H.-T.} \bibnamefont{{Janka}}},
  \bibinfo{journal}{\mnras} \textbf{\bibinfo{volume}{472}},
  \bibinfo{pages}{491} (\bibinfo{year}{2017}), \eprint{1705.00620}.

\bibitem[{\citenamefont{{Kuroda} et~al.}(2018)\citenamefont{{Kuroda}, {Kotake},
  {Takiwaki}, and {Thielemann}}}]{2018MNRAS.477L..80K}
\bibinfo{author}{\bibfnamefont{T.}~\bibnamefont{{Kuroda}}},
  \bibinfo{author}{\bibfnamefont{K.}~\bibnamefont{{Kotake}}},
  \bibinfo{author}{\bibfnamefont{T.}~\bibnamefont{{Takiwaki}}},
  \bibnamefont{and} \bibinfo{author}{\bibfnamefont{F.-K.}
  \bibnamefont{{Thielemann}}}, \bibinfo{journal}{\mnras}
  \textbf{\bibinfo{volume}{477}}, \bibinfo{pages}{L80} (\bibinfo{year}{2018}),
  \eprint{1801.01293}.

\bibitem[{\citenamefont{{Vartanyan}
  et~al.}(2019{\natexlab{a}})\citenamefont{{Vartanyan}, {Burrows}, {Radice},
  {Skinner}, and {Dolence}}}]{vartanyan2018b}
\bibinfo{author}{\bibfnamefont{D.}~\bibnamefont{{Vartanyan}}},
  \bibinfo{author}{\bibfnamefont{A.}~\bibnamefont{{Burrows}}},
  \bibinfo{author}{\bibfnamefont{D.}~\bibnamefont{{Radice}}},
  \bibinfo{author}{\bibfnamefont{M.~A.} \bibnamefont{{Skinner}}},
  \bibnamefont{and}
  \bibinfo{author}{\bibfnamefont{J.}~\bibnamefont{{Dolence}}},
  \bibinfo{journal}{\mnras} \textbf{\bibinfo{volume}{482}},
  \bibinfo{pages}{351} (\bibinfo{year}{2019}{\natexlab{a}}),
  \eprint{1809.05106}.

\bibitem[{\citenamefont{{Summa} et~al.}(2018)\citenamefont{{Summa}, {Janka},
  {Melson}, and {Marek}}}]{summa2018}
\bibinfo{author}{\bibfnamefont{A.}~\bibnamefont{{Summa}}},
  \bibinfo{author}{\bibfnamefont{H.-T.} \bibnamefont{{Janka}}},
  \bibinfo{author}{\bibfnamefont{T.}~\bibnamefont{{Melson}}}, \bibnamefont{and}
  \bibinfo{author}{\bibfnamefont{A.}~\bibnamefont{{Marek}}},
  \bibinfo{journal}{\apj} \textbf{\bibinfo{volume}{852}}, \bibinfo{eid}{28}
  (\bibinfo{year}{2018}), \eprint{1708.04154}.

\bibitem[{\citenamefont{{O'Connor} and {Couch}}(2018)}]{oconnor_couch2018b}
\bibinfo{author}{\bibfnamefont{E.~P.} \bibnamefont{{O'Connor}}}
  \bibnamefont{and} \bibinfo{author}{\bibfnamefont{S.~M.}
  \bibnamefont{{Couch}}}, \bibinfo{journal}{\apj}
  \textbf{\bibinfo{volume}{865}}, \bibinfo{eid}{81} (\bibinfo{year}{2018}),
  \eprint{1807.07579}.

\bibitem[{\citenamefont{{Glas} et~al.}(2019)\citenamefont{{Glas}, {Just},
  {Janka}, and {Obergaulinger}}}]{glas2019}
\bibinfo{author}{\bibfnamefont{R.}~\bibnamefont{{Glas}}},
  \bibinfo{author}{\bibfnamefont{O.}~\bibnamefont{{Just}}},
  \bibinfo{author}{\bibfnamefont{H.~T.} \bibnamefont{{Janka}}},
  \bibnamefont{and}
  \bibinfo{author}{\bibfnamefont{M.}~\bibnamefont{{Obergaulinger}}},
  \bibinfo{journal}{\apj} \textbf{\bibinfo{volume}{873}}, \bibinfo{eid}{45}
  (\bibinfo{year}{2019}), \eprint{1809.10146}.

\bibitem[{\citenamefont{{Burrows} et~al.}(2020)\citenamefont{{Burrows},
  {Radice}, {Vartanyan}, {Nagakura}, {Skinner}, and {Dolence}}}]{burrows_2020}
\bibinfo{author}{\bibfnamefont{A.}~\bibnamefont{{Burrows}}},
  \bibinfo{author}{\bibfnamefont{D.}~\bibnamefont{{Radice}}},
  \bibinfo{author}{\bibfnamefont{D.}~\bibnamefont{{Vartanyan}}},
  \bibinfo{author}{\bibfnamefont{H.}~\bibnamefont{{Nagakura}}},
  \bibinfo{author}{\bibfnamefont{M.~A.} \bibnamefont{{Skinner}}},
  \bibnamefont{and} \bibinfo{author}{\bibfnamefont{J.~C.}
  \bibnamefont{{Dolence}}}, \bibinfo{journal}{\mnras}
  \textbf{\bibinfo{volume}{491}}, \bibinfo{pages}{2715} (\bibinfo{year}{2020}),
  \eprint{1909.04152}.

\bibitem[{\citenamefont{{Vartanyan} et~al.}(2022)\citenamefont{{Vartanyan},
  {Coleman}, and {Burrows}}}]{2022MNRAS.510.4689V}
\bibinfo{author}{\bibfnamefont{D.}~\bibnamefont{{Vartanyan}}},
  \bibinfo{author}{\bibfnamefont{M.~S.~B.} \bibnamefont{{Coleman}}},
  \bibnamefont{and}
  \bibinfo{author}{\bibfnamefont{A.}~\bibnamefont{{Burrows}}},
  \bibinfo{journal}{\mnras} \textbf{\bibinfo{volume}{510}},
  \bibinfo{pages}{4689} (\bibinfo{year}{2022}), \eprint{2109.10920}.

\bibitem[{\citenamefont{{Herant} et~al.}(1992)\citenamefont{{Herant}, {Benz},
  and {Colgate}}}]{herant:92}
\bibinfo{author}{\bibfnamefont{M.}~\bibnamefont{{Herant}}},
  \bibinfo{author}{\bibfnamefont{W.}~\bibnamefont{{Benz}}}, \bibnamefont{and}
  \bibinfo{author}{\bibfnamefont{S.}~\bibnamefont{{Colgate}}},
  \bibinfo{journal}{\apj} \textbf{\bibinfo{volume}{395}}, \bibinfo{pages}{642}
  (\bibinfo{year}{1992}).

\bibitem[{\citenamefont{{Burrows} et~al.}(1995)\citenamefont{{Burrows},
  {Hayes}, and {Fryxell}}}]{burrows:95}
\bibinfo{author}{\bibfnamefont{A.}~\bibnamefont{{Burrows}}},
  \bibinfo{author}{\bibfnamefont{J.}~\bibnamefont{{Hayes}}}, \bibnamefont{and}
  \bibinfo{author}{\bibfnamefont{B.~A.} \bibnamefont{{Fryxell}}},
  \bibinfo{journal}{\apj} \textbf{\bibinfo{volume}{450}}, \bibinfo{pages}{830}
  (\bibinfo{year}{1995}), \eprint{astro-ph/9506061}.

\bibitem[{\citenamefont{{Couch} and {Ott}}(2015)}]{2015ApJ...799....5C}
\bibinfo{author}{\bibfnamefont{S.~M.} \bibnamefont{{Couch}}} \bibnamefont{and}
  \bibinfo{author}{\bibfnamefont{C.~D.} \bibnamefont{{Ott}}},
  \bibinfo{journal}{\apj} \textbf{\bibinfo{volume}{799}}, \bibinfo{eid}{5}
  (\bibinfo{year}{2015}), \eprint{1408.1399}.

\bibitem[{\citenamefont{{Nagakura} et~al.}(2019)\citenamefont{{Nagakura},
  {Burrows}, {Radice}, and {Vartanyan}}}]{hiroki_2019}
\bibinfo{author}{\bibfnamefont{H.}~\bibnamefont{{Nagakura}}},
  \bibinfo{author}{\bibfnamefont{A.}~\bibnamefont{{Burrows}}},
  \bibinfo{author}{\bibfnamefont{D.}~\bibnamefont{{Radice}}}, \bibnamefont{and}
  \bibinfo{author}{\bibfnamefont{D.}~\bibnamefont{{Vartanyan}}},
  \bibinfo{journal}{\mnras} \textbf{\bibinfo{volume}{490}},
  \bibinfo{pages}{4622} (\bibinfo{year}{2019}), \eprint{1905.03786}.

\bibitem[{\citenamefont{{Bollig} et~al.}(2021)\citenamefont{{Bollig}, {Yadav},
  {Kresse}, {Janka}, {M{\"u}ller}, and {Heger}}}]{2021ApJ...915...28B}
\bibinfo{author}{\bibfnamefont{R.}~\bibnamefont{{Bollig}}},
  \bibinfo{author}{\bibfnamefont{N.}~\bibnamefont{{Yadav}}},
  \bibinfo{author}{\bibfnamefont{D.}~\bibnamefont{{Kresse}}},
  \bibinfo{author}{\bibfnamefont{H.-T.} \bibnamefont{{Janka}}},
  \bibinfo{author}{\bibfnamefont{B.}~\bibnamefont{{M{\"u}ller}}},
  \bibnamefont{and} \bibinfo{author}{\bibfnamefont{A.}~\bibnamefont{{Heger}}},
  \bibinfo{journal}{\apj} \textbf{\bibinfo{volume}{915}}, \bibinfo{eid}{28}
  (\bibinfo{year}{2021}), \eprint{2010.10506}.

\bibitem[{\citenamefont{{Wang} et~al.}(2022)\citenamefont{{Wang}, {Vartanyan},
  {Burrows}, and {Coleman}}}]{wang}
\bibinfo{author}{\bibfnamefont{T.}~\bibnamefont{{Wang}}},
  \bibinfo{author}{\bibfnamefont{D.}~\bibnamefont{{Vartanyan}}},
  \bibinfo{author}{\bibfnamefont{A.}~\bibnamefont{{Burrows}}},
  \bibnamefont{and} \bibinfo{author}{\bibfnamefont{M.~S.~B.}
  \bibnamefont{{Coleman}}}, \bibinfo{journal}{\mnras}
  \textbf{\bibinfo{volume}{517}}, \bibinfo{pages}{543} (\bibinfo{year}{2022}),
  \eprint{2207.02231}.

\bibitem[{\citenamefont{{Tsang} et~al.}(2022)\citenamefont{{Tsang},
  {Vartanyan}, and {Burrows}}}]{tsang2022}
\bibinfo{author}{\bibfnamefont{B.~T.~H.} \bibnamefont{{Tsang}}},
  \bibinfo{author}{\bibfnamefont{D.}~\bibnamefont{{Vartanyan}}},
  \bibnamefont{and}
  \bibinfo{author}{\bibfnamefont{A.}~\bibnamefont{{Burrows}}},
  \bibinfo{journal}{\apjl} \textbf{\bibinfo{volume}{937}}, \bibinfo{eid}{L15}
  (\bibinfo{year}{2022}), \eprint{2208.01661}.

\bibitem[{\citenamefont{M\"osta et~al.}(2014)\citenamefont{M\"osta, Richers,
  Ott, Haas, Piro, Boydstun, Abdikamalov, Reisswig, and Schnetter}}]{mosta2014}
\bibinfo{author}{\bibfnamefont{P.}~\bibnamefont{M\"osta}},
  \bibinfo{author}{\bibfnamefont{S.}~\bibnamefont{Richers}},
  \bibinfo{author}{\bibfnamefont{C.~D.} \bibnamefont{Ott}},
  \bibinfo{author}{\bibfnamefont{R.}~\bibnamefont{Haas}},
  \bibinfo{author}{\bibfnamefont{A.~L.} \bibnamefont{Piro}},
  \bibinfo{author}{\bibfnamefont{K.}~\bibnamefont{Boydstun}},
  \bibinfo{author}{\bibfnamefont{E.}~\bibnamefont{Abdikamalov}},
  \bibinfo{author}{\bibfnamefont{C.}~\bibnamefont{Reisswig}}, \bibnamefont{and}
  \bibinfo{author}{\bibfnamefont{E.}~\bibnamefont{Schnetter}},
  \bibinfo{journal}{The Astrophysical Journal} \textbf{\bibinfo{volume}{785}},
  \bibinfo{pages}{L29} (\bibinfo{year}{2014}), ISSN \bibinfo{issn}{0004-637X}.

\bibitem[{\citenamefont{{Kuroda} et~al.}(2020)\citenamefont{{Kuroda},
  {Arcones}, {Takiwaki}, and {Kotake}}}]{2020ApJ...896..102K}
\bibinfo{author}{\bibfnamefont{T.}~\bibnamefont{{Kuroda}}},
  \bibinfo{author}{\bibfnamefont{A.}~\bibnamefont{{Arcones}}},
  \bibinfo{author}{\bibfnamefont{T.}~\bibnamefont{{Takiwaki}}},
  \bibnamefont{and} \bibinfo{author}{\bibfnamefont{K.}~\bibnamefont{{Kotake}}},
  \bibinfo{journal}{\apj} \textbf{\bibinfo{volume}{896}}, \bibinfo{eid}{102}
  (\bibinfo{year}{2020}), \eprint{2003.02004}.

\bibitem[{\citenamefont{{Nagakura} et~al.}(2020)\citenamefont{{Nagakura},
  {Burrows}, {Radice}, and {Vartanyan}}}]{nagakura_pns}
\bibinfo{author}{\bibfnamefont{H.}~\bibnamefont{{Nagakura}}},
  \bibinfo{author}{\bibfnamefont{A.}~\bibnamefont{{Burrows}}},
  \bibinfo{author}{\bibfnamefont{D.}~\bibnamefont{{Radice}}}, \bibnamefont{and}
  \bibinfo{author}{\bibfnamefont{D.}~\bibnamefont{{Vartanyan}}},
  \bibinfo{journal}{\mnras} \textbf{\bibinfo{volume}{492}},
  \bibinfo{pages}{5764} (\bibinfo{year}{2020}), \eprint{1912.07615}.

\bibitem[{\citenamefont{Obergaulinger and Aloy}(2021)}]{Obergaulinger2021}
\bibinfo{author}{\bibfnamefont{M.}~\bibnamefont{Obergaulinger}}
  \bibnamefont{and} \bibinfo{author}{\bibfnamefont{M.~A.} \bibnamefont{Aloy}},
  \bibinfo{journal}{Monthly Notices of the Royal Astronomical Society}
  \textbf{\bibinfo{volume}{503}}, \bibinfo{pages}{4942–4963}
  (\bibinfo{year}{2021}), ISSN \bibinfo{issn}{0035-8711}.

\bibitem[{\citenamefont{{Aloy} and {Obergaulinger}}(2021)}]{Aloy2021}
\bibinfo{author}{\bibfnamefont{M.~{\'A}.} \bibnamefont{{Aloy}}}
  \bibnamefont{and}
  \bibinfo{author}{\bibfnamefont{M.}~\bibnamefont{{Obergaulinger}}},
  \bibinfo{journal}{\mnras} \textbf{\bibinfo{volume}{500}},
  \bibinfo{pages}{4365} (\bibinfo{year}{2021}), \eprint{2008.03779}.

\bibitem[{\citenamefont{{White} et~al.}(2022)\citenamefont{{White}, {Burrows},
  {Coleman}, and {Vartanyan}}}]{2022ApJ...926..111W}
\bibinfo{author}{\bibfnamefont{C.~J.} \bibnamefont{{White}}},
  \bibinfo{author}{\bibfnamefont{A.}~\bibnamefont{{Burrows}}},
  \bibinfo{author}{\bibfnamefont{M.~S.~B.} \bibnamefont{{Coleman}}},
  \bibnamefont{and}
  \bibinfo{author}{\bibfnamefont{D.}~\bibnamefont{{Vartanyan}}},
  \bibinfo{journal}{\apj} \textbf{\bibinfo{volume}{926}}, \bibinfo{eid}{111}
  (\bibinfo{year}{2022}), \eprint{2111.01814}.

\bibitem[{\citenamefont{{Powell} et~al.}(2022)\citenamefont{{Powell},
  {Mueller}, {Aguilera-Dena}, and {Langer}}}]{2022arXiv221200200P}
\bibinfo{author}{\bibfnamefont{J.}~\bibnamefont{{Powell}}},
  \bibinfo{author}{\bibfnamefont{B.}~\bibnamefont{{Mueller}}},
  \bibinfo{author}{\bibfnamefont{D.~R.} \bibnamefont{{Aguilera-Dena}}},
  \bibnamefont{and} \bibinfo{author}{\bibfnamefont{N.}~\bibnamefont{{Langer}}},
  \bibinfo{journal}{arXiv e-prints} \bibinfo{eid}{arXiv:2212.00200}
  (\bibinfo{year}{2022}), \eprint{2212.00200}.

\bibitem[{\citenamefont{{Summa} et~al.}(2016)\citenamefont{{Summa}, {Hanke},
  {Janka}, {Melson}, {Marek}, and {M{\"u}ller}}}]{summa2016}
\bibinfo{author}{\bibfnamefont{A.}~\bibnamefont{{Summa}}},
  \bibinfo{author}{\bibfnamefont{F.}~\bibnamefont{{Hanke}}},
  \bibinfo{author}{\bibfnamefont{H.-T.} \bibnamefont{{Janka}}},
  \bibinfo{author}{\bibfnamefont{T.}~\bibnamefont{{Melson}}},
  \bibinfo{author}{\bibfnamefont{A.}~\bibnamefont{{Marek}}}, \bibnamefont{and}
  \bibinfo{author}{\bibfnamefont{B.}~\bibnamefont{{M{\"u}ller}}},
  \bibinfo{journal}{\apj} \textbf{\bibinfo{volume}{825}}, \bibinfo{eid}{6}
  (\bibinfo{year}{2016}), \eprint{1511.07871}.

\bibitem[{\citenamefont{{Vartanyan} et~al.}(2018)\citenamefont{{Vartanyan},
  {Burrows}, {Radice}, {Skinner}, and {Dolence}}}]{vartanyan2018a}
\bibinfo{author}{\bibfnamefont{D.}~\bibnamefont{{Vartanyan}}},
  \bibinfo{author}{\bibfnamefont{A.}~\bibnamefont{{Burrows}}},
  \bibinfo{author}{\bibfnamefont{D.}~\bibnamefont{{Radice}}},
  \bibinfo{author}{\bibfnamefont{M.~A.} \bibnamefont{{Skinner}}},
  \bibnamefont{and}
  \bibinfo{author}{\bibfnamefont{J.}~\bibnamefont{{Dolence}}},
  \bibinfo{journal}{\mnras} \textbf{\bibinfo{volume}{477}},
  \bibinfo{pages}{3091} (\bibinfo{year}{2018}), \eprint{1801.08148}.

\bibitem[{\citenamefont{Bionta et~al.}(1987)}]{Bionta:1987qt}
\bibinfo{author}{\bibfnamefont{R.~M.} \bibnamefont{Bionta}}
  \bibnamefont{et~al.}, \bibinfo{journal}{Phys. Rev. Lett.}
  \textbf{\bibinfo{volume}{58}}, \bibinfo{pages}{1494} (\bibinfo{year}{1987}).

\bibitem[{\citenamefont{Hirata et~al.}(1987)}]{Hirata:1987hu}
\bibinfo{author}{\bibfnamefont{K.}~\bibnamefont{Hirata}} \bibnamefont{et~al.}
  (\bibinfo{collaboration}{Kamiokande-II}), \bibinfo{journal}{Phys. Rev. Lett.}
  \textbf{\bibinfo{volume}{58}}, \bibinfo{pages}{1490} (\bibinfo{year}{1987}),
  \bibinfo{note}{[,727(1987)]}.

\bibitem[{\citenamefont{{Murphy} et~al.}(2009)\citenamefont{{Murphy}, {Ott},
  and {Burrows}}}]{murphy:09}
\bibinfo{author}{\bibfnamefont{J.~W.} \bibnamefont{{Murphy}}},
  \bibinfo{author}{\bibfnamefont{C.~D.} \bibnamefont{{Ott}}}, \bibnamefont{and}
  \bibinfo{author}{\bibfnamefont{A.}~\bibnamefont{{Burrows}}},
  \bibinfo{journal}{\apj} \textbf{\bibinfo{volume}{707}}, \bibinfo{pages}{1173}
  (\bibinfo{year}{2009}), \eprint{0907.4762}.

\bibitem[{\citenamefont{{Yakunin} et~al.}(2010)\citenamefont{{Yakunin},
  {Marronetti}, {Mezzacappa}, {Bruenn}, {Lee}, {Chertkow}, {Hix}, {Blondin},
  {Lentz}, {Messer} et~al.}}]{yakunin:10}
\bibinfo{author}{\bibfnamefont{K.~N.} \bibnamefont{{Yakunin}}},
  \bibinfo{author}{\bibfnamefont{P.}~\bibnamefont{{Marronetti}}},
  \bibinfo{author}{\bibfnamefont{A.}~\bibnamefont{{Mezzacappa}}},
  \bibinfo{author}{\bibfnamefont{S.~W.} \bibnamefont{{Bruenn}}},
  \bibinfo{author}{\bibfnamefont{C.-T.} \bibnamefont{{Lee}}},
  \bibinfo{author}{\bibfnamefont{M.~A.} \bibnamefont{{Chertkow}}},
  \bibinfo{author}{\bibfnamefont{W.~R.} \bibnamefont{{Hix}}},
  \bibinfo{author}{\bibfnamefont{J.~M.} \bibnamefont{{Blondin}}},
  \bibinfo{author}{\bibfnamefont{E.~J.} \bibnamefont{{Lentz}}},
  \bibinfo{author}{\bibfnamefont{O.~E.~B.} \bibnamefont{{Messer}}},
  \bibnamefont{et~al.}, \bibinfo{journal}{Classical and Quantum Gravity}
  \textbf{\bibinfo{volume}{27}}, \bibinfo{eid}{194005} (\bibinfo{year}{2010}),
  \eprint{1005.0779}.

\bibitem[{\citenamefont{{M{\"u}ller} et~al.}(2013)\citenamefont{{M{\"u}ller},
  {Janka}, and {Marek}}}]{2013ApJ...766...43M}
\bibinfo{author}{\bibfnamefont{B.}~\bibnamefont{{M{\"u}ller}}},
  \bibinfo{author}{\bibfnamefont{H.-T.} \bibnamefont{{Janka}}},
  \bibnamefont{and} \bibinfo{author}{\bibfnamefont{A.}~\bibnamefont{{Marek}}},
  \bibinfo{journal}{\apj} \textbf{\bibinfo{volume}{766}}, \bibinfo{eid}{43}
  (\bibinfo{year}{2013}), \eprint{1210.6984}.

\bibitem[{\citenamefont{{Kuroda} et~al.}(2014)\citenamefont{{Kuroda},
  {Takiwaki}, and {Kotake}}}]{kuroda:14}
\bibinfo{author}{\bibfnamefont{T.}~\bibnamefont{{Kuroda}}},
  \bibinfo{author}{\bibfnamefont{T.}~\bibnamefont{{Takiwaki}}},
  \bibnamefont{and} \bibinfo{author}{\bibfnamefont{K.}~\bibnamefont{{Kotake}}},
  \bibinfo{journal}{\prd} \textbf{\bibinfo{volume}{89}}, \bibinfo{eid}{044011}
  (\bibinfo{year}{2014}), \eprint{1304.4372}.

\bibitem[{\citenamefont{{Yakunin} et~al.}(2015)\citenamefont{{Yakunin},
  {Mezzacappa}, {Marronetti}, {Yoshida}, {Bruenn}, {Hix}, {Lentz}, {Bronson
  Messer}, {Harris}, {Endeve} et~al.}}]{yakunin:15}
\bibinfo{author}{\bibfnamefont{K.~N.} \bibnamefont{{Yakunin}}},
  \bibinfo{author}{\bibfnamefont{A.}~\bibnamefont{{Mezzacappa}}},
  \bibinfo{author}{\bibfnamefont{P.}~\bibnamefont{{Marronetti}}},
  \bibinfo{author}{\bibfnamefont{S.}~\bibnamefont{{Yoshida}}},
  \bibinfo{author}{\bibfnamefont{S.~W.} \bibnamefont{{Bruenn}}},
  \bibinfo{author}{\bibfnamefont{W.~R.} \bibnamefont{{Hix}}},
  \bibinfo{author}{\bibfnamefont{E.~J.} \bibnamefont{{Lentz}}},
  \bibinfo{author}{\bibfnamefont{O.~E.} \bibnamefont{{Bronson Messer}}},
  \bibinfo{author}{\bibfnamefont{J.~A.} \bibnamefont{{Harris}}},
  \bibinfo{author}{\bibfnamefont{E.}~\bibnamefont{{Endeve}}},
  \bibnamefont{et~al.}, \bibinfo{journal}{\prd} \textbf{\bibinfo{volume}{92}},
  \bibinfo{eid}{084040} (\bibinfo{year}{2015}), \eprint{1505.05824}.

\bibitem[{\citenamefont{{Kuroda} et~al.}(2016)\citenamefont{{Kuroda}, {Kotake},
  and {Takiwaki}}}]{kuroda2016}
\bibinfo{author}{\bibfnamefont{T.}~\bibnamefont{{Kuroda}}},
  \bibinfo{author}{\bibfnamefont{K.}~\bibnamefont{{Kotake}}}, \bibnamefont{and}
  \bibinfo{author}{\bibfnamefont{T.}~\bibnamefont{{Takiwaki}}},
  \bibinfo{journal}{\apjl} \textbf{\bibinfo{volume}{829}}, \bibinfo{eid}{L14}
  (\bibinfo{year}{2016}), \eprint{1605.09215}.

\bibitem[{\citenamefont{{Andresen} et~al.}(2017)\citenamefont{{Andresen},
  {M{\"u}ller}, {M{\"u}ller}, and {Janka}}}]{andresen2017}
\bibinfo{author}{\bibfnamefont{H.}~\bibnamefont{{Andresen}}},
  \bibinfo{author}{\bibfnamefont{B.}~\bibnamefont{{M{\"u}ller}}},
  \bibinfo{author}{\bibfnamefont{E.}~\bibnamefont{{M{\"u}ller}}},
  \bibnamefont{and} \bibinfo{author}{\bibfnamefont{H.~T.}
  \bibnamefont{{Janka}}}, \bibinfo{journal}{\mnras}
  \textbf{\bibinfo{volume}{468}}, \bibinfo{pages}{2032} (\bibinfo{year}{2017}),
  \eprint{1607.05199}.

\bibitem[{\citenamefont{{M{\"u}ller}}(2017)}]{muller2017b}
\bibinfo{author}{\bibfnamefont{B.}~\bibnamefont{{M{\"u}ller}}},
  \bibinfo{journal}{arXiv e-prints} \bibinfo{eid}{arXiv:1703.04633}
  (\bibinfo{year}{2017}), \eprint{1703.04633}.

\bibitem[{\citenamefont{{Kuroda} et~al.}(2017)\citenamefont{{Kuroda}, {Kotake},
  {Hayama}, and {Takiwaki}}}]{kuroda2017}
\bibinfo{author}{\bibfnamefont{T.}~\bibnamefont{{Kuroda}}},
  \bibinfo{author}{\bibfnamefont{K.}~\bibnamefont{{Kotake}}},
  \bibinfo{author}{\bibfnamefont{K.}~\bibnamefont{{Hayama}}}, \bibnamefont{and}
  \bibinfo{author}{\bibfnamefont{T.}~\bibnamefont{{Takiwaki}}},
  \bibinfo{journal}{\apj} \textbf{\bibinfo{volume}{851}}, \bibinfo{eid}{62}
  (\bibinfo{year}{2017}), \eprint{1708.05252}.

\bibitem[{\citenamefont{{Takiwaki} and {Kotake}}(2018)}]{tk18}
\bibinfo{author}{\bibfnamefont{T.}~\bibnamefont{{Takiwaki}}} \bibnamefont{and}
  \bibinfo{author}{\bibfnamefont{K.}~\bibnamefont{{Kotake}}},
  \bibinfo{journal}{\mnras}  (\bibinfo{year}{2018}), \eprint{1711.01905}.

\bibitem[{\citenamefont{{Morozova}
  et~al.}(2018{\natexlab{a}})\citenamefont{{Morozova}, {Radice}, {Burrows}, and
  {Vartanyan}}}]{2018ApJ...861...10M}
\bibinfo{author}{\bibfnamefont{V.}~\bibnamefont{{Morozova}}},
  \bibinfo{author}{\bibfnamefont{D.}~\bibnamefont{{Radice}}},
  \bibinfo{author}{\bibfnamefont{A.}~\bibnamefont{{Burrows}}},
  \bibnamefont{and}
  \bibinfo{author}{\bibfnamefont{D.}~\bibnamefont{{Vartanyan}}},
  \bibinfo{journal}{\apj} \textbf{\bibinfo{volume}{861}}, \bibinfo{eid}{10}
  (\bibinfo{year}{2018}{\natexlab{a}}), \eprint{1801.01914}.

\bibitem[{\citenamefont{{Radice} et~al.}(2019)\citenamefont{{Radice},
  {Morozova}, {Burrows}, {Vartanyan}, and {Nagakura}}}]{radice2019}
\bibinfo{author}{\bibfnamefont{D.}~\bibnamefont{{Radice}}},
  \bibinfo{author}{\bibfnamefont{V.}~\bibnamefont{{Morozova}}},
  \bibinfo{author}{\bibfnamefont{A.}~\bibnamefont{{Burrows}}},
  \bibinfo{author}{\bibfnamefont{D.}~\bibnamefont{{Vartanyan}}},
  \bibnamefont{and}
  \bibinfo{author}{\bibfnamefont{H.}~\bibnamefont{{Nagakura}}},
  \bibinfo{journal}{\apj} \textbf{\bibinfo{volume}{876}}, \bibinfo{eid}{L9}
  (\bibinfo{year}{2019}), \eprint{1812.07703}.

\bibitem[{\citenamefont{{Shibagaki} et~al.}(2021)\citenamefont{{Shibagaki},
  {Kuroda}, {Kotake}, and {Takiwaki}}}]{2021MNRAS.502.3066S}
\bibinfo{author}{\bibfnamefont{S.}~\bibnamefont{{Shibagaki}}},
  \bibinfo{author}{\bibfnamefont{T.}~\bibnamefont{{Kuroda}}},
  \bibinfo{author}{\bibfnamefont{K.}~\bibnamefont{{Kotake}}}, \bibnamefont{and}
  \bibinfo{author}{\bibfnamefont{T.}~\bibnamefont{{Takiwaki}}},
  \bibinfo{journal}{\mnras} \textbf{\bibinfo{volume}{502}},
  \bibinfo{pages}{3066} (\bibinfo{year}{2021}), \eprint{2010.03882}.

\bibitem[{\citenamefont{{Mezzacappa} et~al.}(2023)\citenamefont{{Mezzacappa},
  {Marronetti}, {Landfield}, {Lentz}, {Murphy}, {Hix}, {Harris}, {Bruenn},
  {Blondin}, {Bronson Messer} et~al.}}]{mezzacappa2022}
\bibinfo{author}{\bibfnamefont{A.}~\bibnamefont{{Mezzacappa}}},
  \bibinfo{author}{\bibfnamefont{P.}~\bibnamefont{{Marronetti}}},
  \bibinfo{author}{\bibfnamefont{R.~E.} \bibnamefont{{Landfield}}},
  \bibinfo{author}{\bibfnamefont{E.~J.} \bibnamefont{{Lentz}}},
  \bibinfo{author}{\bibfnamefont{R.~D.} \bibnamefont{{Murphy}}},
  \bibinfo{author}{\bibfnamefont{W.~R.} \bibnamefont{{Hix}}},
  \bibinfo{author}{\bibfnamefont{J.~A.} \bibnamefont{{Harris}}},
  \bibinfo{author}{\bibfnamefont{S.~W.} \bibnamefont{{Bruenn}}},
  \bibinfo{author}{\bibfnamefont{J.~M.} \bibnamefont{{Blondin}}},
  \bibinfo{author}{\bibfnamefont{O.~E.} \bibnamefont{{Bronson Messer}}},
  \bibnamefont{et~al.}, \bibinfo{journal}{\prd} \textbf{\bibinfo{volume}{107}},
  \bibinfo{eid}{043008} (\bibinfo{year}{2023}), \eprint{2208.10643}.

\bibitem[{\citenamefont{{Dessart} et~al.}(2006)\citenamefont{{Dessart},
  {Burrows}, {Livne}, and {Ott}}}]{2006ApJ...645..534D}
\bibinfo{author}{\bibfnamefont{L.}~\bibnamefont{{Dessart}}},
  \bibinfo{author}{\bibfnamefont{A.}~\bibnamefont{{Burrows}}},
  \bibinfo{author}{\bibfnamefont{E.}~\bibnamefont{{Livne}}}, \bibnamefont{and}
  \bibinfo{author}{\bibfnamefont{C.~D.} \bibnamefont{{Ott}}},
  \bibinfo{journal}{\apj} \textbf{\bibinfo{volume}{645}}, \bibinfo{pages}{534}
  (\bibinfo{year}{2006}), \eprint{astro-ph/0510229}.

\bibitem[{\citenamefont{{Roberts}}(2012)}]{roberts_12c}
\bibinfo{author}{\bibfnamefont{L.~F.} \bibnamefont{{Roberts}}},
  \bibinfo{journal}{\apj} \textbf{\bibinfo{volume}{755}}, \bibinfo{eid}{126}
  (\bibinfo{year}{2012}), \eprint{1205.3228}.

\bibitem[{\citenamefont{{Hayama} et~al.}(2018)\citenamefont{{Hayama}, {Kuroda},
  {Kotake}, and {Takiwaki}}}]{2018hayama}
\bibinfo{author}{\bibfnamefont{K.}~\bibnamefont{{Hayama}}},
  \bibinfo{author}{\bibfnamefont{T.}~\bibnamefont{{Kuroda}}},
  \bibinfo{author}{\bibfnamefont{K.}~\bibnamefont{{Kotake}}}, \bibnamefont{and}
  \bibinfo{author}{\bibfnamefont{T.}~\bibnamefont{{Takiwaki}}},
  \bibinfo{journal}{\mnras} \textbf{\bibinfo{volume}{477}},
  \bibinfo{pages}{L96} (\bibinfo{year}{2018}), \eprint{1802.03842}.

\bibitem[{\citenamefont{{Pajkos} et~al.}(2019)\citenamefont{{Pajkos}, {Couch},
  {Pan}, and {O{\textquoteright}Connor}}}]{pajkos2019}
\bibinfo{author}{\bibfnamefont{M.~A.} \bibnamefont{{Pajkos}}},
  \bibinfo{author}{\bibfnamefont{S.~M.} \bibnamefont{{Couch}}},
  \bibinfo{author}{\bibfnamefont{K.-C.} \bibnamefont{{Pan}}}, \bibnamefont{and}
  \bibinfo{author}{\bibfnamefont{E.~P.}
  \bibnamefont{{O{\textquoteright}Connor}}}, \bibinfo{journal}{\apj}
  \textbf{\bibinfo{volume}{878}}, \bibinfo{eid}{13} (\bibinfo{year}{2019}),
  \eprint{1901.09055}.

\bibitem[{\citenamefont{{Marek} et~al.}(2009)\citenamefont{{Marek}, {Janka},
  and {M{\"u}ller}}}]{marek:09}
\bibinfo{author}{\bibfnamefont{A.}~\bibnamefont{{Marek}}},
  \bibinfo{author}{\bibfnamefont{H.-T.} \bibnamefont{{Janka}}},
  \bibnamefont{and}
  \bibinfo{author}{\bibfnamefont{E.}~\bibnamefont{{M{\"u}ller}}},
  \bibinfo{journal}{\aap} \textbf{\bibinfo{volume}{496}}, \bibinfo{pages}{475}
  (\bibinfo{year}{2009}), \eprint{0808.4136}.

\bibitem[{\citenamefont{{Burrows} and {Hayes}}(1996)}]{burrows1996}
\bibinfo{author}{\bibfnamefont{A.}~\bibnamefont{{Burrows}}} \bibnamefont{and}
  \bibinfo{author}{\bibfnamefont{J.}~\bibnamefont{{Hayes}}},
  \bibinfo{journal}{\prl} \textbf{\bibinfo{volume}{76}}, \bibinfo{pages}{352}
  (\bibinfo{year}{1996}), \eprint{astro-ph/9511106}.

\bibitem[{\citenamefont{{Vartanyan} and
  {Burrows}}(2020{\natexlab{a}})}]{2020ApJ...901..108V}
\bibinfo{author}{\bibfnamefont{D.}~\bibnamefont{{Vartanyan}}} \bibnamefont{and}
  \bibinfo{author}{\bibfnamefont{A.}~\bibnamefont{{Burrows}}},
  \bibinfo{journal}{\apj} \textbf{\bibinfo{volume}{901}}, \bibinfo{eid}{108}
  (\bibinfo{year}{2020}{\natexlab{a}}), \eprint{2007.07261}.

\bibitem[{\citenamefont{{Richardson} et~al.}(2022)\citenamefont{{Richardson},
  {Zanolin}, {Andresen}, {Szczepa{\'n}czyk}, {Gill}, and
  {Wongwathanarat}}}]{2022PhRvD.105j3008R}
\bibinfo{author}{\bibfnamefont{C.~J.} \bibnamefont{{Richardson}}},
  \bibinfo{author}{\bibfnamefont{M.}~\bibnamefont{{Zanolin}}},
  \bibinfo{author}{\bibfnamefont{H.}~\bibnamefont{{Andresen}}},
  \bibinfo{author}{\bibfnamefont{M.~J.} \bibnamefont{{Szczepa{\'n}czyk}}},
  \bibinfo{author}{\bibfnamefont{K.}~\bibnamefont{{Gill}}}, \bibnamefont{and}
  \bibinfo{author}{\bibfnamefont{A.}~\bibnamefont{{Wongwathanarat}}},
  \bibinfo{journal}{\prd} \textbf{\bibinfo{volume}{105}}, \bibinfo{eid}{103008}
  (\bibinfo{year}{2022}), \eprint{2109.01582}.

\bibitem[{\citenamefont{{Mukhopadhyay}
  et~al.}(2022)\citenamefont{{Mukhopadhyay}, {Lin}, and
  {Lunardini}}}]{2022PhRvD.106d3020M}
\bibinfo{author}{\bibfnamefont{M.}~\bibnamefont{{Mukhopadhyay}}},
  \bibinfo{author}{\bibfnamefont{Z.}~\bibnamefont{{Lin}}}, \bibnamefont{and}
  \bibinfo{author}{\bibfnamefont{C.}~\bibnamefont{{Lunardini}}},
  \bibinfo{journal}{\prd} \textbf{\bibinfo{volume}{106}}, \bibinfo{eid}{043020}
  (\bibinfo{year}{2022}), \eprint{2110.14657}.

\bibitem[{\citenamefont{Sumiyoshi et~al.}(2005)\citenamefont{Sumiyoshi, Yamada,
  Suzuki, Shen, Chiba, and Toki}}]{SuYaSu05}
\bibinfo{author}{\bibfnamefont{K.}~\bibnamefont{Sumiyoshi}},
  \bibinfo{author}{\bibfnamefont{S.}~\bibnamefont{Yamada}},
  \bibinfo{author}{\bibfnamefont{H.}~\bibnamefont{Suzuki}},
  \bibinfo{author}{\bibfnamefont{H.}~\bibnamefont{Shen}},
  \bibinfo{author}{\bibfnamefont{S.}~\bibnamefont{Chiba}}, \bibnamefont{and}
  \bibinfo{author}{\bibfnamefont{H.}~\bibnamefont{Toki}}, \bibinfo{journal}{The
  Astrophysical Journal} \textbf{\bibinfo{volume}{629}}, \bibinfo{pages}{922}
  (\bibinfo{year}{2005}).

\bibitem[{\citenamefont{{Pan} et~al.}(2018)\citenamefont{{Pan},
  {Liebend{\"o}rfer}, {Couch}, and {Thielemann}}}]{pan:17}
\bibinfo{author}{\bibfnamefont{K.-C.} \bibnamefont{{Pan}}},
  \bibinfo{author}{\bibfnamefont{M.}~\bibnamefont{{Liebend{\"o}rfer}}},
  \bibinfo{author}{\bibfnamefont{S.~M.} \bibnamefont{{Couch}}},
  \bibnamefont{and} \bibinfo{author}{\bibfnamefont{F.-K.}
  \bibnamefont{{Thielemann}}}, \bibinfo{journal}{\apj}
  \textbf{\bibinfo{volume}{857}}, \bibinfo{eid}{13} (\bibinfo{year}{2018}),
  \eprint{1710.01690}.

\bibitem[{\citenamefont{{Richers} et~al.}(2017)\citenamefont{{Richers}, {Ott},
  {Abdikamalov}, {O'Connor}, and {Sullivan}}}]{richers:17}
\bibinfo{author}{\bibfnamefont{S.}~\bibnamefont{{Richers}}},
  \bibinfo{author}{\bibfnamefont{C.~D.} \bibnamefont{{Ott}}},
  \bibinfo{author}{\bibfnamefont{E.}~\bibnamefont{{Abdikamalov}}},
  \bibinfo{author}{\bibfnamefont{E.}~\bibnamefont{{O'Connor}}},
  \bibnamefont{and}
  \bibinfo{author}{\bibfnamefont{C.}~\bibnamefont{{Sullivan}}},
  \bibinfo{journal}{\prd} \textbf{\bibinfo{volume}{95}}, \bibinfo{eid}{063019}
  (\bibinfo{year}{2017}), \eprint{1701.02752}.

\bibitem[{\citenamefont{{Burrows} and {Lattimer}}(1986)}]{burlat86}
\bibinfo{author}{\bibfnamefont{A.}~\bibnamefont{{Burrows}}} \bibnamefont{and}
  \bibinfo{author}{\bibfnamefont{J.~M.} \bibnamefont{{Lattimer}}},
  \bibinfo{journal}{\apj} \textbf{\bibinfo{volume}{307}}, \bibinfo{pages}{178}
  (\bibinfo{year}{1986}).

\bibitem[{\citenamefont{{Burrows}}(1986)}]{1986ApJ...300..488B}
\bibinfo{author}{\bibfnamefont{A.}~\bibnamefont{{Burrows}}},
  \bibinfo{journal}{\apj} \textbf{\bibinfo{volume}{300}}, \bibinfo{pages}{488}
  (\bibinfo{year}{1986}).

\bibitem[{\citenamefont{{Blondin} et~al.}(2003)\citenamefont{{Blondin},
  {Mezzacappa}, and {DeMarino}}}]{blondin:03}
\bibinfo{author}{\bibfnamefont{J.~M.} \bibnamefont{{Blondin}}},
  \bibinfo{author}{\bibfnamefont{A.}~\bibnamefont{{Mezzacappa}}},
  \bibnamefont{and}
  \bibinfo{author}{\bibfnamefont{C.}~\bibnamefont{{DeMarino}}},
  \bibinfo{journal}{\apj} \textbf{\bibinfo{volume}{584}}, \bibinfo{pages}{971}
  (\bibinfo{year}{2003}), \eprint{astro-ph/0210634}.

\bibitem[{\citenamefont{{Foglizzo} et~al.}(2007)\citenamefont{{Foglizzo},
  {Galletti}, {Scheck}, and {Janka}}}]{foglizzo:07}
\bibinfo{author}{\bibfnamefont{T.}~\bibnamefont{{Foglizzo}}},
  \bibinfo{author}{\bibfnamefont{P.}~\bibnamefont{{Galletti}}},
  \bibinfo{author}{\bibfnamefont{L.}~\bibnamefont{{Scheck}}}, \bibnamefont{and}
  \bibinfo{author}{\bibfnamefont{H.-T.} \bibnamefont{{Janka}}},
  \bibinfo{journal}{\apj} \textbf{\bibinfo{volume}{654}}, \bibinfo{pages}{1006}
  (\bibinfo{year}{2007}), \eprint{astro-ph/0606640}.

\bibitem[{\citenamefont{{Blondin} and {Shaw}}(2007)}]{blondin_shaw}
\bibinfo{author}{\bibfnamefont{J.~M.} \bibnamefont{{Blondin}}}
  \bibnamefont{and} \bibinfo{author}{\bibfnamefont{S.}~\bibnamefont{{Shaw}}},
  \bibinfo{journal}{\apj} \textbf{\bibinfo{volume}{656}}, \bibinfo{pages}{366}
  (\bibinfo{year}{2007}), \eprint{astro-ph/0611698}.

\bibitem[{\citenamefont{{Vartanyan}
  et~al.}(2019{\natexlab{b}})\citenamefont{{Vartanyan}, {Burrows}, and
  {Radice}}}]{vartanyan2019}
\bibinfo{author}{\bibfnamefont{D.}~\bibnamefont{{Vartanyan}}},
  \bibinfo{author}{\bibfnamefont{A.}~\bibnamefont{{Burrows}}},
  \bibnamefont{and} \bibinfo{author}{\bibfnamefont{D.}~\bibnamefont{{Radice}}},
  \bibinfo{journal}{\mnras} \textbf{\bibinfo{volume}{489}},
  \bibinfo{pages}{2227} (\bibinfo{year}{2019}{\natexlab{b}}),
  \eprint{1906.08787}.

\bibitem[{\citenamefont{{Coleman} and {Burrows}}(2022)}]{coleman}
\bibinfo{author}{\bibfnamefont{M.~S.~B.} \bibnamefont{{Coleman}}}
  \bibnamefont{and}
  \bibinfo{author}{\bibfnamefont{A.}~\bibnamefont{{Burrows}}},
  \bibinfo{journal}{\mnras} \textbf{\bibinfo{volume}{517}},
  \bibinfo{pages}{3938} (\bibinfo{year}{2022}), \eprint{2209.02711}.

\bibitem[{\citenamefont{{O'Connor} and {Ott}}(2011)}]{2011ApJ...730...70O}
\bibinfo{author}{\bibfnamefont{E.}~\bibnamefont{{O'Connor}}} \bibnamefont{and}
  \bibinfo{author}{\bibfnamefont{C.~D.} \bibnamefont{{Ott}}},
  \bibinfo{journal}{\apj} \textbf{\bibinfo{volume}{730}}, \bibinfo{eid}{70}
  (\bibinfo{year}{2011}), \eprint{1010.5550}.

\bibitem[{\citenamefont{{Skinner} et~al.}(2019)\citenamefont{{Skinner},
  {Dolence}, {Burrows}, {Radice}, and {Vartanyan}}}]{skinner2019}
\bibinfo{author}{\bibfnamefont{M.~A.} \bibnamefont{{Skinner}}},
  \bibinfo{author}{\bibfnamefont{J.~C.} \bibnamefont{{Dolence}}},
  \bibinfo{author}{\bibfnamefont{A.}~\bibnamefont{{Burrows}}},
  \bibinfo{author}{\bibfnamefont{D.}~\bibnamefont{{Radice}}}, \bibnamefont{and}
  \bibinfo{author}{\bibfnamefont{D.}~\bibnamefont{{Vartanyan}}},
  \bibinfo{journal}{\apjs} \textbf{\bibinfo{volume}{241}}, \bibinfo{eid}{7}
  (\bibinfo{year}{2019}), \eprint{1806.07390}.

\bibitem[{\citenamefont{{Finn} and {Evans}}(1990)}]{finn}
\bibinfo{author}{\bibfnamefont{L.~S.} \bibnamefont{{Finn}}} \bibnamefont{and}
  \bibinfo{author}{\bibfnamefont{C.~R.} \bibnamefont{{Evans}}},
  \bibinfo{journal}{\apj} \textbf{\bibinfo{volume}{351}}, \bibinfo{pages}{588}
  (\bibinfo{year}{1990}).

\bibitem[{\citenamefont{Oohara et~al.}(1997)\citenamefont{Oohara, Nakamura, and
  Shibata}}]{oohara97}
\bibinfo{author}{\bibfnamefont{K.}~\bibnamefont{Oohara}},
  \bibinfo{author}{\bibfnamefont{T.}~\bibnamefont{Nakamura}}, \bibnamefont{and}
  \bibinfo{author}{\bibfnamefont{M.}~\bibnamefont{Shibata}},
  \bibinfo{journal}{Progress of Theoretical Physics Supplement}
  \textbf{\bibinfo{volume}{128}}, \bibinfo{pages}{183} (\bibinfo{year}{1997}).

\bibitem[{\citenamefont{{Sukhbold} et~al.}(2018)\citenamefont{{Sukhbold},
  {Woosley}, and {Heger}}}]{sukhbold2018}
\bibinfo{author}{\bibfnamefont{T.}~\bibnamefont{{Sukhbold}}},
  \bibinfo{author}{\bibfnamefont{S.~E.} \bibnamefont{{Woosley}}},
  \bibnamefont{and} \bibinfo{author}{\bibfnamefont{A.}~\bibnamefont{{Heger}}},
  \bibinfo{journal}{\apj} \textbf{\bibinfo{volume}{860}}, \bibinfo{eid}{93}
  (\bibinfo{year}{2018}), \eprint{1710.03243}.

\bibitem[{\citenamefont{{Sukhbold} et~al.}(2016)\citenamefont{{Sukhbold},
  {Ertl}, {Woosley}, {Brown}, and {Janka}}}]{swbj16}
\bibinfo{author}{\bibfnamefont{T.}~\bibnamefont{{Sukhbold}}},
  \bibinfo{author}{\bibfnamefont{T.}~\bibnamefont{{Ertl}}},
  \bibinfo{author}{\bibfnamefont{S.~E.} \bibnamefont{{Woosley}}},
  \bibinfo{author}{\bibfnamefont{J.~M.} \bibnamefont{{Brown}}},
  \bibnamefont{and} \bibinfo{author}{\bibfnamefont{H.-T.}
  \bibnamefont{{Janka}}}, \bibinfo{journal}{\apj}
  \textbf{\bibinfo{volume}{821}}, \bibinfo{eid}{38} (\bibinfo{year}{2016}),
  \eprint{1510.04643}.

\bibitem[{\citenamefont{{Steiner} et~al.}(2013)\citenamefont{{Steiner},
  {Hempel}, and {Fischer}}}]{2013ApJ...774...17S}
\bibinfo{author}{\bibfnamefont{A.~W.} \bibnamefont{{Steiner}}},
  \bibinfo{author}{\bibfnamefont{M.}~\bibnamefont{{Hempel}}}, \bibnamefont{and}
  \bibinfo{author}{\bibfnamefont{T.}~\bibnamefont{{Fischer}}},
  \bibinfo{journal}{\apj} \textbf{\bibinfo{volume}{774}}, \bibinfo{eid}{17}
  (\bibinfo{year}{2013}), \eprint{1207.2184}.

\bibitem[{\citenamefont{{Burrows} et~al.}(2019)\citenamefont{{Burrows},
  {Radice}, and {Vartanyan}}}]{burrows_2019}
\bibinfo{author}{\bibfnamefont{A.}~\bibnamefont{{Burrows}}},
  \bibinfo{author}{\bibfnamefont{D.}~\bibnamefont{{Radice}}}, \bibnamefont{and}
  \bibinfo{author}{\bibfnamefont{D.}~\bibnamefont{{Vartanyan}}},
  \bibinfo{journal}{\mnras} \textbf{\bibinfo{volume}{485}},
  \bibinfo{pages}{3153} (\bibinfo{year}{2019}), \eprint{1902.00547}.

\bibitem[{\citenamefont{{Fryer}}(1999)}]{fryer1999}
\bibinfo{author}{\bibfnamefont{C.~L.} \bibnamefont{{Fryer}}},
  \bibinfo{journal}{\apj} \textbf{\bibinfo{volume}{522}}, \bibinfo{pages}{413}
  (\bibinfo{year}{1999}), \eprint{astro-ph/9902315}.

\bibitem[{\citenamefont{{Vartanyan} et~al.}(2021)\citenamefont{{Vartanyan},
  {Laplace}, {Renzo}, {G{\"o}tberg}, {Burrows}, and {de
  Mink}}}]{2021ApJ...916L...5V}
\bibinfo{author}{\bibfnamefont{D.}~\bibnamefont{{Vartanyan}}},
  \bibinfo{author}{\bibfnamefont{E.}~\bibnamefont{{Laplace}}},
  \bibinfo{author}{\bibfnamefont{M.}~\bibnamefont{{Renzo}}},
  \bibinfo{author}{\bibfnamefont{Y.}~\bibnamefont{{G{\"o}tberg}}},
  \bibinfo{author}{\bibfnamefont{A.}~\bibnamefont{{Burrows}}},
  \bibnamefont{and} \bibinfo{author}{\bibfnamefont{S.~E.} \bibnamefont{{de
  Mink}}}, \bibinfo{journal}{\apjl} \textbf{\bibinfo{volume}{916}},
  \bibinfo{eid}{L5} (\bibinfo{year}{2021}), \eprint{2104.03317}.

\bibitem[{\citenamefont{{Wanajo} et~al.}(2018)\citenamefont{{Wanajo},
  {M{\"u}ller}, {Janka}, and {Heger}}}]{wanajo2018}
\bibinfo{author}{\bibfnamefont{S.}~\bibnamefont{{Wanajo}}},
  \bibinfo{author}{\bibfnamefont{B.}~\bibnamefont{{M{\"u}ller}}},
  \bibinfo{author}{\bibfnamefont{H.-T.} \bibnamefont{{Janka}}},
  \bibnamefont{and} \bibinfo{author}{\bibfnamefont{A.}~\bibnamefont{{Heger}}},
  \bibinfo{journal}{\apj} \textbf{\bibinfo{volume}{852}}, \bibinfo{eid}{40}
  (\bibinfo{year}{2018}), \eprint{1701.06786}.

\bibitem[{\citenamefont{{Torres-Forn{\'e}}
  et~al.}(2019)\citenamefont{{Torres-Forn{\'e}}, {Cerd{\'a}-Dur{\'a}n},
  {Obergaulinger}, {M{\"u}ller}, and {Font}}}]{torres2019}
\bibinfo{author}{\bibfnamefont{A.}~\bibnamefont{{Torres-Forn{\'e}}}},
  \bibinfo{author}{\bibfnamefont{P.}~\bibnamefont{{Cerd{\'a}-Dur{\'a}n}}},
  \bibinfo{author}{\bibfnamefont{M.}~\bibnamefont{{Obergaulinger}}},
  \bibinfo{author}{\bibfnamefont{B.}~\bibnamefont{{M{\"u}ller}}},
  \bibnamefont{and} \bibinfo{author}{\bibfnamefont{J.~A.}
  \bibnamefont{{Font}}}, \bibinfo{journal}{\prl}
  \textbf{\bibinfo{volume}{123}}, \bibinfo{eid}{051102} (\bibinfo{year}{2019}),
  \eprint{1902.10048}.

\bibitem[{\citenamefont{{Eggenberger Andersen}
  et~al.}(2021)\citenamefont{{Eggenberger Andersen}, {Zha}, {da Silva
  Schneider}, {Betranhandy}, {Couch}, and {O'Connor}}}]{eggenberger2021}
\bibinfo{author}{\bibfnamefont{O.}~\bibnamefont{{Eggenberger Andersen}}},
  \bibinfo{author}{\bibfnamefont{S.}~\bibnamefont{{Zha}}},
  \bibinfo{author}{\bibfnamefont{A.}~\bibnamefont{{da Silva Schneider}}},
  \bibinfo{author}{\bibfnamefont{A.}~\bibnamefont{{Betranhandy}}},
  \bibinfo{author}{\bibfnamefont{S.~M.} \bibnamefont{{Couch}}},
  \bibnamefont{and} \bibinfo{author}{\bibfnamefont{E.~P.}
  \bibnamefont{{O'Connor}}}, \bibinfo{journal}{\apj}
  \textbf{\bibinfo{volume}{923}}, \bibinfo{eid}{201} (\bibinfo{year}{2021}),
  \eprint{2106.09734}.

\bibitem[{\citenamefont{{Bruel} et~al.}(2023)\citenamefont{{Bruel}, {Bizouard},
  {Obergaulinger}, {Maturana-Russel}, {Torres-Forn{\'e}},
  {Cerd{\'a}-Dur{\'a}n}, {Christensen}, {Font}, and {Meyer}}}]{bruel2023}
\bibinfo{author}{\bibfnamefont{T.}~\bibnamefont{{Bruel}}},
  \bibinfo{author}{\bibfnamefont{M.-A.} \bibnamefont{{Bizouard}}},
  \bibinfo{author}{\bibfnamefont{M.}~\bibnamefont{{Obergaulinger}}},
  \bibinfo{author}{\bibfnamefont{P.}~\bibnamefont{{Maturana-Russel}}},
  \bibinfo{author}{\bibfnamefont{A.}~\bibnamefont{{Torres-Forn{\'e}}}},
  \bibinfo{author}{\bibfnamefont{P.}~\bibnamefont{{Cerd{\'a}-Dur{\'a}n}}},
  \bibinfo{author}{\bibfnamefont{N.}~\bibnamefont{{Christensen}}},
  \bibinfo{author}{\bibfnamefont{J.~A.} \bibnamefont{{Font}}},
  \bibnamefont{and} \bibinfo{author}{\bibfnamefont{R.}~\bibnamefont{{Meyer}}},
  \bibinfo{journal}{\prd} \textbf{\bibinfo{volume}{107}}, \bibinfo{eid}{083029}
  (\bibinfo{year}{2023}), \eprint{2301.10019}.

\bibitem[{\citenamefont{{Aizenman} et~al.}(1977)\citenamefont{{Aizenman},
  {Smeyers}, and {Weigert}}}]{aizenman}
\bibinfo{author}{\bibfnamefont{M.}~\bibnamefont{{Aizenman}}},
  \bibinfo{author}{\bibfnamefont{P.}~\bibnamefont{{Smeyers}}},
  \bibnamefont{and}
  \bibinfo{author}{\bibfnamefont{A.}~\bibnamefont{{Weigert}}},
  \bibinfo{journal}{\aap} \textbf{\bibinfo{volume}{58}}, \bibinfo{pages}{41}
  (\bibinfo{year}{1977}).

\bibitem[{\citenamefont{{Jakobus} et~al.}(2023)\citenamefont{{Jakobus},
  {M{\"u}ller}, {Heger}, {Zha}, {Powell}, {Motornenko}, {Steinheimer}, and
  {Stoecker}}}]{jakobus}
\bibinfo{author}{\bibfnamefont{P.}~\bibnamefont{{Jakobus}}},
  \bibinfo{author}{\bibfnamefont{B.}~\bibnamefont{{M{\"u}ller}}},
  \bibinfo{author}{\bibfnamefont{A.}~\bibnamefont{{Heger}}},
  \bibinfo{author}{\bibfnamefont{S.}~\bibnamefont{{Zha}}},
  \bibinfo{author}{\bibfnamefont{J.}~\bibnamefont{{Powell}}},
  \bibinfo{author}{\bibfnamefont{A.}~\bibnamefont{{Motornenko}}},
  \bibinfo{author}{\bibfnamefont{J.}~\bibnamefont{{Steinheimer}}},
  \bibnamefont{and}
  \bibinfo{author}{\bibfnamefont{H.}~\bibnamefont{{Stoecker}}},
  \bibinfo{journal}{arXiv e-prints} \bibinfo{eid}{arXiv:2301.06515}
  (\bibinfo{year}{2023}), \eprint{2301.06515}.

\bibitem[{\citenamefont{{Epstein}}(1978)}]{epstein1978}
\bibinfo{author}{\bibfnamefont{R.}~\bibnamefont{{Epstein}}},
  \bibinfo{journal}{\apj} \textbf{\bibinfo{volume}{223}}, \bibinfo{pages}{1037}
  (\bibinfo{year}{1978}).

\bibitem[{\citenamefont{{M{\"u}ller} et~al.}(2012)\citenamefont{{M{\"u}ller},
  {Janka}, and {Wongwathanarat}}}]{emuller}
\bibinfo{author}{\bibfnamefont{E.}~\bibnamefont{{M{\"u}ller}}},
  \bibinfo{author}{\bibfnamefont{H.~T.} \bibnamefont{{Janka}}},
  \bibnamefont{and}
  \bibinfo{author}{\bibfnamefont{A.}~\bibnamefont{{Wongwathanarat}}},
  \bibinfo{journal}{\aap} \textbf{\bibinfo{volume}{537}}, \bibinfo{eid}{A63}
  (\bibinfo{year}{2012}), \eprint{1106.6301}.

\bibitem[{\citenamefont{{Vartanyan} and
  {Burrows}}(2020{\natexlab{b}})}]{vartanyan2020}
\bibinfo{author}{\bibfnamefont{D.}~\bibnamefont{{Vartanyan}}} \bibnamefont{and}
  \bibinfo{author}{\bibfnamefont{A.}~\bibnamefont{{Burrows}}},
  \bibinfo{journal}{\apj} \textbf{\bibinfo{volume}{901}}, \bibinfo{eid}{108}
  (\bibinfo{year}{2020}{\natexlab{b}}), \eprint{2007.07261}.

\bibitem[{\citenamefont{Aasi et~al.}(2015)\citenamefont{Aasi, Abbott, Abbott,
  Abbott, Abernathy, Ackley, Adams, Adams, Addesso, Adhikari et~al.}}]{aligo}
\bibinfo{author}{\bibfnamefont{J.}~\bibnamefont{Aasi}},
  \bibinfo{author}{\bibfnamefont{B.~P.} \bibnamefont{Abbott}},
  \bibinfo{author}{\bibfnamefont{R.}~\bibnamefont{Abbott}},
  \bibinfo{author}{\bibfnamefont{T.}~\bibnamefont{Abbott}},
  \bibinfo{author}{\bibfnamefont{M.~R.} \bibnamefont{Abernathy}},
  \bibinfo{author}{\bibfnamefont{K.}~\bibnamefont{Ackley}},
  \bibinfo{author}{\bibfnamefont{C.}~\bibnamefont{Adams}},
  \bibinfo{author}{\bibfnamefont{T.}~\bibnamefont{Adams}},
  \bibinfo{author}{\bibfnamefont{P.}~\bibnamefont{Addesso}},
  \bibinfo{author}{\bibfnamefont{R.~X.} \bibnamefont{Adhikari}},
  \bibnamefont{et~al.}, \bibinfo{journal}{Classical and Quantum Gravity}
  \textbf{\bibinfo{volume}{32}}, \bibinfo{pages}{074001}
  (\bibinfo{year}{2015}).

\bibitem[{\citenamefont{{Acernese} et~al.}(2015)\citenamefont{{Acernese},
  {Agathos}, {Agatsuma}, {Aisa}, {Allemandou}, {Allocca}, {Amarni}, {Astone},
  {Balestri}, {Ballardin} et~al.}}]{virgo}
\bibinfo{author}{\bibfnamefont{F.}~\bibnamefont{{Acernese}}},
  \bibinfo{author}{\bibfnamefont{M.}~\bibnamefont{{Agathos}}},
  \bibinfo{author}{\bibfnamefont{K.}~\bibnamefont{{Agatsuma}}},
  \bibinfo{author}{\bibfnamefont{D.}~\bibnamefont{{Aisa}}},
  \bibinfo{author}{\bibfnamefont{N.}~\bibnamefont{{Allemandou}}},
  \bibinfo{author}{\bibfnamefont{A.}~\bibnamefont{{Allocca}}},
  \bibinfo{author}{\bibfnamefont{J.}~\bibnamefont{{Amarni}}},
  \bibinfo{author}{\bibfnamefont{P.}~\bibnamefont{{Astone}}},
  \bibinfo{author}{\bibfnamefont{G.}~\bibnamefont{{Balestri}}},
  \bibinfo{author}{\bibfnamefont{G.}~\bibnamefont{{Ballardin}}},
  \bibnamefont{et~al.}, \bibinfo{journal}{Classical and Quantum Gravity}
  \textbf{\bibinfo{volume}{32}}, \bibinfo{eid}{024001} (\bibinfo{year}{2015}),
  \eprint{1408.3978}.

\bibitem[{\citenamefont{{Akutsu} et~al.}(2021)\citenamefont{{Akutsu}, {Ando},
  {Arai}, {Arai}, {Araki}, {Araya}, {Aritomi}, {Aso}, {Bae}, {Bae}
  et~al.}}]{kagra}
\bibinfo{author}{\bibfnamefont{T.}~\bibnamefont{{Akutsu}}},
  \bibinfo{author}{\bibfnamefont{M.}~\bibnamefont{{Ando}}},
  \bibinfo{author}{\bibfnamefont{K.}~\bibnamefont{{Arai}}},
  \bibinfo{author}{\bibfnamefont{Y.}~\bibnamefont{{Arai}}},
  \bibinfo{author}{\bibfnamefont{S.}~\bibnamefont{{Araki}}},
  \bibinfo{author}{\bibfnamefont{A.}~\bibnamefont{{Araya}}},
  \bibinfo{author}{\bibfnamefont{N.}~\bibnamefont{{Aritomi}}},
  \bibinfo{author}{\bibfnamefont{Y.}~\bibnamefont{{Aso}}},
  \bibinfo{author}{\bibfnamefont{S.}~\bibnamefont{{Bae}}},
  \bibinfo{author}{\bibfnamefont{Y.}~\bibnamefont{{Bae}}},
  \bibnamefont{et~al.}, \bibinfo{journal}{Progress of Theoretical and
  Experimental Physics} \textbf{\bibinfo{volume}{2021}}, \bibinfo{eid}{05A101}
  (\bibinfo{year}{2021}), \eprint{2005.05574}.

\bibitem[{\citenamefont{{Punturo} et~al.}(2010)\citenamefont{{Punturo},
  {Abernathy}, {Acernese}, {Allen}, {Andersson}, {Arun}, {Barone}, {Barr},
  {Barsuglia}, {Beker} et~al.}}]{ET_1}
\bibinfo{author}{\bibfnamefont{M.}~\bibnamefont{{Punturo}}},
  \bibinfo{author}{\bibfnamefont{M.}~\bibnamefont{{Abernathy}}},
  \bibinfo{author}{\bibfnamefont{F.}~\bibnamefont{{Acernese}}},
  \bibinfo{author}{\bibfnamefont{B.}~\bibnamefont{{Allen}}},
  \bibinfo{author}{\bibfnamefont{N.}~\bibnamefont{{Andersson}}},
  \bibinfo{author}{\bibfnamefont{K.}~\bibnamefont{{Arun}}},
  \bibinfo{author}{\bibfnamefont{F.}~\bibnamefont{{Barone}}},
  \bibinfo{author}{\bibfnamefont{B.}~\bibnamefont{{Barr}}},
  \bibinfo{author}{\bibfnamefont{M.}~\bibnamefont{{Barsuglia}}},
  \bibinfo{author}{\bibfnamefont{M.}~\bibnamefont{{Beker}}},
  \bibnamefont{et~al.}, \bibinfo{journal}{Classical and Quantum Gravity}
  \textbf{\bibinfo{volume}{27}}, \bibinfo{eid}{194002} (\bibinfo{year}{2010}).

\bibitem[{\citenamefont{Maggiore et~al.}(2020)\citenamefont{Maggiore, Broeck,
  Bartolo, Belgacem, Bertacca, Bizouard, Branchesi, Clesse, Foffa,
  García-Bellido et~al.}}]{ET_3}
\bibinfo{author}{\bibfnamefont{M.}~\bibnamefont{Maggiore}},
  \bibinfo{author}{\bibfnamefont{C.~V.~D.} \bibnamefont{Broeck}},
  \bibinfo{author}{\bibfnamefont{N.}~\bibnamefont{Bartolo}},
  \bibinfo{author}{\bibfnamefont{E.}~\bibnamefont{Belgacem}},
  \bibinfo{author}{\bibfnamefont{D.}~\bibnamefont{Bertacca}},
  \bibinfo{author}{\bibfnamefont{M.~A.} \bibnamefont{Bizouard}},
  \bibinfo{author}{\bibfnamefont{M.}~\bibnamefont{Branchesi}},
  \bibinfo{author}{\bibfnamefont{S.}~\bibnamefont{Clesse}},
  \bibinfo{author}{\bibfnamefont{S.}~\bibnamefont{Foffa}},
  \bibinfo{author}{\bibfnamefont{J.}~\bibnamefont{García-Bellido}},
  \bibnamefont{et~al.}, \bibinfo{journal}{Journal of Cosmology and
  Astroparticle Physics} \textbf{\bibinfo{volume}{2020}},
  \bibinfo{pages}{050–050} (\bibinfo{year}{2020}), ISSN
  \bibinfo{issn}{1475-7516}.

\bibitem[{\citenamefont{{Srivastava} et~al.}(2022)\citenamefont{{Srivastava},
  {Davis}, {Kuns}, {Landry}, {Ballmer}, {Evans}, {Hall}, {Read}, and
  {Sathyaprakash}}}]{ce}
\bibinfo{author}{\bibfnamefont{V.}~\bibnamefont{{Srivastava}}},
  \bibinfo{author}{\bibfnamefont{D.}~\bibnamefont{{Davis}}},
  \bibinfo{author}{\bibfnamefont{K.}~\bibnamefont{{Kuns}}},
  \bibinfo{author}{\bibfnamefont{P.}~\bibnamefont{{Landry}}},
  \bibinfo{author}{\bibfnamefont{S.}~\bibnamefont{{Ballmer}}},
  \bibinfo{author}{\bibfnamefont{M.}~\bibnamefont{{Evans}}},
  \bibinfo{author}{\bibfnamefont{E.~D.} \bibnamefont{{Hall}}},
  \bibinfo{author}{\bibfnamefont{J.}~\bibnamefont{{Read}}}, \bibnamefont{and}
  \bibinfo{author}{\bibfnamefont{B.~S.} \bibnamefont{{Sathyaprakash}}},
  \bibinfo{journal}{\apj} \textbf{\bibinfo{volume}{931}}, \bibinfo{eid}{22}
  (\bibinfo{year}{2022}), \eprint{2201.10668}.

\bibitem[{\citenamefont{{Cutler} and {Holz}}(2009)}]{BBO}
\bibinfo{author}{\bibfnamefont{C.}~\bibnamefont{{Cutler}}} \bibnamefont{and}
  \bibinfo{author}{\bibfnamefont{D.~E.} \bibnamefont{{Holz}}},
  \bibinfo{journal}{\prd} \textbf{\bibinfo{volume}{80}}, \bibinfo{eid}{104009}
  (\bibinfo{year}{2009}), \eprint{0906.3752}.

\bibitem[{\citenamefont{{Yagi} and {Seto}}(2011)}]{decigo1}
\bibinfo{author}{\bibfnamefont{K.}~\bibnamefont{{Yagi}}} \bibnamefont{and}
  \bibinfo{author}{\bibfnamefont{N.}~\bibnamefont{{Seto}}},
  \bibinfo{journal}{\prd} \textbf{\bibinfo{volume}{83}}, \bibinfo{eid}{044011}
  (\bibinfo{year}{2011}), \eprint{1101.3940}.

\bibitem[{\citenamefont{{Sato} et~al.}(2017)\citenamefont{{Sato}, {Kawamura},
  {Ando}, {Nakamura}, {Tsubono}, {Araya}, {Funaki}, {Ioka}, {Kanda}, {Moriwaki}
  et~al.}}]{decigo2}
\bibinfo{author}{\bibfnamefont{S.}~\bibnamefont{{Sato}}},
  \bibinfo{author}{\bibfnamefont{S.}~\bibnamefont{{Kawamura}}},
  \bibinfo{author}{\bibfnamefont{M.}~\bibnamefont{{Ando}}},
  \bibinfo{author}{\bibfnamefont{T.}~\bibnamefont{{Nakamura}}},
  \bibinfo{author}{\bibfnamefont{K.}~\bibnamefont{{Tsubono}}},
  \bibinfo{author}{\bibfnamefont{A.}~\bibnamefont{{Araya}}},
  \bibinfo{author}{\bibfnamefont{I.}~\bibnamefont{{Funaki}}},
  \bibinfo{author}{\bibfnamefont{K.}~\bibnamefont{{Ioka}}},
  \bibinfo{author}{\bibfnamefont{N.}~\bibnamefont{{Kanda}}},
  \bibinfo{author}{\bibfnamefont{S.}~\bibnamefont{{Moriwaki}}},
  \bibnamefont{et~al.}, in \emph{\bibinfo{booktitle}{Journal of Physics
  Conference Series}} (\bibinfo{year}{2017}), vol. \bibinfo{volume}{840} of
  \emph{\bibinfo{series}{Journal of Physics Conference Series}}, p.
  \bibinfo{pages}{012010}.

\bibitem[{\citenamefont{{Arca Sedda} et~al.}(2020)\citenamefont{{Arca Sedda},
  {Berry}, {Jani}, {Amaro-Seoane}, {Auclair}, {Baird}, {Baker}, {Berti},
  {Breivik}, {Burrows} et~al.}}]{sedda}
\bibinfo{author}{\bibfnamefont{M.}~\bibnamefont{{Arca Sedda}}},
  \bibinfo{author}{\bibfnamefont{C.~P.~L.} \bibnamefont{{Berry}}},
  \bibinfo{author}{\bibfnamefont{K.}~\bibnamefont{{Jani}}},
  \bibinfo{author}{\bibfnamefont{P.}~\bibnamefont{{Amaro-Seoane}}},
  \bibinfo{author}{\bibfnamefont{P.}~\bibnamefont{{Auclair}}},
  \bibinfo{author}{\bibfnamefont{J.}~\bibnamefont{{Baird}}},
  \bibinfo{author}{\bibfnamefont{T.}~\bibnamefont{{Baker}}},
  \bibinfo{author}{\bibfnamefont{E.}~\bibnamefont{{Berti}}},
  \bibinfo{author}{\bibfnamefont{K.}~\bibnamefont{{Breivik}}},
  \bibinfo{author}{\bibfnamefont{A.}~\bibnamefont{{Burrows}}},
  \bibnamefont{et~al.}, \bibinfo{journal}{Classical and Quantum Gravity}
  \textbf{\bibinfo{volume}{37}}, \bibinfo{eid}{215011} (\bibinfo{year}{2020}),
  \eprint{1908.11375}.

\bibitem[{\citenamefont{Stanzione et~al.}(2020)\citenamefont{Stanzione, West,
  Evans, Minyard, Ghattas, and Panda}}]{Stanzione2020}
\bibinfo{author}{\bibfnamefont{D.}~\bibnamefont{Stanzione}},
  \bibinfo{author}{\bibfnamefont{J.}~\bibnamefont{West}},
  \bibinfo{author}{\bibfnamefont{R.~T.} \bibnamefont{Evans}},
  \bibinfo{author}{\bibfnamefont{T.}~\bibnamefont{Minyard}},
  \bibinfo{author}{\bibfnamefont{O.}~\bibnamefont{Ghattas}}, \bibnamefont{and}
  \bibinfo{author}{\bibfnamefont{D.~K.} \bibnamefont{Panda}}, in
  \emph{\bibinfo{booktitle}{{PEARC '20}}} (\bibinfo{address}{Portland, OR},
  \bibinfo{year}{2020}), Practice and Experience in Advanced Research
  Computing, pp. \bibinfo{pages}{106--111}.

\bibitem[{\citenamefont{{Morozova}
  et~al.}(2018{\natexlab{b}})\citenamefont{{Morozova}, {Radice}, {Burrows}, and
  {Vartanyan}}}]{vsg2018}
\bibinfo{author}{\bibfnamefont{V.}~\bibnamefont{{Morozova}}},
  \bibinfo{author}{\bibfnamefont{D.}~\bibnamefont{{Radice}}},
  \bibinfo{author}{\bibfnamefont{A.}~\bibnamefont{{Burrows}}},
  \bibnamefont{and}
  \bibinfo{author}{\bibfnamefont{D.}~\bibnamefont{{Vartanyan}}},
  \bibinfo{journal}{\apj} \textbf{\bibinfo{volume}{861}}, \bibinfo{eid}{10}
  (\bibinfo{year}{2018}{\natexlab{b}}), \eprint{1801.01914}.

\end{thebibliography}

\begin{table*}
\center\caption{\Large{Model Properties}}
\begin{tabular}{p{15 mm} p{17 mm} p{19 mm} p{25 mm} p{27mm} p{15 mm}}

    \hline\hline
  Model        & Run Time & {Explosion?} &  Shock Velocity & f/g-mode  &Nyquist  \\
  & & & & Energy Fraction & Frequency \\
(M$_{\odot}$) & (s, pb) & & (km s$^{-1}$)  &Total (Late) &(Hz)\\ \hline
9$\mathrm{\Theta}$    &  1.47  & \checkmark\    & 16,000 & 56.08\% (N/A) &5000\\
9BW  &  1.1 & \checkmark      & 14,000 & 68.11\% (N/A) &8000\\
9F  & 1.6 & \checkmark       &  16,000 & 60.92\% (5.75\%)  &3000\\
9F,\,l & 2.1 &  \checkmark       &  16,000 & 76.68\% (6.63\%) &5000\\
9.25  & 2.7 & \checkmark       & 13,000 &72.55\% (3.06\%) &8000\\
9.5  & 2.1 & \checkmark       & 11,000 & 76.35\% (0.72\%) &8000\\
11 & 4.5 & \checkmark       &  11,000 & 85.02\% (13.55\%) &6000\\ 
12.25  & 2.0 & \xmark       & \,\,\,\, & 94.55\% (25.71\%) &8000\\
14 & 1.0 &  \xmark        & \,\,\,  & 75.82\% (N/A) &8000\\
15.01 & 2.0 &  \checkmark       & 6,000 & 84.46\% (5.12\%) &8000\\
23  & 6.2 & \checkmark\      &  6,000   & 85.74\% (8.50\%) &8000\\
\hline
  \end{tabular}

  \begin{flushleft}\large{Table of some the features of the 3D simulation. We include the simulation time, in seconds post-bounce, the explosion outcome, the asymptotic shock velocity, and the fraction of the GW energy radiated via the f/g-mode. Models with a checkmark explode, and models with an \xmark\, do not explode. The various 9 solar mass models were run on Theta (ALCC), Blue Waters (NCSA), and Frontera (TACC), respectively. 9F,l is a longer simulation of the 9-M$_{\odot}$ progenitor, also done on Frontera. The f/g-mode energy is the gravitational-wave energy within $\pm100$ Hz of its central frequency. ``Total" means for all time, while ``Late" means after 1.5 seconds after bounce. After the mode repulsion, almost all the power is the an unmixed f-mode. Also in the table are the lowest Nyquist sampling frequencies for each run. The behavior at frequencies below these should not be compromised by Nyquist errors.} \end{flushleft}
  \label{sn_tab}
\end{table*}

\begin{figure*}
    \centering
    \includegraphics[width=.75\textwidth]{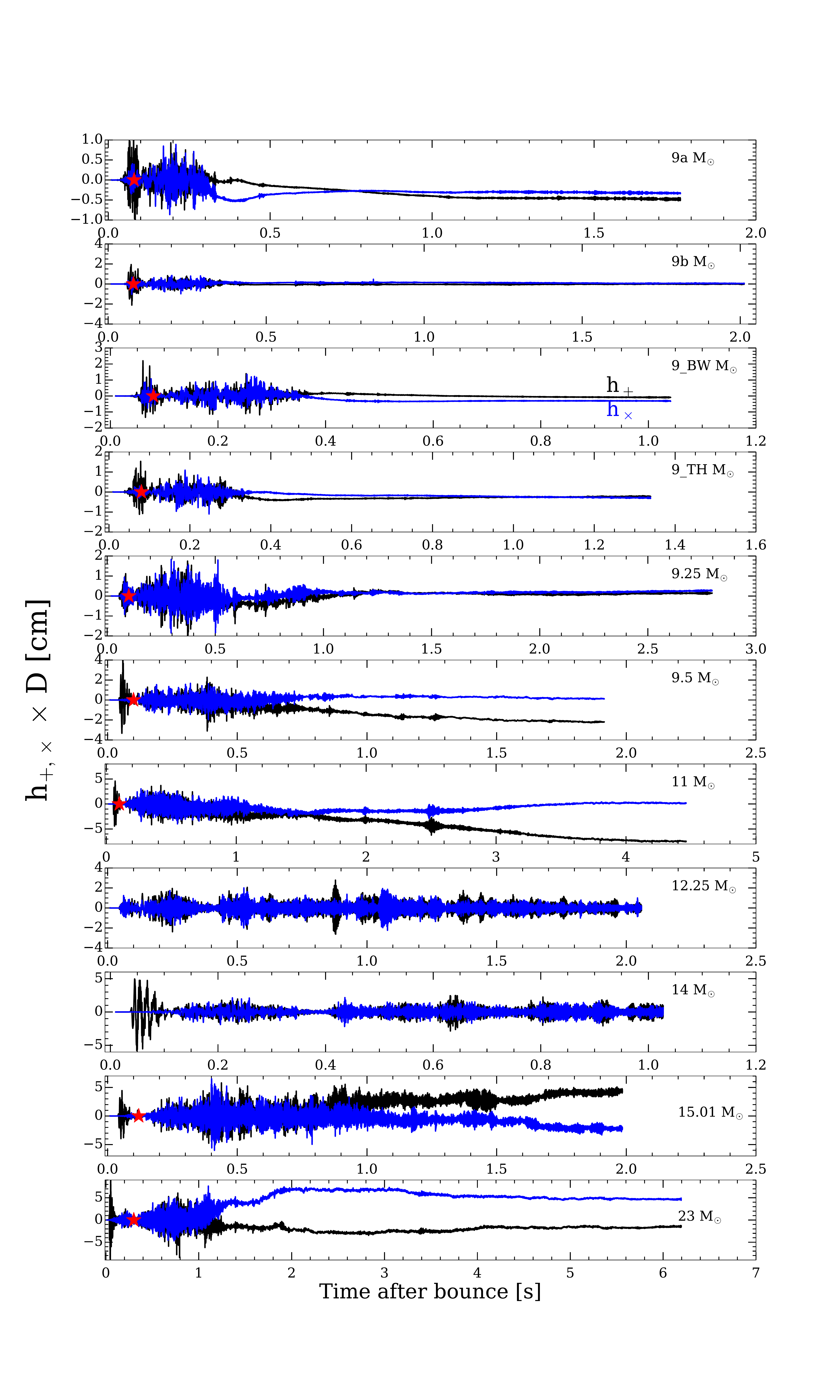}\vspace{-50pt}
    \caption{The gravitational-wave strain, from matter motions, times distance (product in cm) for all the 3D models studied in this paper as a function of time after bounce (in seconds) observed along the positive x-axis in the simulation frame. Both cross (blue) and plus (black) polarizations are illustrated. The red star indicates the approximate time of the onset of explosion, defined here as when the stalled shock radius changes concavity at $\sim$150 km and accelerates. See Figure \ref{fig:rs1}. }
    \label{fig:strain}
\end{figure*}

\begin{figure*}
    \centering
    \includegraphics[width=.75\textwidth]{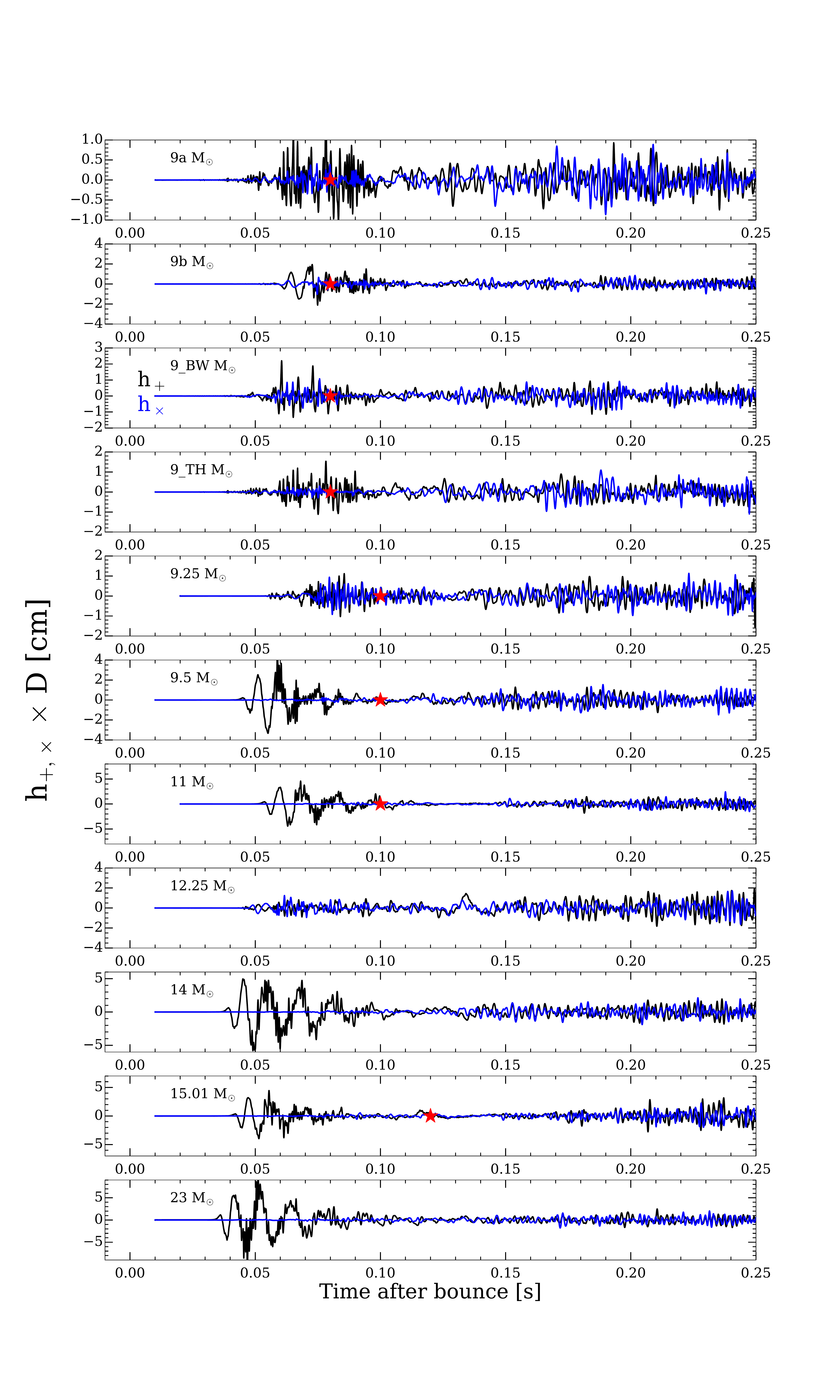}\vspace{-50pt}
    \caption{Same as Figure\,\ref{fig:strain}, but now focusing on the matter strain during the first 250 ms after core bounce to illustrate the prompt convection. Note the stately oscillation present in most models of $\sim$10 ms periods visible in the $+$ polarization and absent in the $\times$ polarization. This is a generic feature indicative of the strain polarization geometry. Note that only model 9a has any perturbations on bounce. }
    \label{fig:strain_early}
\end{figure*}

\begin{figure*}
    \centering
    \includegraphics[width=.8\textwidth]{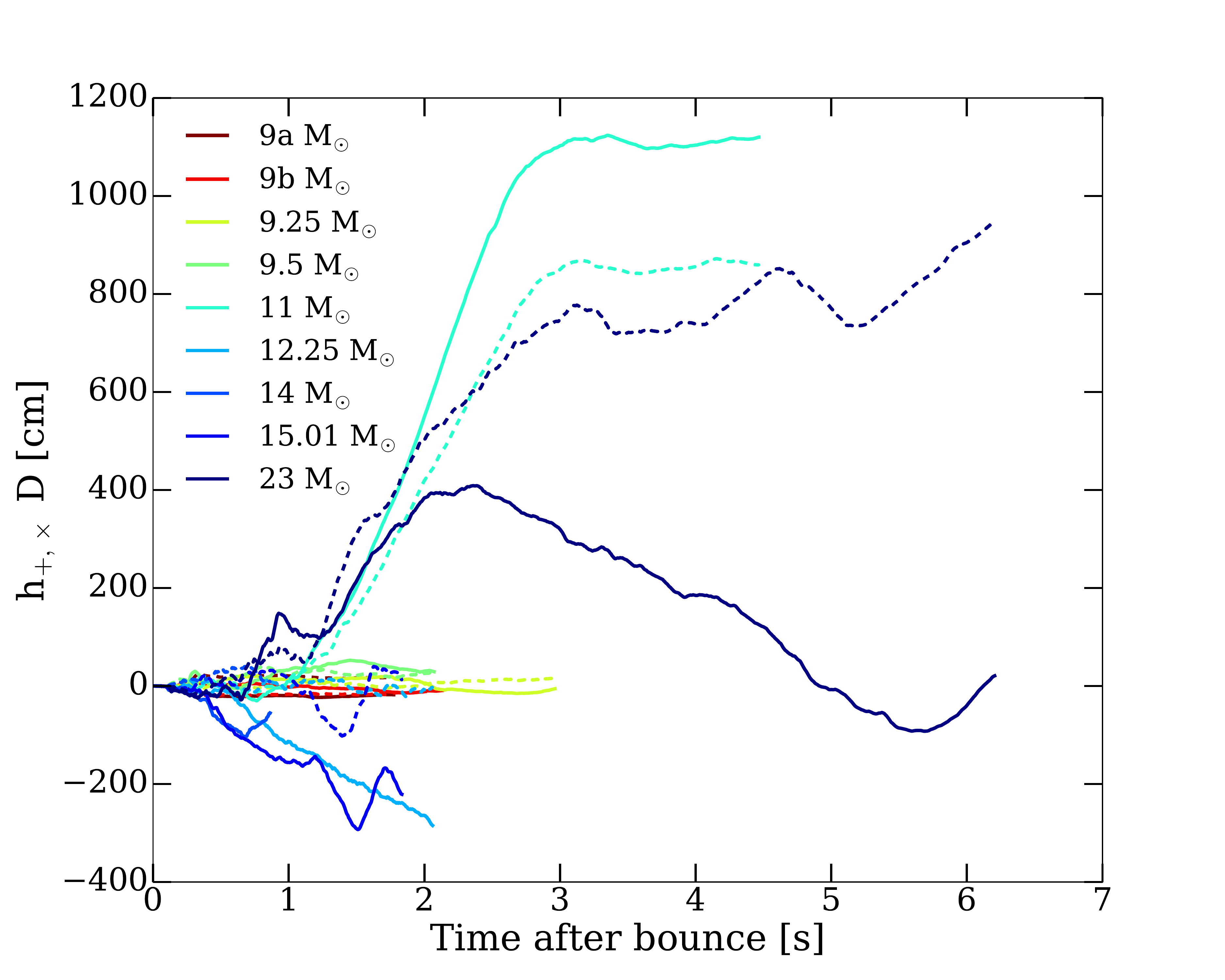}
    \caption{Same as Figure\,\ref{fig:strain}, but for the gravitational-wave memory due to neutrino emission anisotropy. Because of the cumulative nature of the neutrino GW memory, the strain is both larger and shows less variation on small timescales when compared to the matter-driven strain. We reach strains as high as 1000 cm for the 11- and 23-M$_{\odot}$ models. Note that this is consistent with earlier work \protect\citep{vartanyan2020} showing continued growth with time of the memory strain. The strain growth rate also accelerates, showing upwards of 600 cm growth from $\sim$2 to $\sim$3 seconds for the 11-M$_{\odot}$ progenitor. This emphasizes our essential point that models should be carried out to late times to fully capture not only the explosion dynamics, but the emerging GW signatures. The 11-M$_{\odot}$ progenitor has the highest neutrino memory strain among those models we present in this paper. Otherwise, we see a general correlation with mass, with the low-mass progenitors having the smallest neutrino memory. The non-exploding 12.25- and 14-M$_{\odot}$ models show significant evolution and magnitude in their memory signatures.}
    \label{fig:strain_nu}
\end{figure*}

\begin{figure*}
    \centering
    \includegraphics[width=0.9\textwidth]{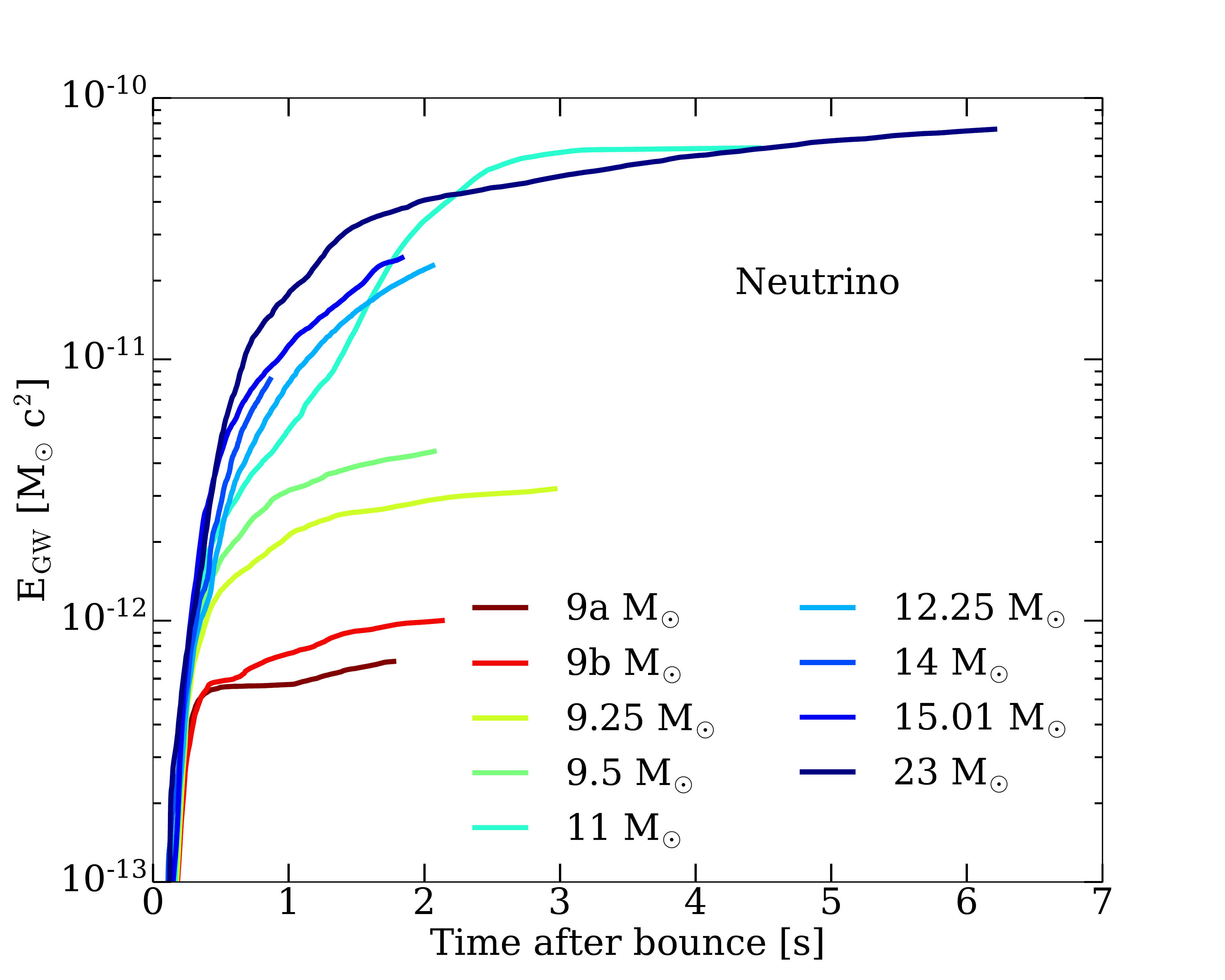}
    \caption{Same as Figure\,\ref{fig:EGW}, but for the radiated gravitational-wave energy due to the neutrino anisotropy. As seen in Figure\,\ref{fig:strain_nu}, unlike the matter contribution, the neutrino anisotropy continues to contribute to the gravitational signal past six seconds (for our longest simulation), emphasizing the presence of sustained neutrino emission asymmetry. We see again at least two orders of magnitude of growth with time after $\sim$50 ms in the associated GW energy, this time in the neutrino memory component. The 23-M$_{\odot}$ progenitor has neutrino GW memory energies only slightly higher than the matter GW components for the least energetic, lowest-mass progenitors. Importantly, for the same progenitor model, the late time neutrino-sourced GW energies, albeit still growing, are more than two orders of magnitude smaller than the corresponding matter components.}
    \label{fig:EGW_neutrino}
\end{figure*}

\begin{figure*}
    \centering
    \includegraphics[width=0.95\textwidth]{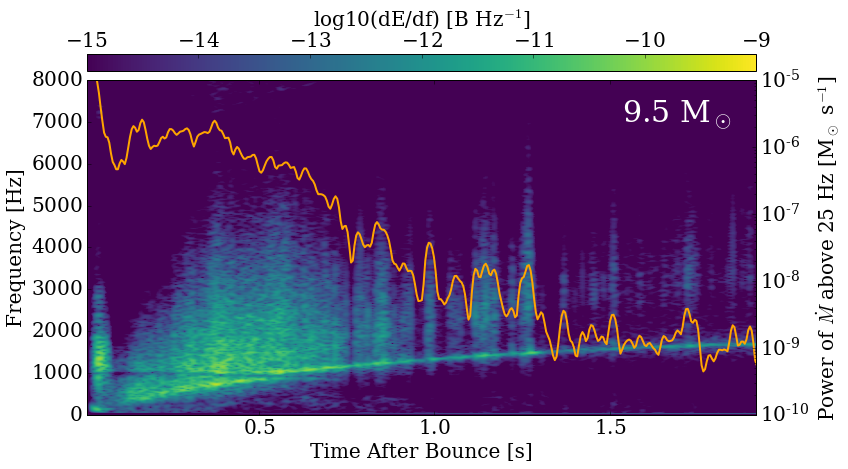}
    \includegraphics[width=0.95\textwidth]{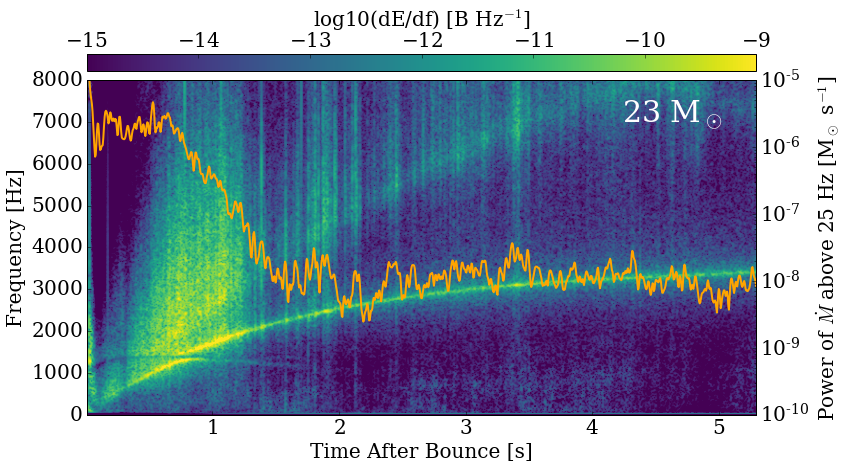}
    \caption{In this figure, we show the correlation between the gravitational-wave energy spectrogram and the accretion rate power in frequency components with $f>25$ Hz for two representative models. The background spectra in this figure are the same as those plotted in Figure \ref{fig:EGW_spec} and \ref{fig:EGW_spec2}, but with different color-bar ranges so that the late time structures can be more clearly seen. The orange curve shows the accretion rate ``power" in frequency components with $f>25$ Hz. The 25-Hz filter is chosen here because the spectrogram is calculated using a 0.04s-wide sliding window. The accretion rates are measured at a radius of 25 km, and the PNS radii of the two models fall below this value at $\sim$0.9 seconds and $\sim$1.0 seconds after bounce, respectively. Note also that this colormap for the 23-M$_{\odot}$ model spectrogram reveals a weak signal below the f-mode associated with the avoided crossing that may be the bumped g-mode and that seems to continue beyond $\sim$1.5 seconds. See text in \S\ref{excitation} for a discussion.} 
    \label{fig:GW-mdot-corr}
\end{figure*}

\end{document}